\documentclass[10pt]{article}
\makeindex
\usepackage{epsf}
\usepackage{axodraw}
\usepackage{epsfig}                            
\usepackage{graphicx}
\usepackage{rotate}
\usepackage{latexsym}
\renewcommand{\cal}{\mathcal}
\begin{document}
\marginparwidth 3cm
%
%
%
%
\newcommand{\nl}{\nonumber\\}
\newcommand{\nn}{\nonumber}
\newcommand{\ds}{\displaystyle}
\newcommand{\mpar}[1]{{\marginpar{\hbadness10000%
                      \sloppy\hfuzz10pt\boldmath\bf#1}}%
                      \typeout{marginpar: #1}\ignorespaces}
\def\mnew{\mpar{\hfil NEW \hfil}\ignorespaces}
\newcommand{\lpar}{\left(}                            
\newcommand{\rpar}{\right)} 
\newcommand{\lrbr}{\left[}
\newcommand{\rrbr}{\right]}
\newcommand{\lcbr}{\left\{}
\newcommand{\rcbr}{\right\}} 
\newcommand{\rbrak}[1]{\lrbr#1\rrbr}
\newcommand{\bq}{\begin{equation}}                    
\newcommand{\eq}{\end{equation}}
\newcommand{\bqa}{\begin{eqnarray}}
\newcommand{\eqa}{\end{eqnarray}}
\newcommand{\ba}[1]{\begin{array}{#1}}
\newcommand{\ea}{\end{array}}
\newcommand{\ben}{\begin{enumerate}}
\newcommand{\een}{\end{enumerate}}
\newcommand{\bei}{\begin{itemize}}
\newcommand{\eei}{\end{itemize}}
\newcommand{\bec}{\begin{center}}
\newcommand{\eec}{\end{center}}
\newcommand{\eqn}[1]{Eq.(\ref{#1})}
\newcommand{\eqns}[2]{Eqs.(\ref{#1}--\ref{#2})}
\newcommand{\eqnss}[1]{Eqs.(\ref{#1})}
\newcommand{\eqnsc}[2]{Eqs.(\ref{#1},~\ref{#2})}
\newcommand{\tbn}[1]{Tab.~\ref{#1}}
\newcommand{\tbns}[2]{Tabs.~\ref{#1}--\ref{#2}}
\newcommand{\tbnsc}[2]{Tabs.~\ref{#1},~\ref{#2}}
\newcommand{\fig}[1]{Fig.~\ref{#1}}
\newcommand{\figs}[2]{Figs.~\ref{#1}--\ref{#2}}
\newcommand{\sect}[1]{Sect.~\ref{#1}}
\newcommand{\subsect}[1]{Sub-Sect.~\ref{#1}}
%
%
\newcommand{\TeV}{\;\mathrm{TeV}}                     
\newcommand{\GeV}{\;\mathrm{GeV}}
\newcommand{\MeV}{\;\mathrm{MeV}}
\newcommand{\nb}{\;\mathrm{nb}}
\newcommand{\pb}{\;\mathrm{pb}}
\newcommand{\fb}{\;\mathrm{fb}}
\def\Re{\mathop{\operator@font Re}\nolimits}
\def\Im{\mathop{\operator@font Im}\nolimits}
\newcommand{\ord}[1]{{\cal O}\lpar#1\rpar}
\newcommand{\group}{SU(2)\otimes U(1)}
\newcommand{\ib}{i}
\newcommand{\asums}[1]{\sum_{#1}}
\newcommand{\asumt}[2]{\sum_{#1}^{#2}}
\newcommand{\asum}[3]{\sum_{#1=#2}^{#3}}
%
%
\newcommand{\tmi}{\times 10^{-1}}
\newcommand{\tmii}{\times 10^{-2}}
\newcommand{\tmiii}{\times 10^{-3}}
\newcommand{\tmiv}{\times 10^{-4}}
\newcommand{\tmfv}{\times 10^{-5}}
\newcommand{\tmfvi}{\times 10^{-6}}
\newcommand{\tmfvii}{\times 10^{-7}}
\newcommand{\tmfviii}{\times 10^{-8}}
\newcommand{\tmfix}{\times 10^{-9}}
\newcommand{\tmfx}{\times 10^{-10}}
%
%
\newcommand{\fer}{{\rm{fer}}}
\newcommand{\bos}{{\rm{bos}}}
\newcommand{\lep}{{l}}
\newcommand{\had}{{h}}
\newcommand{\gen}{\rm{g}}
\newcommand{\dbl}{\rm{d}}
\newcommand{\philone}{\phi}
\newcommand{\philoneb}{\phi_{0}}
\newcommand{\phiind}[1]{\phi_{#1}}
\newcommand{\gBi}[2]{B_{#1}^{#2}}
\newcommand{\gBn}[1]{B_{#1}}
%
%
\newcommand{\ph}{\gamma}
\newcommand{\ab}{A}
\newcommand{\abr}{A^r}
\newcommand{\abb}{A^{0}}
\newcommand{\abi}[1]{A_{#1}}
\newcommand{\abri}[1]{A^r_{#1}}
\newcommand{\abbi}[1]{A^{0}_{#1}}
\newcommand{\wb}{W}            
\newcommand{\wbi}[1]{W_{#1}}           
\newcommand{\wbp}{W^{+}}
\newcommand{\wbm}{W^{-}}
\newcommand{\wbpm}{W^{\pm}}
\newcommand{\wbpi}[1]{W^{+}_{#1}}
\newcommand{\wbmi}[1]{W^{-}_{#1}}
\newcommand{\wbpmi}[1]{W^{\pm}_{#1}}
\newcommand{\wbli}[1]{W^{[+}_{#1}}
\newcommand{\wbri}[1]{W^{-]}_{#1}}
\newcommand{\zb}{Z}
\newcommand{\zbi}[1]{Z_{#1}}
\newcommand{\vb}{V}
\newcommand{\vbi}[1]{V_{#1}}      
\newcommand{\vbiv}[1]{V^{*}_{#1}}      
\newcommand{\Pb}{P}
\newcommand{\Sb}{S}
\newcommand{\Bb}{B}
%
%
\newcommand{\hk}{K}
\newcommand{\hKi}[1]{K_{#1}}
\newcommand{\hkg}{\phi}
\newcommand{\hkn}{\phi^{0}}                 
\newcommand{\hkp}{\phi^{+}}
\newcommand{\hkm}{\phi^{-}}
\newcommand{\hkpm}{\phi^{\pm}}
\newcommand{\hkmp}{\phi^{\mp}}
\newcommand{\hki}[1]{\phi^{#1}}
\newcommand{\hb}{H}
\newcommand{\hbi}[1]{H_{#1}}
\newcommand{\hkl}{\phi^{[+\cgfi\cgfi}}
\newcommand{\hkr}{\phi^{-]}}
%
%
\newcommand{\fpx}{X}
\newcommand{\fpy}{Y}
\newcommand{\fpxp}{X^+}
\newcommand{\fpxm}{X^-}
\newcommand{\fpxpm}{X^{\pm}}
\newcommand{\fpxi}[1]{X^{#1}}
\newcommand{\fpyZ}{Y^{\ssZ}}
\newcommand{\fpyA}{Y^{\ssA}}
\newcommand{\fpyZA}{Y_{\ssZ,\ssA}}
\newcommand{\fpbxi}[1]{{\overline{X}}^{#1}}
\newcommand{\fpbyZ}{{\overline{Y}}^{\ssZ}}
\newcommand{\fpbyA}{{\overline{Y}}^{\ssA}}
\newcommand{\fpbyZA}{{\overline{Y}}^{\ssZ,\ssA}}
%
%
\newcommand{\Flone}{F}
\newcommand{\fpsi}{\psi}
\newcommand{\fpsii}[1]{\psi^{#1}}
\newcommand{\fpsib}{\psi^{0}}
\newcommand{\fpsir}{\psi^r}
\newcommand{\fpsiL}{\psi_{_L}}
\newcommand{\fpsiR}{\psi_{_R}}
\newcommand{\fpsiLi}[1]{\psi_{_L}^{#1}}
\newcommand{\fpsiRi}[1]{\psi_{_R}^{#1}}
\newcommand{\fpsiLbi}[1]{\psi_{_{0L}}^{#1}}
\newcommand{\fpsiRbi}[1]{\psi_{_{0R}}^{#1}}
\newcommand{\fpsiLR}{\psi_{_{L,R}}}
\newcommand{\fbpsi}{{\overline{\psi}}}
\newcommand{\fbpsii}[1]{{\overline{\psi}}^{#1}}
\newcommand{\fbpsir}{{\overline{\psi}}^r}
\newcommand{\fbpsiL}{{\overline{\psi}}_{_L}}
\newcommand{\fbpsiR}{{\overline{\psi}}_{_R}}
\newcommand{\fbpsiLi}[1]{\overline{\psi_{_L}}^{#1}}
\newcommand{\fbpsiRi}[1]{\overline{\psi_{_R}}^{#1}}
\newcommand{\fe}{e}
\newcommand{\ff}{f}
\newcommand{\fep}{e^{+}}
\newcommand{\fem}{e^{-}}
\newcommand{\fepm}{e^{\pm}}
\newcommand{\fp}{f^{+}}
\newcommand{\fm}{f^{-}}
\newcommand{\fhp}{h^{+}}
\newcommand{\fhm}{h^{-}}
\newcommand{\fh}{h}
\newcommand{\flm}{\mu}
\newcommand{\flmp}{\mu^{+}}
\newcommand{\flmm}{\mu^{-}}
\newcommand{\fll}{l}
\newcommand{\fllp}{l^{+}}
\newcommand{\fllm}{l^{-}}
\newcommand{\flt}{\tau}
\newcommand{\fltp}{\tau^{+}}
\newcommand{\fltm}{\tau^{-}}
\newcommand{\fq}{q}
\newcommand{\fqi}[1]{\fq_{#1}}
\newcommand{\bfqi}[1]{\barq_{#1}}
\newcommand{\ffQ}{Q}
\newcommand{\fu}{u}
\newcommand{\fd}{d}
\newcommand{\fc}{c}
\newcommand{\fs}{s}
\newcommand{\fqp}{q'}
\newcommand{\fup}{u'}
\newcommand{\fdp}{d'}
\newcommand{\fcp}{c'}
\newcommand{\fsp}{s'}
\newcommand{\fdpp}{d''}
\newcommand{\ffi}[1]{f_{#1}}
\newcommand{\bffi}[1]{{\overline{f}}_{#1}}
\newcommand{\ffpi}[1]{f'_{#1}}
\newcommand{\bffpi}[1]{{\overline{f}}'_{#1}}
\newcommand{\ft}{t}
\newcommand{\ffb}{b}
\newcommand{\ffp}{f'}
\newcommand{\fft}{{\tilde{f}}}
\newcommand{\fl}{l}
\newcommand{\fli}[1]{\fl_{#1}}
\newcommand{\fnu}{\nu}
\newcommand{\fU}{U}
\newcommand{\fD}{D}
\newcommand{\fUc}{\overline{U}}
\newcommand{\fDc}{\overline{D}}
\newcommand{\fnul}{\nu_l}
\newcommand{\fnue}{\nu_e}
\newcommand{\fnum}{\nu_{\mu}}
\newcommand{\fnut}{\nu_{\tau}}
\newcommand{\fbe}{{\overline{e}}}
\newcommand{\fbu}{{\overline{u}}}
\newcommand{\fbd}{{\overline{d}}}
\newcommand{\fbf}{{\overline{f}}}
\newcommand{\fbfp}{{\overline{f}}'}
\newcommand{\fbl}{{\overline{l}}}
\newcommand{\fbnu}{{\overline{\nu}}}
\newcommand{\fbnul}{{\overline{\nu}}_{\fl}}
\newcommand{\fbnue}{{\overline{\nu}}_{\fe}}
\newcommand{\fbnum}{{\overline{\nu}}_{\flm}}
\newcommand{\fbnut}{{\overline{\nu}}_{\flt}}
\newcommand{\fuL}{u_{_L}}
\newcommand{\fdL}{d_{_L}}
\newcommand{\ffL}{f_{_L}}
\newcommand{\fbuL}{{\overline{u}}_{_L}}
\newcommand{\fbdL}{{\overline{d}}_{_L}}
\newcommand{\fbfL}{{\overline{f}}_{_L}}
\newcommand{\fuR}{u_{_R}}
\newcommand{\fdR}{d_{_R}}
\newcommand{\ffR}{f_{_R}}
\newcommand{\fbuR}{{\overline{u}}_{_R}}
\newcommand{\fbdR}{{\overline{d}}_{_R}}
\newcommand{\fbfR}{{\overline{f}}_{_R}}
%
%
\newcommand{\barf}{\overline f}                
\newcommand{\barl}{\overline l}
\newcommand{\barq}{\overline q}
\newcommand{\barqp}{\overline{q}'}
\newcommand{\barb}{\overline b}
\newcommand{\bart}{\overline t}
\newcommand{\barc}{\overline c}
\newcommand{\baru}{\overline u}
\newcommand{\bard}{\overline d}
\newcommand{\bars}{\overline s}
\newcommand{\barv}{\overline v}
\newcommand{\barnu}{\overline{\nu}}
\newcommand{\barne}{\overline{\nu}_{\fe}}
\newcommand{\barnm}{\overline{\nu}_{\flm}}
\newcommand{\barnt}{\overline{\nu}_{\flt}}
%
%
\newcommand{\glu}{g}
%
%
\newcommand{\prot}{p}
\newcommand{\aprot}{{\bar{p}}}
\newcommand{\Nucln}{N}
%
%
\newcommand{\tM}{{\tilde M}}
\newcommand{\tMs}{{\tilde M}^2}
\newcommand{\tW}{{\tilde \Gamma}}
\newcommand{\tWs}{{\tilde\Gamma}^2}
\newcommand{\fphi}{\phi}
\newcommand{\fJpsi}{J/\psi}
\newcommand{\fgpsi}{\psi}
\newcommand{\Glone}{\Gamma}
\newcommand{\Gloni}[1]{\Gamma_{#1}}
\newcommand{\Glones}{\Gamma^2}
\newcommand{\Glonec}{\Gamma^3}
\newcommand{\glone}{\gamma}
\newcommand{\glones}{\gamma^2}
\newcommand{\gloneq}{\gamma^4}
\newcommand{\gloni}[1]{\gamma_{#1}}
\newcommand{\glonis}[1]{\gamma^2_{#1}}
\newcommand{\Grest}[2]{\Gamma_{#1}^{#2}}
\newcommand{\grest}[2]{\gamma_{#1}^{#2}}
\newcommand{\resampl}{A_{_R}}
\newcommand{\resasyi}[1]{{\cal{A}}_{#1}}
\newcommand{\sSrest}[1]{\sigma_{#1}}
\newcommand{\Srest}[2]{\sigma_{#1}\lpar{#2}\rpar}
\newcommand{\Gdist}[1]{{\cal{G}}\lpar{#1}\rpar}
\newcommand{\sGdist}{{\cal{G}}}
\newcommand{\Aarea}{A_{0}}
\newcommand{\Aareai}[1]{{\cal{A}}\lpar{#1}\rpar}
\newcommand{\sAarea}{{\cal{A}}}
\newcommand{\resolw}{\sigma_{\Energ}}
\newcommand{\chizer}{\chi_{_0}}
\newcommand{\ini}{\rm{in}}
\newcommand{\fin}{\rm{fin}}
\newcommand{\ifi}{\rm{if}}
\newcommand{\ipf}{\rm{i+f}}
\newcommand{\tot}{\rm{tot}}
\newcommand{\Bac}{Q}
\newcommand{\Res}{R}
\newcommand{\Int}{I}
\newcommand{\NRe}{NR}
\newcommand{\ratoe}{\delta}
\newcommand{\ratoes}{\delta^2}
%
%
\newcommand{\Fbox}[2]{f^{\rm{box}}_{#1}\lpar{#2}\rpar}
\newcommand{\Dbox}[2]{\delta^{\rm{box}}_{#1}\lpar{#2}\rpar}
\newcommand{\Bbox}[3]{{\cal{B}}_{#1}^{#2}\lpar{#3}\rpar}
%
%
\newcommand{\phm}{\lambda}
\newcommand{\phms}{\lambda^2}
\newcommand{\mV}{M_{_V}}
\newcommand{\mw}{M_{_W}}
\newcommand{\mX}{M_{_X}}
\newcommand{\mY}{M_{_Y}}
\newcommand{\LM}{M}
\newcommand{\mz}{M_{_Z}}
\newcommand{\bzm}{M_{_0}}
\newcommand{\mh}{M_{_H}}
\newcommand{\bhm}{M_{_{0H}}}
\newcommand{\mf}{m_f}
\newcommand{\mfp}{m_{f'}}
\newcommand{\mfh}{m_{h}}
\newcommand{\mt}{m_t}
\newcommand{\me}{m_e}
\newcommand{\mm}{m_{\mu}}
\newcommand{\mtau}{m_{\tau}}
\newcommand{\muq}{m_u}
\newcommand{\md}{m_d}
\newcommand{\muqp}{m'_u}
\newcommand{\mdqp}{m'_d}
\newcommand{\mc}{m_c}
\newcommand{\ms}{m_s}
\newcommand{\mb}{m_b}
\newcommand{\mup}{M_u}                              
\newcommand{\mdp}{M_d}
\newcommand{\mcp}{M_c}
\newcommand{\msp}{M_s}
\newcommand{\mbp}{M_b}
%
%
\newcommand{\mls}{m^2_l}
\newcommand{\mVs}{M^2_{_V}}
\newcommand{\mws}{M^2_{_W}}
\newcommand{\mwc}{M^3_{_W}}
\newcommand{\LMs}{M^2}
\newcommand{\LMc}{M^3}
\newcommand{\mzs}{M^2_{_Z}}
\newcommand{\mzc}{M^3_{_Z}}
\newcommand{\bzms}{M^2_{_0}}
\newcommand{\bzmc}{M^3_{_0}}
\newcommand{\bhms}{M^2_{_{0H}}}
\newcommand{\mhs}{M^2_{_H}}
\newcommand{\mfs}{m^2_f}
\newcommand{\mfc}{m^3_f}
\newcommand{\mfps}{m^2_{f'}}
\newcommand{\mfhs}{m^2_{h}}
\newcommand{\mfpc}{m^3_{f'}}
\newcommand{\mts}{m^2_t}
\newcommand{\mes}{m^2_e}
\newcommand{\mms}{m^2_{\mu}}
\newcommand{\mmc}{m^3_{\mu}}
\newcommand{\mmfour}{m^4_{\mu}}
\newcommand{\mmf}{m^5_{\mu}}
\newcommand{\mmfive}{m^5_{\mu}}
\newcommand{\mmsix}{m^6_{\mu}}
\newcommand{\mminv}{\frac{1}{m_{\mu}}}
\newcommand{\mtaus}{m^2_{\tau}}
\newcommand{\mus}{m^2_u}
\newcommand{\mds}{m^2_d}
\newcommand{\muqps}{m'^2_u}
\newcommand{\mdqps}{m'^2_d}
\newcommand{\mcs}{m^2_c}
\newcommand{\mss}{m^2_s}
\newcommand{\mbs}{m^2_b}
\newcommand{\mups}{M^2_u}
\newcommand{\mdps}{M^2_d}
\newcommand{\mcps}{M^2_c}
\newcommand{\msps}{M^2_s}
\newcommand{\mbps}{M^2_b}
%
%
\newcommand{\muf}{\mu_{\ff}}
\newcommand{\mufs}{\mu^2_{\ff}}
\newcommand{\mufq}{\mu^4_{\ff}}
\newcommand{\mufx}{\mu^6_{\ff}}
\newcommand{\muz}{\mu_{_{\zb}}}
\newcommand{\muw}{\mu_{_{\wb}}}
\newcommand{\mut}{\mu_{\ft}}
\newcommand{\muzs}{\mu^2_{_{\zb}}}
\newcommand{\muws}{\mu^2_{_{\wb}}}
\newcommand{\muts}{\mu^2_{\ft}}
\newcommand{\muSW}{\mu^2_{_{\wb}}}
\newcommand{\muwq}{\mu^4_{_{\wb}}}
\newcommand{\muwsx}{\mu^6_{_{\wb}}}
\newcommand{\muwms}{\mu^{-2}_{_{\wb}}}
\newcommand{\muhs}{\mu^2_{_{\hb}}}
\newcommand{\muhq}{\mu^4_{_{\hb}}}
\newcommand{\muhsx}{\mu^6_{_{\hb}}}
\newcommand{\mutq}{\mu^4_{_{\hb}}}   
\newcommand{\mutsx}{\mu^6_{_{\hb}}}  
\newcommand{\muL}{\mu}
\newcommand{\muS}{\mu^2}
\newcommand{\muQ}{\mu^4}
\newcommand{\muizs}{\mu^2_{0}}
\newcommand{\muizq}{\mu^4_{0}}
\newcommand{\muis}{\mu^2_{1}}
\newcommand{\muiis}{\mu^2_{2}}
\newcommand{\muiiis}{\mu^2_{3}}
\newcommand{\muii}[1]{\mu_{#1}}
\newcommand{\muisi}[1]{\mu^2_{#1}}
\newcommand{\muiqi}[1]{\mu^4_{#1}}
\newcommand{\muixi}[1]{\mu^6_{#1}}
\newcommand{\zm}{z_m}
\newcommand{\ri}[1]{r_{#1}}
\newcommand{\xw}{x_w}
\newcommand{\xws}{x^2_w}
\newcommand{\xwc}{x^3_w}
\newcommand{\xth}{x_t}
\newcommand{\xths}{x^2_t}
\newcommand{\xthc}{x^3_t}
\newcommand{\xthf}{x^4_t}
\newcommand{\xthv}{x^5_t}
\newcommand{\xthx}{x^6_t}
\newcommand{\xh}{x_h}
\newcommand{\xhs}{x^2_h}
\newcommand{\xhc}{x^3_h}
\newcommand{\Rl}{R_{\fl}}
\newcommand{\Rb}{R_{\ffb}}
\newcommand{\Rc}{R_{\fc}}
%
%
\newcommand{\mwq}{M^4_{_\wb}}
\newcommand{\mwf}{M^4_{_\wb}}
\newcommand{\LMq}{M^4}
\newcommand{\mzq}{M^4_{_Z}}
\newcommand{\bzmq}{M^4_{_0}}
\newcommand{\mhq}{M^4_{_H}}
\newcommand{\mfq}{m^4_f}
\newcommand{\mfpq}{m^4_{f'}}
\newcommand{\mtq}{m^4_t}
\newcommand{\meq}{m^4_e}
\newcommand{\mmq}{m^4_{\mu}}
\newcommand{\mtauq}{m^4_{\tau}}
\newcommand{\muqq}{m^4_u}
\newcommand{\mdq}{m^4_d}
\newcommand{\mcq}{m^4_c}
\newcommand{\msq}{m^4_s}
\newcommand{\mbq}{m^4_b}
\newcommand{\mupq}{M^4_u}
\newcommand{\mdpq}{M^4_d}
\newcommand{\mcpq}{M^4_c}
\newcommand{\mspq}{M^4_s}
\newcommand{\mbpq}{M^4_b}
%
%
\newcommand{\mwx}{M^6_{_W}}
\newcommand{\mzx}{M^6_{_Z}}
\newcommand{\mfx}{m^6_f}
\newcommand{\mfpx}{m^6_{f'}}
\newcommand{\LMx}{M^6}
%
%
\newcommand{\mer}{m_{er}}
\newcommand{\mlep}{m_l}
\newcommand{\mleps}{m^2_l}
\newcommand{\mone}{m_1}
\newcommand{\mtwo}{m_2}
\newcommand{\mtre}{m_3}
\newcommand{\mfor}{m_4}
\newcommand{\mlone}{m}
\newcommand{\mloneb}{\bar{m}}
\newcommand{\mind}[1]{m_{#1}}
\newcommand{\mones}{m^2_1}
\newcommand{\mtwos}{m^2_2}
\newcommand{\mtres}{m^2_3}
\newcommand{\mfors}{m^2_4}
\newcommand{\mlones}{m^2}
\newcommand{\minds}[1]{m^2_{#1}}
\newcommand{\moneq}{m^4_1}
\newcommand{\mtwoq}{m^4_2}
\newcommand{\mtreq}{m^4_3}
\newcommand{\mforq}{m^4_4}
\newcommand{\mloneq}{m^4}
\newcommand{\mindq}[1]{m^4_{#1}}
\newcommand{\mlonev}{m^5}
\newcommand{\mindv}[1]{m^5_{#1}}
\newcommand{\monex}{m^6_1}
\newcommand{\mtwox}{m^6_2}
\newcommand{\mtrex}{m^6_3}
\newcommand{\mforx}{m^6_4}
\newcommand{\mlonex}{m^6}
\newcommand{\mindx}[1]{m^6_{#1}}
\newcommand{\Mone}{M_1}
\newcommand{\Mtwo}{M_2}
\newcommand{\Mtre}{M_3}
\newcommand{\Mfor}{M_4}
\newcommand{\Mlone}{M}
\newcommand{\Mlonep}{M'}
\newcommand{\Miind}{M_i}
\newcommand{\Mind}[1]{M_{#1}}
\newcommand{\Minds}[1]{M^2_{#1}}
\newcommand{\Mindc}[1]{M^3_{#1}}
\newcommand{\Mindf}[1]{M^4_{#1}}
\newcommand{\Mones}{M^2_1}
\newcommand{\Mtwos}{M^2_2}
\newcommand{\Mtres}{M^2_3}
\newcommand{\Mfors}{M^2_4}
\newcommand{\Mlones}{M^2}
\newcommand{\Mloneps}{M'^2}
\newcommand{\Miinds}{M^2_i}
\newcommand{\Mlonec}{M^3}
\newcommand{\Monec}{M^3_1}
\newcommand{\Mtwoc}{M^3_2}
\newcommand{\Moneq}{M^4_1}
\newcommand{\Mtwoq}{M^4_2}
\newcommand{\Mtreq}{M^4_3}
\newcommand{\Mforq}{M^4_4}
\newcommand{\Mloneq}{M^4}
\newcommand{\Miindq}{M^4_i}
\newcommand{\Monex}{M^6_1}
\newcommand{\Mtwox}{M^6_2}
\newcommand{\Mtrex}{M^6_3}
\newcommand{\Mforx}{M^6_4}
\newcommand{\Mlonex}{M^6}
\newcommand{\Miindx}{M^6_i}
\newcommand{\meb}{m_0}
\newcommand{\mebs}{m^2_0}
%
%
\newcommand{\Mq }{M_q  }
\newcommand{\MqS}{M^2_q}
\newcommand{\Ms }{M_s  }
\newcommand{\MsS}{M^2_s}
\newcommand{\Mc }{M_c  }
\newcommand{\McS}{M^2_c}
\newcommand{\Mb }{M_b  }
\newcommand{\MbS}{M^2_b}
\newcommand{\Mt }{M_t  }
\newcommand{\MtS}{M^2_t}
%
%
\newcommand{\mq}{m_q}
\newcommand{\mqs}{m^2_q}
\newcommand{\mqS}{m^2_q}
\newcommand{\mqQ}{m^4_q}
\newcommand{\mqX}{m^6_q}
\newcommand{\mqp}{m'_q }
\newcommand{\mqpS}{m'^2_q}
\newcommand{\mqpQ}{m'^4_q}
%
%
\newcommand{\lL}{l}
\newcommand{\ls}{l^2}
\newcommand{\LL}{L}
\newcommand{\LcalL}{\cal{L}}
\newcommand{\LS}{L^2}
\newcommand{\LC}{L^3}
\newcommand{\LQ}{L^4}
\newcommand{\lw}{l_w}
\newcommand{\Lw}{L_w}
\newcommand{\Lws}{L^2_w}
\newcommand{\Lz}{L_z}
\newcommand{\Lzs}{L^2_z}
\newcommand{\Li}[1]{L_{#1}}
\newcommand{\Lis}[1]{L^2_{#1}}
\newcommand{\Lic}[1]{L^3_{#1}}
%
%
\newcommand{\sman}{s}
\newcommand{\tman}{t}
\newcommand{\uman}{u}
\newcommand{\smani}[1]{s_{#1}}
\newcommand{\bsmani}[1]{{\bar{s}}_{#1}}
\newcommand{\smans}{s^2}
\newcommand{\tmans}{t^2}
\newcommand{\umans}{u^2}
\newcommand{\shat}{{\hat s}}
\newcommand{\that}{{\hat t}}
\newcommand{\uhat}{{\hat u}}
\newcommand{\hq}{{\hat Q}}
%
%
\newcommand{\smanp}{s'}
\newcommand{\smanpi}[1]{s'_{#1}}
\newcommand{\tmanp}{t'}
\newcommand{\umanp}{u'}
\newcommand{\kappi}[1]{\kappa_{#1}}
\newcommand{\zetai}[1]{\zeta_{#1}}
%
%
%
\newcommand{\Phaspi}[1]{\Gamma_{#1}}
\newcommand{\rbetai}[1]{\beta_{#1}}
\newcommand{\ralphai}[1]{\alpha_{#1}}
\newcommand{\rbetais}[1]{\beta^2_{#1}}
\newcommand{\Lambdi}[1]{\Lambda_{#1}}
\newcommand{\Nomini}[1]{N_{#1}}
\newcommand{\smlone}{\frac{-\sman-\ib\ep}{\mlones}}
%
%
\newcommand{\theti}[1]{\theta_{#1}}
\newcommand{\delti}[1]{\delta_{#1}}
\newcommand{\phigi}[1]{\phi_{#1}}
\newcommand{\acoli}[1]{\xi_{#1}}
\newcommand{\scats}{s}
\newcommand{\scatss}{s^2}
\newcommand{\scatsi}[1]{s_{#1}}
\newcommand{\scatsis}[1]{s^2_{#1}}
\newcommand{\scatst}[2]{s_{#1}^{#2}}
\newcommand{\scatc}{c}
\newcommand{\scatcs}{c^2}
\newcommand{\scatci}[1]{c_{#1}}
\newcommand{\scatcis}[1]{c^2_{#1}}
\newcommand{\scatct}[2]{c_{#1}^{#2}}
\newcommand{\angamt}[2]{\gamma_{#1}^{#2}}
%
%
\newcommand{\Regia}{{\cal{R}}}
\newcommand{\Iconi}[2]{{\cal{I}}_{#1}\lpar{#2}\rpar}
\newcommand{\sIcon}[1]{{\cal{I}}_{#1}}
\newcommand{\betaf}{\beta_{\ff}}
\newcommand{\betafs}{\beta^2_{\ff}}
\newcommand{\Kfact}[2]{{\cal{K}}_{#1}\lpar{#2}\rpar}
%
%
\newcommand{\Struf}[4]{{\cal D}^{#1}_{#2}\lpar{#3;#4}\rpar}
\newcommand{\sStruf}[2]{{\cal D}^{#1}_{#2}}
\newcommand{\Fluxf}[2]{H\lpar{#1;#2}\rpar}
\newcommand{\Fluxfi}[4]{H_{#1}^{#2}\lpar{#3;#4}\rpar}
\newcommand{\sFluxf}{H}
\newcommand{\Bflux}[2]{{\cal{B}}_{#1}\lpar{#2}\rpar}
\newcommand{\bflux}[2]{{\cal{B}}_{#1}\lpar{#2}\rpar}
\newcommand{\Fluxd}[2]{D_{#1}\lpar{#2}\rpar}
\newcommand{\fluxd}[2]{C_{#1}\lpar{#2}\rpar}
\newcommand{\Fluxh}[4]{{\cal{H}}_{#1}^{#2}\lpar{#3;#4}\rpar}
\newcommand{\Sluxh}[4]{{\cal{S}}_{#1}^{#2}\lpar{#3;#4}\rpar}
\newcommand{\Fluxhb}[4]{{\overline{{\cal{H}}}}_{#1}^{#2}\lpar{#3;#4}\rpar}
\newcommand{\sFluxhb}{{\overline{{\cal{H}}}}}
\newcommand{\Sluxhb}[4]{{\overline{{\cal{S}}}}_{#1}^{#2}\lpar{#3;#4}\rpar}
\newcommand{\sSluxhb}[2]{{\overline{{\cal{S}}}}_{#1}^{#2}}
\newcommand{\fluxh}[4]{h_{#1}^{#2}\lpar{#3;#4}\rpar}
\newcommand{\fluxhs}[3]{h_{#1}^{#2}\lpar{#3}\rpar}
\newcommand{\sfluxhs}[2]{h_{#1}^{#2}}
\newcommand{\fluxhb}[4]{{\overline{h}}_{#1}^{#2}\lpar{#3;#4}\rpar}
\newcommand{\Strufd}[2]{D\lpar{#1;#2}\rpar}
%
%
\newcommand{\rMQ}[1]{r^2_{#1}}
\newcommand{\rMQs}[1]{r^4_{#1}}
\newcommand{\rf}{w_{\ff}}
\newcommand{\zf}{z_{\ff}}
\newcommand{\rfs}{w^2_{\ff}}
\newcommand{\zfs}{z^2_{\ff}}
\newcommand{\rfc}{w^3_{\ff}}
\newcommand{\zfc}{z^3_{\ff}}
\newcommand{\df}{d_{\ff}}
\newcommand{\rfp}{w_{\ffp}}
\newcommand{\rfps}{w^2_{\ffp}}
\newcommand{\rfpc}{w^3_{\ffp}}
\newcommand{\rt}{w_{\ft}}
\newcommand{\rts}{w^2_{\ft}}
\newcommand{\dt}{d_{\ft}}
\newcommand{\dts}{d^2_{\ft}}
\newcommand{\rh}{r_{h}}
\newcommand{\Lnrt}{\ln{\rt}}
\newcommand{\Rw}{R_{_{\wb}}}
\newcommand{\Rws}{R^2_{_{\wb}}}
\newcommand{\Rz}{R_{_{\zb}}}
\newcommand{\Rzp}{R^{+}_{_{\zb}}}
\newcommand{\Rzm}{R^{-}_{_{\zb}}}
\newcommand{\Rzs}{R^2_{_{\zb}}}
\newcommand{\Rzc}{R^3_{_{\zb}}}
\newcommand{\Rv}{R_{_{\vb}}}
\newcommand{\rhw}{r_{_{\wb}}}
\newcommand{\rhz}{r_{_{\zb}}}
\newcommand{\rhws}{r^2_{_{\wb}}}
\newcommand{\rhzs}{r^2_{_{\zb}}}
%
%
\newcommand{\vqrato}{z}
\newcommand{\vqrats}{w}
\newcommand{\vqratq}{w^2}
\newcommand{\seyrat}{z}
\newcommand{\sexrat}{w}
\newcommand{\sehrat}{h}
\newcommand{\sewrat}{w}
\newcommand{\sezrat}{z}
\newcommand{\zetav}{\zeta}
\newcommand{\zetavi}[1]{\zeta_{#1}}
\newcommand{\bpo}{\beta^2}
\newcommand{\bpos}{\beta^4}
\newcommand{\bpt}{{\tilde\beta}^2}
\newcommand{\lap}{\kappa}
\newcommand{\hw}{h_{_{\wb}}}
\newcommand{\hz}{h_{_{\zb}}}
%
%
\newcommand{\ec}{e}
\newcommand{\ecs}{e^2}
\newcommand{\ect}{e^3}
\newcommand{\ecq}{e^4}
\newcommand{\ecb}{e_{_0}}
\newcommand{\ecbs}{e^2_{_0}}
\newcommand{\ecbq}{e^4_{_0}}
\newcommand{\eci}[1]{e_{#1}}
\newcommand{\ecis}[1]{e^2_{#1}}
\newcommand{\hate}{{\hat e}}
\newcommand{\gss}{g_{_S}}
\newcommand{\gsss}{g^2_{_S}}
\newcommand{\gssb}{g^2_{_{S_0}}}
\newcommand{\als}{\alpha_{_S}}
\newcommand{\as}{a_{_S}}
\newcommand{\ass}{a^2_{_S}}
\newcommand{\gf}{G_{\ssF}}
\newcommand{\gfs}{G^2_{\ssF}}
\newcommand{\gb}{g} 
\newcommand{\gbi}[1]{g_{#1}}
\newcommand{\gbb}{g_{0}}
\newcommand{\gbs}{g^2}
\newcommand{\gbc}{g^3}
\newcommand{\gbf}{g^4}
\newcommand{\gpb}{g'}
\newcommand{\gpbs}{g'^2}
\newcommand{\vc}[1]{v_{#1}}
\newcommand{\ac}[1]{a_{#1}}
\newcommand{\vcc}[1]{v^*_{#1}}
\newcommand{\acc}[1]{a^*_{#1}}
\newcommand{\hatv}[1]{{\hat v}_{#1}}
\newcommand{\vcs}[1]{v^2_{#1}}
\newcommand{\acs}[1]{a^2_{#1}}
\newcommand{\gcv}[1]{g^{#1}_{\ssV}}
\newcommand{\gca}[1]{g^{#1}_{\ssA}}
\newcommand{\gcp}[1]{g^{+}_{#1}}
\newcommand{\gcm}[1]{g^{-}_{#1}}
\newcommand{\gcpm}[1]{g^{\pm}_{#1}}
\newcommand{\vci}[2]{v^{#2}_{#1}}
\newcommand{\aci}[2]{a^{#2}_{#1}}
\newcommand{\vceff}[1]{v^{#1}_{\rm{eff}}}
\newcommand{\hvc}[1]{\hat{v}_{#1}}
\newcommand{\hvcs}[1]{\hat{v}^2_{#1}}
\newcommand{\Vc}[1]{V_{#1}}
\newcommand{\Ac}[1]{A_{#1}}
\newcommand{\Vcs}[1]{V^2_{#1}}
\newcommand{\Acs}[1]{A^2_{#1}}
\newcommand{\vpa}[2]{\sigma_{#1}^{#2}}
\newcommand{\vma}[2]{\delta_{#1}^{#2}}
\newcommand{\vfw}{\sigma^{a}_{\ff}}
\newcommand{\vfpw}{\sigma^{a}_{\ffp}}
\newcommand{\vfwi}[1]{\sigma^{a}_{#1}}
\newcommand{\vfwsi}[1]{\lpar\sigma^{a}_{#1}\rpar^2}
\newcommand{\vvfw}{\sigma^{a}_{\ff}}
\newcommand{\vvew}{\sigma^{a}_{\fe}}
\newcommand{\gv}{g_{_V}}
\newcommand{\ga}{g_{_A}}
\newcommand{\gve}{g^{\fe}_{_{V}}}
\newcommand{\gae}{g^{\fe}_{_{A}}}
\newcommand{\gvf}{g^{\ff}_{_{V}}}
\newcommand{\gaf}{g^{\ff}_{_{A}}}
\newcommand{\gva}{g_{_{V,A}}}
\newcommand{\gvae}{g^{\fe}_{_{V,A}}}
\newcommand{\gvaf}{g^{\ff}_{_{V,A}}}
\newcommand{\sGv}{{\cal{G}}_{_V}}
\newcommand{\cGa}{{\cal{G}}^{*}_{_A}}
\newcommand{\cGv}{{\cal{G}}^{*}_{_V}}
\newcommand{\sGa}{{\cal{G}}_{_A}}
\newcommand{\Gvf}{{\cal{G}}^{\ff}_{_{V}}}
\newcommand{\Gaf}{{\cal{G}}^{\ff}_{_{A}}}
\newcommand{\Gvaf}{{\cal{G}}^{\ff}_{_{V,A}}}
\newcommand{\Gve}{{\cal{G}}^{\fe}_{_{V}}}
\newcommand{\Gae}{{\cal{G}}^{\fe}_{_{A}}}
\newcommand{\Gvae}{{\cal{G}}^{\fe}_{_{V,A}}}
\newcommand{\gvl}{g^{\fl}_{_{V}}}
\newcommand{\gal}{g^{\fl}_{_{A}}}
\newcommand{\gval}{g^{\fl}_{_{V,A}}}
\newcommand{\gvb}{g^{\ffb}_{_{V}}}
\newcommand{\gab}{g^{\ffb}_{_{A}}}
\newcommand{\fvf}{F_{_V}^{\ff}}
\newcommand{\faf}{F_{_A}^{\ff}}
\newcommand{\fvl}{F_{_V}^{\fl}}
\newcommand{\fal}{F_{_A}^{\fl}}
\newcommand{\corat}{\kappa}
\newcommand{\corats}{\kappa^2}
%
%
\newcommand{\dr}{\Delta r}
\newcommand{\drl}{\Delta r_{_L}}
\newcommand{\drh}{\Delta{\hat r}}
\newcommand{\drhw}{\Delta{\hat r}_{_W}}
\newcommand{\rhou}{\rho_{_U}}
\newcommand{\rhoz}{\rho_{_\zb}}
\newcommand{\rZ}{\rho_{_\zb}}
\newcommand{\rhob}{\rho_{_0}}
\newcommand{\rZf}{\rho^{\ff}_{_\zb}}
\newcommand{\rhoe}{\rho_{\fe}}
\newcommand{\rhof}{\rho_{\ff}}
\newcommand{\rhoi}[1]{\rho_{#1}}
\newcommand{\kZf}{\kappa^{\ff}_{_\zb}}
\newcommand{\rWf}{\rho^{\ff}_{_\wb}}
\newcommand{\brWf}{{\bar{\rho}}^{\ff}_{_\wb}}
\newcommand{\rHf}{\rho^{\ff}_{_\hb}}
\newcommand{\brHf}{{\bar{\rho}}^{\ff}_{_\hb}}
\newcommand{\rhoR}{\rho^R_{_{\zb}}}
\newcommand{\hatrh}{{\hat\rho}}
\newcommand{\ku}{\kappa_{_U}}
\newcommand{\rZdf}[1]{\rho^{#1}_{_\zb}}
\newcommand{\kZdf}[1]{\kappa^{#1}_{_\zb}}
\newcommand{\rdfL}[1]{\rho^{#1}_{_L}}
\newcommand{\kdfL}[1]{\kappa^{#1}_{_L}}
\newcommand{\rdfR}[1]{\rho^{#1}_{\rm{rem}}}
\newcommand{\kdfR}[1]{\kappa^{#1}_{\rm{rem}}}
\newcommand{\bark}{\overline\kappa}
\newcommand{\kbar}{\overline k}
%
%
\newcommand{\stw}{s_{\theta}}             
\newcommand{\ctw}{c_{\theta}}
\newcommand{\stws}{s_{\theta}^2}
\newcommand{\stwc}{s_{\theta}^3}
\newcommand{\stwf}{s_{\theta}^4}
\newcommand{\stwx}{s_{\theta}^6}
\newcommand{\ctws}{c_{\theta}^2}
\newcommand{\ctwc}{c_{\theta}^3}
\newcommand{\ctwf}{c_{\theta}^4}
\newcommand{\ctwx}{c_{\theta}^6}
\newcommand{\stwfiv}{s_{\theta}^5}
\newcommand{\ctwfiv}{c_{\theta}^5}
\newcommand{\stwsix}{s_{\theta}^6}
\newcommand{\ctwsix}{c_{\theta}^6}
%
%
\newcommand{\siw}{s_{_W}}           
\newcommand{\cow}{c_{_W}}
\newcommand{\siws}{s^2_{_W}}
\newcommand{\cows}{c^2_{_W}}
\newcommand{\siwc}{s^3_{_W}}
\newcommand{\cowc}{c^3_{_W}}
\newcommand{\siwf}{s^4_{_W}}
\newcommand{\cowf}{c^4_{_W}}
\newcommand{\siwx}{s^6_{_W}}
\newcommand{\cowx}{c^6_{_W}}
\newcommand{\sons}{s_{_W}}
\newcommand{\sonss}{s^2_{_W}}
\newcommand{\cons}{c_{_W}}
\newcommand{\cooss}{c^2_{_W}}
%
%
\newcommand{\szs}{{\overline s}^2}
\newcommand{\szq}{{\overline s}^4}
\newcommand{\czs}{{\overline c}^2}
\newcommand{\sbs}{s_{_0}^2}
\newcommand{\cbs}{c_{_0}^2}
\newcommand{\dss}{\Delta s^2}
\newcommand{\snes}{s_{\nu e}^2}
\newcommand{\cnes}{c_{\nu e}^2}
\newcommand{\shs}{{\hat s}^2}
\newcommand{\chs}{{\hat c}^2}
\newcommand{\chl}{{\hat c}}
\newcommand{\seffs}{s^2_{\rm{eff}}}
\newcommand{\seffsf}[1]{\sin^2\theta^{#1}_{\rm{eff}}}
\newcommand{\sress}{s^2_{\rm res}}                
\newcommand{\sR}{s_{_R}}
\newcommand{\sRs}{s^2_{_R}}
\newcommand{\ctwe}{c_{\theta}^6}
\newcommand{\sany}{s}
\newcommand{\cany}{c}
\newcommand{\sanys}{s^2}
\newcommand{\canys}{c^2}
%
%
\newcommand{\sip}{u}                             
\newcommand{\siap}{{\bar{v}}}                    
\newcommand{\sop}{{\bar{u}}}                     
\newcommand{\soap}{v}                            
\newcommand{\ip}[1]{u\lpar{#1}\rpar}             
\newcommand{\iap}[1]{{\bar{v}}\lpar{#1}\rpar}    
\newcommand{\op}[1]{{\bar{u}}\lpar{#1}\rpar}     
\newcommand{\oap}[1]{v\lpar{#1}\rpar}            
%
%
\newcommand{\ipp}[2]{u\lpar{#1,#2}\rpar}         
\newcommand{\ipap}[2]{{\bar v}\lpar{#1,#2}\rpar} 
\newcommand{\opp}[2]{{\bar u}\lpar{#1,#2}\rpar}  
\newcommand{\opap}[2]{v\lpar{#1,#2}\rpar}        
\newcommand{\upspi}[1]{u\lpar{#1}\rpar}
\newcommand{\vpspi}[1]{v\lpar{#1}\rpar}
\newcommand{\wpspi}[1]{w\lpar{#1}\rpar}
\newcommand{\ubpspi}[1]{{\bar{u}}\lpar{#1}\rpar}
\newcommand{\vbpspi}[1]{{\bar{v}}\lpar{#1}\rpar}
\newcommand{\wbpspi}[1]{{\bar{w}}\lpar{#1}\rpar}
\newcommand{\udpspi}[1]{u^{\dagger}\lpar{#1}\rpar}
\newcommand{\vdpspi}[1]{v^{\dagger}\lpar{#1}\rpar}
\newcommand{\wdpspi}[1]{w^{\dagger}\lpar{#1}\rpar}
\newcommand{\Ubilin}[1]{U\lpar{#1}\rpar}
\newcommand{\Vbilin}[1]{V\lpar{#1}\rpar}
\newcommand{\Xbilin}[1]{X\lpar{#1}\rpar}
\newcommand{\Ybilin}[1]{Y\lpar{#1}\rpar}
\newcommand{\up}[2]{u_{#1}\lpar #2\rpar}
\newcommand{\vp}[2]{v_{#1}\lpar #2\rpar}
\newcommand{\ubp}[2]{{\overline u}_{#1}\lpar #2\rpar}
\newcommand{\vbp}[2]{{\overline v}_{#1}\lpar #2\rpar}
\newcommand{\Pje}[1]{\frac{1}{2}\lpar 1 + #1\,\gfd\rpar}
\newcommand{\Pj}[1]{\Pi_{#1}}
\newcommand{\trace}{\mbox{Tr}}
%
%
\newcommand{\Poper}[2]{P_{#1}\lpar{#2}\rpar}
\newcommand{\Loper}[2]{\Lambda_{#1}\lpar{#2}\rpar}
\newcommand{\proj}[3]{P_{#1}\lpar{#2,#3}\rpar}
\newcommand{\sproj}[1]{P_{#1}}
\newcommand{\Nden}[3]{N_{#1}^{#2}\lpar{#3}\rpar}
\newcommand{\sNden}[1]{N_{#1}}
\newcommand{\nden}[2]{n_{#1}^{#2}}
%
%
\newcommand{\vwf}[2]{e_{#1}\lpar#2\rpar}             
\newcommand{\vwfb}[2]{{\overline e}_{#1}\lpar#2\rpar}
\newcommand{\pwf}[2]{\epsilon_{#1}\lpar#2\rpar}      
\newcommand{\sla}[1]{/\!\!\!#1}
\newcommand{\slac}[1]{/\!\!\!\!#1}
%
%
\newcommand{\iemom}{p_{_-}}                    
\newcommand{\ipmom}{p_{_+}}
\newcommand{\oemom}{q_{_-}}                    
\newcommand{\opmom}{q_{_+}}
%
%
\newcommand{\spro}[2]{{#1}\cdot{#2}}
%
%
\newcommand{\gfour}{\gamma_4}                    
\newcommand{\gfd}{\gamma_5}                    
\newcommand{\gap}{\lpar 1+\gamma_5\rpar}
\newcommand{\gam}{\lpar 1-\gamma_5\rpar}
\newcommand{\gdp}{\gamma_+}
\newcommand{\gdm}{\gamma_-}
\newcommand{\gdpm}{\gamma_{\pm}}
\newcommand{\gad}{\gamma}
\newcommand{\gapm}{\lpar 1\pm\gamma_5\rpar}
\newcommand{\gadi}[1]{\gamma_{#1}}
\newcommand{\gadu}[1]{\gamma_{#1}}
\newcommand{\gapu}[1]{\gamma^{#1}}
\newcommand{\sigd}[2]{\sigma_{#1#2}}
\newcommand{\sumsp}{\overline{\sum_{\mbox{spins}}}}
%
%
\newcommand{\li}[2]{\mathrm{Li}_{#1}\lpar\displaystyle{#2}\rpar} 
\newcommand{\etaf}[2]{\eta\lpar#1,#2\rpar}
\newcommand{\lkall}[3]{\lambda\lpar#1,#2,#3\rpar}       
\newcommand{\slkall}[3]{\lambda^{1/2}\lpar#1,#2,#3\rpar}
\newcommand{\segam}{\Gamma}                             
\newcommand{\egam}[1]{\Gamma\lpar#1\rpar}               
\newcommand{\ebe}[2]{B\lpar#1,#2\rpar}                  
\newcommand{\ddel}[1]{\delta\lpar#1\rpar}               
\newcommand{\drii}[2]{\delta_{#1#2}}                    
\newcommand{\driv}[4]{\delta_{#1#2#3#4}}                
\newcommand{\intmomi}[2]{\int\,d^{#1}#2}
\newcommand{\intmomii}[3]{\int\,d^{#1}#2\,\int\,d^{#1}#3}
\newcommand{\intfx}[1]{\int_{\scriptstyle 0}^{\scriptstyle 1}\,d#1}
\newcommand{\intfxy}[2]{\int_{\scriptstyle 0}^{\scriptstyle 1}\,d#1\,
                        \int_{\scriptstyle 0}^{\scriptstyle #1}\,d#2}
\newcommand{\intfxyz}[3]{\int_{\scriptstyle 0}^{\scriptstyle 1}\,d#1\,
                         \int_{\scriptstyle 0}^{\scriptstyle #1}\,d#2\,
                         \int_{\scriptstyle 0}^{\scriptstyle #2}\,d#3}
\newcommand{\Beta}[2]{{\rm{B}}\lpar #1,#2\rpar}
\newcommand{\sBeta}{\rm{B}}
\newcommand{\sign}[1]{{\rm{sign}}\lpar{#1}\rpar}
%
%
\newcommand{\gn}{\Gamma_{\nu}}
\newcommand{\gel}{\Gamma_{\fe}}
\newcommand{\gmu}{\Gamma_{\mu}}
\newcommand{\gff}{\Gamma_{\ff}}
\newcommand{\gt}{\Gamma_{\tau}}
\newcommand{\gl}{\Gamma_{\fl}}
\newcommand{\gq}{\Gamma_{\fq}}
\newcommand{\gu}{\Gamma_{\fu}}
\newcommand{\gd}{\Gamma_{\fd}}
\newcommand{\gc}{\Gamma_{\fc}}
\newcommand{\gs}{\Gamma_{\fs}}
\newcommand{\gbq}{\Gamma_{\ffb}}
\newcommand{\gz}{\Gamma_{_{\zb}}}
\newcommand{\gw}{\Gamma_{_{\wb}}}
\newcommand{\gh}{\Gamma_{_{h}}}
\newcommand{\ghb}{\Gamma_{_{\hb}}}
\newcommand{\gi}{\Gamma_{\rm{inv}}}
\newcommand{\gzs}{\Gamma^2_{_{\zb}}}
%
%
\newcommand{\tcie}{I^{(3)}_{\fe}}
\newcommand{\tcim}{I^{(3)}_{\flm}}
\newcommand{\tcif}{I^{(3)}_{\ff}}
\newcommand{\tciq}{I^{(3)}_{\fq}}
\newcommand{\tcib}{I^{(3)}_{\ffb}}
\newcommand{\tcih}{I^{(3)}_h}
\newcommand{\tcii}{I^{(3)}_i}
\newcommand{\tcift}{I^{(3)}_{\tilde f}}
\newcommand{\tcifp}{I^{(3)}_{f'}}
\newcommand{\wispt}[1]{I^{(3)}_{#1}}
\newcommand{\ql}{Q_l}
\newcommand{\qe}{Q_e}
\newcommand{\qu}{Q_u}
\newcommand{\qd}{Q_d}
\newcommand{\qb}{Q_b}
\newcommand{\qt}{Q_t}
\newcommand{\qup}{Q'_u}
\newcommand{\qdp}{Q'_d}
\newcommand{\qmu}{Q_{\mu}}
\newcommand{\qes}{Q^2_e}
\newcommand{\qec}{Q^3_e}
\newcommand{\qus}{Q^2_u}
\newcommand{\qds}{Q^2_d}
\newcommand{\qbs}{Q^2_b}
\newcommand{\qts}{Q^2_t}
\newcommand{\qbc}{Q^3_b}
\newcommand{\qf}{Q_f}
\newcommand{\qfs}{Q^2_f}
\newcommand{\qfc}{Q^3_f}
\newcommand{\qff}{Q^4_f}
\newcommand{\qep}{Q_{e'}}
\newcommand{\qfp}{Q_{f'}}
\newcommand{\qfps}{Q^2_{f'}}
\newcommand{\qfpc}{Q^3_{f'}}
\newcommand{\qq}{Q_q}
\newcommand{\qqs}{Q^2_q}
\newcommand{\qi}{Q_i}
\newcommand{\qis}{Q^2_i}
\newcommand{\qj}{Q_j}
\newcommand{\qjs}{Q^2_j}
\newcommand{\QW}{Q_{_\wb}}
\newcommand{\QWs}{Q^2_{_\wb}}
\newcommand{\Qd}{Q_d}
\newcommand{\Qds}{Q^2_d}
\newcommand{\Qu}{Q_u}
\newcommand{\Qus}{Q^2_u}
\newcommand{\vi}{v_i}
\newcommand{\vis}{v^2_i}
\newcommand{\ai}{a_i}
\newcommand{\ais}{a^2_i}
%
%
\newcommand{\piv}{\Pi_{_V}}
\newcommand{\pia}{\Pi_{_A}}
\newcommand{\piva}{\Pi_{_{V,A}}}
\newcommand{\pivi}[1]{\Pi^{({#1})}_{_V}}
\newcommand{\piai}[1]{\Pi^{({#1})}_{_A}}
\newcommand{\pivai}[1]{\Pi^{({#1})}_{_{V,A}}}
\newcommand{\pih}{{\hat\Pi}}
\newcommand{\sgh}{{\hat\Sigma}}
\newcommand{\Pgg}{\Pi_{\ph\ph}}
\newcommand{\Ptg}{\Pi_{_{3Q}}}
\newcommand{\Ptt}{\Pi_{_{33}}}
\newcommand{\Pzg}{\Pi_{_{\zb\ab}}}
\newcommand{\Pzga}[2]{\Pi^{#1}_{_{\zb\ab}}\lpar#2\rpar}
\newcommand{\Pf}{\Pi_{_F}}
\newcommand{\Sgg}{\Sigma_{_{\ab\ab}}}
\newcommand{\Szg}{\Sigma_{_{\zb\ab}}}
\newcommand{\SVV}{\Sigma_{_{\vb\vb}}}
\newcommand{\USvv}{{\hat\Sigma}_{_{\vb\vb}}}
\newcommand{\Sww}{\Sigma_{_{\wb\wb}}}
\newcommand{\Swwg}{\Sigma^{_G}_{_{\wb\wb}}}
\newcommand{\Szz}{\Sigma_{_{\zb\zb}}}
\newcommand{\Shh}{\Sigma_{_{\hb\hb}}}
\newcommand{\Spzz}{\Sigma'_{_{\zb\zb}}}
\newcommand{\Stg}{\Sigma_{_{3Q}}}
\newcommand{\Stt}{\Sigma_{_{33}}}
\newcommand{\bSww}{{\overline\Sigma}_{_{WW}}}
\newcommand{\bStg}{{\overline\Sigma}_{_{3Q}}}
\newcommand{\bStt}{{\overline\Sigma}_{_{33}}}
\newcommand{\Sssn}{\Sigma_{_{\hkn\hkn}}}
\newcommand{\Sssc}{\Sigma_{_{\phi\phi}}}
\newcommand{\Szn}{\Sigma_{_{\zb\hkn}}}
\newcommand{\Swc}{\Sigma_{_{\wb\hkg}}}
\newcommand{\mix}[2]{{\cal{M}}^{#1}\lpar{#2}\rpar}
\newcommand{\bmix}[2]{\Pi^{{#1},F}_{_{\zb\ab}}\lpar{#2}\rpar}
\newcommand{\hPgg}[2]{{\hat{\Pi}^{{#1},F}}_{_{\ph\ph}}\lpar{#2}\rpar}
\newcommand{\hmix}[2]{{\hat{\Pi}^{{#1},F}}_{_{\zb\ab}}\lpar{#2}\rpar}
\newcommand{\Dz}[2]{{\cal{D}}_{_{\zb}}^{#1}\lpar{#2}\rpar}
\newcommand{\bDz}[2]{{\cal{D}}^{{#1},F}_{_{\zb}}\lpar{#2}\rpar}
\newcommand{\hDz}[2]{{\hat{\cal{D}}}^{{#1},F}_{_{\zb}}\lpar{#2}\rpar}
\newcommand{\Szzd}[2]{\Sigma'^{#1}_{_{\zb\zb}}\lpar{#2}\rpar}
\newcommand{\Swwd}[2]{\Sigma'^{#1}_{_{\wb\wb}}\lpar{#2}\rpar}
\newcommand{\Shhd}[2]{\Sigma'^{#1}_{_{\hb\hb}}\lpar{#2}\rpar}
\newcommand{\ZFren}[2]{{\cal{Z}}^{#1}\lpar{#2}\rpar}
\newcommand{\WFren}[2]{{\cal{W}}^{#1}\lpar{#2}\rpar}
\newcommand{\HFren}[2]{{\cal{H}}^{#1}\lpar{#2}\rpar}
\newcommand{\WI}{\cal{W}}
%
%
\newcommand{\cf}{c_f}
\newcommand{\Cf}{C_{_F}}
\newcommand{\Nf}{N_f}
\newcommand{\Nc}{N_c}
\newcommand{\Ncs}{N^2_c}
\newcommand{\nf }{n_f}
\newcommand{\nfs}{n^2_f}
\newcommand{\nfc}{n^3_f}
\newcommand{\MSB}{\overline{MS}}
\newcommand{\LMSB}{\Lambda_{\overline{\mathrm{MS}}}}
\newcommand{\LMSBp}{\Lambda'_{\overline{\mathrm{MS}}}}
\newcommand{\LMSBS}{\Lambda^2_{\overline{\mathrm{MS}}}}
\newcommand{\LMSBv }{\mbox{$\Lambda^{(5)}_{\overline{\mathrm{MS}}}$}}
\newcommand{\LMSBvS}{\mbox{$\left(\Lambda^{(5)}_{\overline{\mathrm{MS}}}\right)^2$}}
\newcommand{\LMSBt }{\mbox{$\Lambda^{(3)}_{\overline{\mathrm{MS}}}$}}
\newcommand{\LMSBtS}{\mbox{$\left(\Lambda^{(3)}_{\overline{\mathrm{MS}}}\right)^2$}}
\newcommand{\LMSBf }{\mbox{$\Lambda^{(4)}_{\overline{\mathrm{MS}}}$}}
\newcommand{\LMSBfS}{\mbox{$\left(\Lambda^{(4)}_{\overline{\mathrm{MS}}}\right)^2$}}
\newcommand{\LMSBn }{\mbox{$\Lambda^{(\nf)}_{\overline{\mathrm{MS}}}$}}
\newcommand{\LMSBnS}{\mbox{$\left(\Lambda^{(\nf)}_{\overline{\mathrm{MS}}}\right)^2$}}
\newcommand{\LMSBnml }{\mbox{$\Lambda^{(\nf-1)}_{\overline{\mathrm{MS}}}$}}
\newcommand{\LMSBnmlS}{\mbox{$\left(\Lambda^{(\nf-1)}_{\overline{\mathrm{MS}}}\right)^2$}}
\newcommand{\Bnf}{\lpar\nf \rpar}
\newcommand{\Bnfm}{\lpar\nf-1 \rpar}
\newcommand{\LuM}{L_{_M}}
\newcommand{\bef}{\beta_{\ff}}
\newcommand{\befs}{\beta^2_{\ff}}
\newcommand{\befc}{\beta^3_{f}}
\newcommand{\alsp}{\alpha'_{_S}}
\newcommand{\api}{\displaystyle \frac{\als(s)}{\pi}}
\newcommand{\alss}{\alpha^2_{_S}}
\newcommand{\ztwo}{\zeta(2)}
\newcommand{\ztri}{\zeta(3)}
\newcommand{\zfor}{\zeta(4)}
\newcommand{\zfiv}{\zeta(5)}
\newcommand{\bi}[1]{b_{#1}}
\newcommand{\ci}[1]{c_{#1}}
\newcommand{\Ci}[1]{C_{#1}}
\newcommand{\bip}[1]{b'_{#1}}
\newcommand{\cip}[1]{c'_{#1}}
%
%
\newcommand{\osps}{16\,\pi^2}
\newcommand{\srt}{\sqrt{2}}
\newcommand{\ospsi}{\displaystyle{\frac{i}{16\,\pi^2}}}
%
%
\newcommand{\tfpromu}{\mbox{$e^+e^-\to \mu^+\mu^-$}}
\newcommand{\tfprotau}{\mbox{$e^+e^-\to \tau^+\tau^-$}}
\newcommand{\tfproe}{\mbox{$e^+e^-\to e^+e^-$}}
\newcommand{\tfpronu}{\mbox{$e^+e^-\to \barnu\nu$}}
\newcommand{\tfproqq}{\mbox{$e^+e^-\to \barq q$}}
\newcommand{\tfprohad}{\mbox{$e^+e^-\to\,$} hadrons}
%
%
\newcommand{\bpromu}{\mbox{$e^+e^-\to \mu^+\mu^-\ph$}}
\newcommand{\bprotau}{\mbox{$e^+e^-\to \tau^+\tau^-\ph$}}
\newcommand{\bproe}{\mbox{$e^+e^-\to e^+e^-\ph$}}
\newcommand{\bpronu}{\mbox{$e^+e^-\to \barnu\nu\ph$}}
\newcommand{\bproqq}{\mbox{$e^+e^-\to \barq q \ph$}}
%
%
\newcommand{\tbprow} {\mbox{$e^+e^-\to \wbp \wbm $}}
\newcommand{\tbproz} {\mbox{$e^+e^-\to \zb  \zb  $}}
\newcommand{\tbproh} {\mbox{$e^+e^-\to \zb  \hb  $}}
\newcommand{\tbprozg}{\mbox{$e^+e^-\to \zb  \ph  $}}
\newcommand{\tbprog} {\mbox{$e^+e^-\to \ph  \ph  $}}
%
%
\newcommand{\Fermionline}[1]{
\vcenter{\hbox{
  \begin{picture}(60,20)(0,{#1})
  \SetScale{2.}
    \ArrowLine(0,5)(30,5)
  \end{picture}}}
}
\newcommand{\AntiFermionline}[1]{
\vcenter{\hbox{
  \begin{picture}(60,20)(0,{#1})
  \SetScale{2.}
    \ArrowLine(30,5)(0,5)
  \end{picture}}}
}
\newcommand{\Photonline}[1]{
\vcenter{\hbox{
  \begin{picture}(60,20)(0,{#1})
  \SetScale{2.}
    \Photon(0,5)(30,5){2}{6.5}
  \end{picture}}}
}
\newcommand{\Gluonline}[1]{
\vcenter{\hbox{
  \begin{picture}(60,20)(0,{#1})
  \SetScale{2.}
    \Gluon(0,5)(30,5){2}{6.5}
  \end{picture}}}
}
\newcommand{\Wbosline}[1]{
\vcenter{\hbox{
  \begin{picture}(60,20)(0,{#1})
  \SetScale{2.}
    \Photon(0,5)(30,5){2}{4}
    \ArrowLine(13.3,3.1)(16.9,7.2)
  \end{picture}}}
}
\newcommand{\Zbosline}[1]{
\vcenter{\hbox{
  \begin{picture}(60,20)(0,{#1})
  \SetScale{2.}
    \Photon(0,5)(30,5){2}{4}
  \end{picture}}}
}
\newcommand{\Philine}[1]{
\vcenter{\hbox{
  \begin{picture}(60,20)(0,{#1})
  \SetScale{2.}
    \DashLine(0,5)(30,5){2}
  \end{picture}}}
}
\newcommand{\Phicline}[1]{
\vcenter{\hbox{
  \begin{picture}(60,20)(0,{#1})
  \SetScale{2.}
    \DashLine(0,5)(30,5){2}
    \ArrowLine(14,5)(16,5)
  \end{picture}}}
}
\newcommand{\Ghostline}[1]{
\vcenter{\hbox{
  \begin{picture}(60,20)(0,{#1})
  \SetScale{2.}
    \DashLine(0,5)(30,5){.5}
    \ArrowLine(14,5)(16,5)
  \end{picture}}}
}
%
%
\newcommand{\gauge}{g}
\newcommand{\gpar}{\xi}
\newcommand{\gparA}{\xi_{_A}}
\newcommand{\gparZ}{\xi_{_Z}}
\newcommand{\gpari}[1]{\gpar_{#1}}
\newcommand{\gparis}[1]{\gpar^2_{#1}}
\newcommand{\gpariq}[1]{\gpar^4_{#1}}
\newcommand{\gpars}{\xi^2}
\newcommand{\dgpar}{\Delta\gpar}
\newcommand{\dgparA}{\Delta\gparA}
\newcommand{\dgparZ}{\Delta\gparZ}
\newcommand{\gparq}{\xi^4}
\newcommand{\gparAs}{\xi^2_{_A}}
\newcommand{\gparAq}{\xi^4_{_A}}
\newcommand{\gparZs}{\xi^2_{_Z}}
\newcommand{\gparZq}{\xi^4_{_Z}}
\newcommand{\Rxi}{R_{\gpar}}
\newcommand{\hxi}{\chi}
%
%
\newcommand{\LSM}{{\cal{L}}_{_{\rm{SM}}}}
\newcommand{\LSMr}{{\cal{L}}^{\rm{ren}}_{_{\rm{SM}}}}
\newcommand{\LYM}{{\cal{L}}_{_{YM}}}
\newcommand{\Lzer}{{\cal{L}}_{_{0}}}
\newcommand{\Lone}{{\cal{L}}^{{\bos},I}}
\newcommand{\Lpro}{{\cal{L}}_{\rm{prop}}}
\newcommand{\Ls  }{{\cal{L}}_{_{S}}}
\newcommand{\Lsi }{{\cal{L}}^{I}_{_{S}}}
\newcommand{\Lgf }{{\cal{L}}_{gf  }}
\newcommand{\Lgfi}{{\cal{L}}^{I}_{gf}}
\newcommand{\Lf  }{{\cal{L}}^{{\fer},I}_{\ssV}}
\newcommand{\LHf }{{\cal{L}}^{\fer}_{\ssS}}
\newcommand{\LHfi}{{\cal{L}}^{{\fer},I}_{\ssS}}
\newcommand{\Lren}{{\cal{L}}_{\rm{ren}}}
\newcommand{\Lct}{{\cal{L}}_{\rm{ct}}}
\newcommand{\Lcti}[1]{{\cal{L}}^{#1}_{\rm{ct}}}
\newcommand{\LctI}{{\cal{L}}^{(2)}_{\rm{ct}}}
\newcommand{\Llone}{{\cal{L}}}
\newcommand{\LQED}{{\cal{L}}_{_{\rm{QED}}}}
\newcommand{\LQEDr}{{\cal{L}}^{\rm{ren}}_{_{\rm{QED}}}}
\newcommand{\FST}[3]{F_{#1#2}^{#3}}
\newcommand{\cD}[1]{D_{#1}}
\newcommand{\pd}[1]{\partial_{#1}}
\newcommand{\tgen}[1]{\tau^{#1}}
\newcommand{\gbl}{g_1}
\newcommand{\lctt}[3]{\varepsilon_{#1#2#3}}
\newcommand{\lctf}[4]{\varepsilon_{#1#2#3#4}}
\newcommand{\lctfb}[4]{\varepsilon\lpar{#1#2#3#4}\rpar}
\newcommand{\slct}{\varepsilon}
\newcommand{\cgfi}[1]{{\cal{C}}^{#1}}
\newcommand{\cgfZ}{{\cal{C}}_{_Z}}
\newcommand{\cgfA}{{\cal{C}}_{_A}}
\newcommand{\cgfZs}{{\cal{C}}^2_{_Z}}
\newcommand{\cgfAs}{{\cal{C}}^2_{_A}}
\newcommand{\hpms}{\mu^2}
\newcommand{\hpal}{\alpha_{_H}}
\newcommand{\hpals}{\alpha^2_{_H}}
\newcommand{\hpbe}{\beta_{_H}}
\newcommand{\hpbep}{\beta^{'}_{_H}}
\newcommand{\hpla}{\lambda}
\newcommand{\hpalf}{\alpha_{f}}
\newcommand{\hpbef}{\beta_{f}}
\newcommand{\tpar}[1]{\Lambda^{#1}}
\newcommand{\Mop}[2]{{\rm{M}}^{#1#2}}
\newcommand{\Lop}[2]{{\rm{L}}^{#1#2}}
\newcommand{\Lgen}[1]{T^{#1}}
\newcommand{\Rgen}[1]{t^{#1}}
\newcommand{\fpari}[1]{\lambda_{#1}}
\newcommand{\fQ}[1]{Q_{#1}}
\newcommand{\unm}{I}
\newcommand{\cDsla}{/\!\!\!\!D}
%
%
\newcommand{\saff}[1]{A_{#1}}                    
\newcommand{\aff}[2]{A_{#1}\lpar #2\rpar}                   
\newcommand{\sbff}[1]{B_{#1}}                    
\newcommand{\sfbff}[1]{B^{F}_{#1}}
\newcommand{\bff}[4]{B_{#1}\lpar #2;#3,#4\rpar}             
\newcommand{\bfft}[3]{B_{#1}\lpar #2,#3\rpar}             
\newcommand{\fbff}[4]{B^{F}_{#1}\lpar #2;#3,#4\rpar}        
\newcommand{\cdbff}[4]{\Delta B_{#1}\lpar #2;#3,#4\rpar}             
\newcommand{\sdbff}[4]{\delta B_{#1}\lpar #2;#3,#4\rpar}             
\newcommand{\cdbfft}[3]{\Delta B_{#1}\lpar #2,#3\rpar}             
\newcommand{\sdbfft}[3]{\delta B_{#1}\lpar #2,#3\rpar}             
\newcommand{\scff}[1]{C_{#1}}                    
\newcommand{\scffo}[2]{C_{#1}\lpar{#2}\rpar}                
\newcommand{\cff}[7]{C_{#1}\lpar #2,#3,#4;#5,#6,#7\rpar}    
\newcommand{\sccff}[5]{c_{#1}\lpar #2;#3,#4,#5\rpar} 
\newcommand{\sdff}[1]{D_{#1}}                    
\newcommand{\dffp}[7]{D_{#1}\lpar #2,#3,#4,#5,#6,#7;}       
\newcommand{\dffm}[4]{#1,#2,#3,#4\rpar}                     
\newcommand{\bzfa}[2]{B^{F}_{_{#2}}\lpar{#1}\rpar}
\newcommand{\bzfaa}[3]{B^{F}_{_{#2#3}}\lpar{#1}\rpar}
\newcommand{\shcff}[4]{C_{_{#2#3#4}}\lpar{#1}\rpar}
\newcommand{\shdff}[6]{D_{_{#3#4#5#6}}\lpar{#1,#2}\rpar}
\newcommand{\scdff}[3]{d_{#1}\lpar #2,#3\rpar} 
\newcommand{\scaldff}[1]{{\cal{D}}^{#1}}
\newcommand{\caldff}[2]{{\cal{D}}^{#1}\lpar{#2}\rpar}
\newcommand{\caldfft}[3]{{\cal{D}}_{#1}^{#2}\lpar{#3}\rpar}
%
%
\newcommand{\slaff}[1]{a_{#1}}                        
\newcommand{\slbff}[1]{b_{#1}}                        
\newcommand{\slbffh}[1]{{\hat{b}}_{#1}}    
\newcommand{\ssldff}[1]{d_{#1}}                        
\newcommand{\sslcff}[1]{c_{#1}}                        
\newcommand{\slcff}[2]{c_{#1}^{(#2)}}                        
\newcommand{\sldff}[2]{d_{#1}^{(#2)}}                        
\newcommand{\lbff}[3]{b_{#1}\lpar #2;#3\rpar}         
\newcommand{\lbffh}[2]{{\hat{b}}_{#1}\lpar #2\rpar}   
\newcommand{\lcff}[8]{c_{#1}^{(#2)}\lpar  #3,#4,#5;#6,#7,#8\rpar}         
\newcommand{\ldffp}[8]{d_{#1}^{(#2)}\lpar #3,#4,#5,#6,#7,#8;}
\newcommand{\ldffm}[4]{#1,#2,#3,#4\rpar}                   
%
%
\newcommand{\Iff}[4]{I_{#1}\lpar #2;#3,#4 \rpar}
\newcommand{\Jff}[4]{J_{#1}\lpar #2;#3,#4 \rpar}
\newcommand{\Jds}[5]{{\bar{J}}_{#1}\lpar #2,#3;#4,#5 \rpar}
%
\newcommand{\nhmt}{\frac{n}{2}-2}
\newcommand{\nhmts}{{n}/{2}-2}
\newcommand{\omnh}{1-\frac{n}{2}}
\newcommand{\nhmo}{\frac{n}{2}-1}
\newcommand{\fmon}{4-n}
\newcommand{\lpi}{\ln\pi}
\newcommand{\lmass}[1]{\ln #1}
\newcommand{\egnh}{\egam{\frac{n}{2}}}
\newcommand{\egomnh}{\egam{1-\frac{n}{2}}}
\newcommand{\egtmnh}{\egam{2-\frac{n}{2}}}
\newcommand{\Ddr}{{\ds\frac{1}{{\bar{\varepsilon}}}}}
\newcommand{\Ddrs}{{\ds\frac{1}{{\bar{\varepsilon}^2}}}}
\newcommand{\Ddrd}{{\bar{\varepsilon}}}
\newcommand{\ept}{\hat\varepsilon}
\newcommand{\Ddrh}{{\ds\frac{1}{\hat{\varepsilon}}}}
\newcommand{\Ddrp}{{\ds\frac{1}{\varepsilon'}}}
\newcommand{\Ddrps}{\lpar{\ds{\frac{1}{\varepsilon'}}}\rpar^2}
\newcommand{\dre}{\varepsilon}
\newcommand{\drei}[1]{\varepsilon_{#1}}
\newcommand{\epp}{\varepsilon'}
\newcommand{\eps}{\varepsilon^*}
\newcommand{\ep}{\epsilon}
\newcommand{\propbt}[6]{{{#1_{#2}#1_{#3}}\over{\lpar #1^2 + #4 
-\ib\ep\rpar\lpar\lpar #5\rpar^2 + #6 -\ib\ep\rpar}}}
\newcommand{\propbo}[5]{{{#1_{#2}}\over{\lpar #1^2 + #3 - \ib\ep\rpar
\lpar\lpar #4\rpar^2 + #5 -\ib\ep\rpar}}}
\newcommand{\propc}[6]{{1\over{\lpar #1^2 + #2 - \ib\ep\rpar
\lpar\lpar #3\rpar^2 + #4 -\ib\ep\rpar
\lpar\lpar #5\rpar^2 + #6 -\ib\ep\rpar}}}
\newcommand{\propa}[2]{{1\over {#1^2 + #2^2 - \ib\ep}}}
\newcommand{\propb}[4]{{1\over {\lpar #1^2 + #2 - \ib\ep\rpar
\lpar\lpar #3\rpar^2 + #4 -\ib\ep\rpar}}}
\newcommand{\propbs}[4]{{1\over {\lpar\lpar #1\rpar^2 + #2 - \ib\ep\rpar
\lpar\lpar #3\rpar^2 + #4 -\ib\ep\rpar}}}
\newcommand{\propat}[4]{{#3_{#1}#3_{#2}\over {#3^2 + #4^2 - \ib\ep}}}
\newcommand{\propaf}[6]{{#5_{#1}#5_{#2}#5_{#3}#5_{#4}\over 
{#5^2 + #6^2 -\ib\ep}}}
\newcommand{\momeps}[1]{#1^2 - \ib\ep}
\newcommand{\mopeps}[1]{#1^2 + \ib\ep}
\newcommand{\propz}[1]{{1\over{#1^2 + \mzs - \ib\ep}}}
\newcommand{\propw}[1]{{1\over{#1^2 + \mws - \ib\ep}}}
\newcommand{\proph}[1]{{1\over{#1^2 + \mhs - \ib\ep}}}
\newcommand{\propf}[2]{{1\over{#1^2 + #2}}}
\newcommand{\propzrg}[3]{{{\delta_{#1#2}}\over{{#3}^2 + \mzs - \ib\ep}}}
\newcommand{\propwrg}[3]{{{\delta_{#1#2}}\over{{#3}^2 + \mws - \ib\ep}}}
\newcommand{\propzug}[3]{{
      {\delta_{#1#2} + \displaystyle{{{#3}^{#1}{#3}^{#2}}\over{\mzs}}}
                         \over{{#3}^2 + \mzs - \ib\ep}}}
\newcommand{\propwug}[3]{{
      {\delta_{#1#2} + \displaystyle{{{#3}^{#1}{#3}^{#2}}\over{\mws}}}
                        \over{{#3}^2 + \mws - \ib\ep}}}
\newcommand{\thf}[1]{\theta\lpar #1\rpar}
\newcommand{\epf}[1]{\varepsilon\lpar #1\rpar}
\newcommand{\singp}{\stackrel{sing}{\rightarrow}}
\newcommand{\aint}[3]{\int_{#1}^{#2}\,d #3}
\newcommand{\aroot}[1]{\sqrt{#1}}
\newcommand{\gramc}{\Delta_3}
\newcommand{\gramd}{\Delta_4}
\newcommand{\pinch}[2]{P^{(#1)}\lpar #2\rpar}
\newcommand{\pinchc}[2]{C^{(#1)}_{#2}}
\newcommand{\pinchd}[2]{D^{(#1)}_{#2}}
\newcommand{\loarg}[1]{\ln\lpar #1\rpar}
\newcommand{\loargr}[1]{\ln\lrbr #1\rrbr}
\newcommand{\lsoarg}[1]{\ln^2\lpar #1\rpar}
\newcommand{\ltarg}[2]{\ln\lpar #1\rpar\lpar #2\rpar}
\newcommand{\rfun}[2]{R\lpar #1,#2\rpar}
\newcommand{\pinchb}[3]{B_{#1}\lpar #2,#3\rpar}
\newcommand{\lga}{\ph}
\newcommand{\lzga}{\ssZ\ph}
%
%
\newcommand{\afa}[5]{A_{#1}^{#2}\lpar #3;#4,#5\rpar}
\newcommand{\bfa}[5]{B_{#1}^{#2}\lpar #3;#4,#5\rpar} 
\newcommand{\hfa}[5]{H_{#1}^{#2}\lpar #3;#4,#5\rpar}
\newcommand{\rfa}[5]{R_{#1}^{#2}\lpar #3;#4,#5\rpar}
\newcommand{\afao}[3]{A_{#1}^{#2}\lpar #3\rpar}
\newcommand{\bfao}[3]{B_{#1}^{#2}\lpar #3\rpar}
\newcommand{\hfao}[3]{H_{#1}^{#2}\lpar #3\rpar}
\newcommand{\rfao}[3]{R_{#1}^{#2}\lpar #3\rpar}
\newcommand{\afas}[2]{A_{#1}^{#2}}
\newcommand{\bfas}[2]{B_{#1}^{#2}}
\newcommand{\hfas}[2]{H_{#1}^{#2}}
\newcommand{\rfas}[2]{R_{#1}^{#2}}
\newcommand{\tfas}[2]{T_{#1}^{#2}}
\newcommand{\afaR}[6]{A_{#1}^{\gpar}\lpar #2;#3,#4,#5,#6 \rpar}
\newcommand{\bfaR}[6]{B_{#1}^{\gpar}\lpar #2;#3,#4,#5,#6 \rpar}
\newcommand{\hfaR}[6]{H_{#1}^{\gpar}\lpar #2;#3,#4,#5,#6 \rpar}
\newcommand{\shfaR}[1]{H_{#1}^{\gpar}}
\newcommand{\rfaR}[6]{R_{#1}^{\gpar}\lpar #2;#3,#4,#5,#6 \rpar}
\newcommand{\srfaR}[1]{R_{#1}^{\gpar}}
\newcommand{\afaRg}[5]{A_{#1 \lga}^{\gpar}\lpar #2;#3,#4,#5 \rpar}
\newcommand{\bfaRg}[5]{B_{#1 \lga}^{\gpar}\lpar #2;#3,#4,#5 \rpar}
\newcommand{\hfaRg}[5]{H_{#1 \lga}^{\gpar}\lpar #2;#3,#4,#5 \rpar}
\newcommand{\shfaRg}[1]{H_{#1\lga}^{\gpar}}
\newcommand{\rfaRg}[5]{R_{#1 \lga}^{\gpar}\lpar #2;#3,#4,#5 \rpar}
\newcommand{\srfaRg}[1]{R_{#1\lga}^{\gpar}}
\newcommand{\afaRt}[3]{A_{#1}^{\gpar}\lpar #2,#3 \rpar}
\newcommand{\hfaRt}[3]{H_{#1}^{\gpar}\lpar #2,#3 \rpar}
\newcommand{\hfaRf}[4]{H_{#1}^{\gpar}\lpar #2,#3,#4 \rpar}
\newcommand{\afasm}[4]{A_{#1}^{\lpar #2,#3,#4 \rpar}}
\newcommand{\bfasm}[4]{B_{#1}^{\lpar #2,#3,#4 \rpar}}
\newcommand{\color}[1]{c_{#1}}
\newcommand{\htf}[2]{H_2\lpar #1,#2\rpar}
\newcommand{\rof}[2]{R_1\lpar #1,#2\rpar}
\newcommand{\rtf}[2]{R_3\lpar #1,#2\rpar}
\newcommand{\rtrans}[2]{R_{#1}^{#2}}
\newcommand{\momf}[2]{#1^2_{#2}}
\newcommand{\Scalvert}[8][70]{
  \vcenter{\hbox{
  \SetScale{0.8}
  \begin{picture}(#1,50)(15,15)
    \Line(25,25)(50,50)      \Text(34,20)[lc]{#6} \Text(11,20)[lc]{#3}
    \Line(50,50)(25,75)      \Text(34,60)[lc]{#7} \Text(11,60)[lc]{#4}
    \Line(50,50)(90,50)      \Text(11,40)[lc]{#2} \Text(55,33)[lc]{#8}
    \GCirc(50,50){10}{1}          \Text(60,48)[lc]{#5} 
  \end{picture}}}
  }
%
%
\newcommand{\tHs}{\mu}
\newcommand{\tHsz}{\mu_{_0}}
\newcommand{\tHss}{\mu^2}
\newcommand{\Reb}{{\rm{Re}}}
\newcommand{\Imb}{{\rm{Im}}}
%
%
\newcommand{\spd}{\partial}
\newcommand{\fun}[1]{f\lpar{#1}\rpar}
\newcommand{\ffun}[2]{F_{#1}\lpar #2\rpar}
\newcommand{\gfun}[2]{G_{#1}\lpar #2\rpar}
\newcommand{\sffun}[1]{F_{#1}}
\newcommand{\csffun}[1]{{\cal{F}}_{#1}}
\newcommand{\sgfun}[1]{G_{#1}}
\newcommand{\tpfi}{\lpar 2\pi\rpar^4\ib}
\newcommand{\ffv}{F_{_V}}
\newcommand{\fga}{G_{_A}}
\newcommand{\ffm}{F_{_M}}
\newcommand{\ffs}{F_{_S}}
\newcommand{\fgp}{G_{_P}}
\newcommand{\fge}{G_{_E}}
\newcommand{\ffa}{F_{_A}}
\newcommand{\ffps}{F_{_P}}
\newcommand{\ffe}{F_{_E}}
\newcommand{\gacom}[2]{\lpar #1 + #2\gfd\rpar}
\newcommand{\mft}{m_{\tilde f}}
\newcommand{\qft}{Q_{f'}}
\newcommand{\vft}{v_{\tilde f}}
\newcommand{\subb}[2]{b_{#1}\lpar #2 \rpar}
\newcommand{\fwfr}[5]{\Sigma\lpar #1,#2,#3;#4,#5 \rpar}
\newcommand{\slim}[2]{\lim_{#1 \to #2}}
\newcommand{\sprop}[3]{
{#1\over {\lpar q^2\rpar^2\lpar \lpar q+ #2\rpar^2+#3^2\rpar }}}
%
%
\newcommand{\xroot}[1]{x_{#1}}
\newcommand{\yroot}[1]{y_{#1}}
\newcommand{\zroot}[1]{z_{#1}}
\newcommand{\lvar}{l}
\newcommand{\rvar}{r}
\newcommand{\tvar}{t}
\newcommand{\uvar}{u}
\newcommand{\vvar}{v}
\newcommand{\xvar}{x}
\newcommand{\yvar}{y}
\newcommand{\zvar}{z}
\newcommand{\yvarp}{y'}
\newcommand{\rvars}{r^2}
\newcommand{\vvars}{v^2}
\newcommand{\xvars}{x^2}
\newcommand{\yvars}{y^2}
\newcommand{\zvars}{z^2}
\newcommand{\rvarc}{r^3}
\newcommand{\xvarc}{x^3}
\newcommand{\yvarc}{y^3}
\newcommand{\zvarc}{z^3}
\newcommand{\rvarq}{r^4}
\newcommand{\xvarq}{x^4}
\newcommand{\yvarq}{y^4}
\newcommand{\zvarq}{z^4}
\newcommand{\avar}{a}
\newcommand{\avars}{a^2}
\newcommand{\avarc}{a^3}
\newcommand{\avari}[1]{a_{#1}}
\newcommand{\avart}[2]{a_{#1}^{#2}}
\newcommand{\delvari}[1]{\delta_{#1}}
\newcommand{\rvari}[1]{r_{#1}}
\newcommand{\xvari}[1]{x_{#1}}
\newcommand{\yvari}[1]{y_{#1}}
\newcommand{\zvari}[1]{z_{#1}}
\newcommand{\rvart}[2]{r_{#1}^{#2}}
\newcommand{\xvart}[2]{x_{#1}^{#2}}
\newcommand{\yvart}[2]{y_{#1}^{#2}}
\newcommand{\zvart}[2]{z_{#1}^{#2}}
\newcommand{\rvaris}[1]{r^2_{#1}}
\newcommand{\xvaris}[1]{x^2_{#1}}
\newcommand{\yvaris}[1]{y^2_{#1}}
\newcommand{\zvaris}[1]{z^2_{#1}}
\newcommand{\Xvar}{X}
\newcommand{\Xvars}{X^2}
\newcommand{\Xvari}[1]{X_{#1}}
\newcommand{\Xvaris}[1]{X^2_{#1}}
\newcommand{\Yvar}{Y}
\newcommand{\Yvars}{Y^2}
\newcommand{\Yvari}[1]{Y_{#1}}
\newcommand{\Yvaris}[1]{Y^2_{#1}}
\newcommand{\lnx}{\ln\xvar}
\newcommand{\lnz}{\ln\zvar}
\newcommand{\lnsx}{\ln^2\xvar}
\newcommand{\lnsz}{\ln^2\zvar}
\newcommand{\lncz}{\ln^3\zvar}
\newcommand{\lnomz}{\ln\lpar 1-\zvar\rpar}
\newcommand{\lnsomz}{\ln^2\lpar 1-\zvar\rpar}
\newcommand{\ccoefi}[1]{c_{#1}}
\newcommand{\ccoeft}[2]{c^{#1}_{#2}}
%
%
\newcommand{\Smat}{{\cal{S}}}
\newcommand{\Mmat}{{\cal{M}}}
\newcommand{\Xmat}[1]{X_{#1}}
\newcommand{\XmatI}[1]{X^{-1}_{#1}}
\newcommand{\unitmat}{I}
\newcommand{\Kmat}{{C}}
\newcommand{\Kmatc}{{C}^{\dagger}}
\newcommand{\Kmati}[1]{{C}_{#1}}
\newcommand{\Kmatci}[1]{{C}^{\dagger}_{#1}}
\newcommand{\ffac}[2]{f_{#1}^{#2}}
\newcommand{\Ffac}[1]{F_{#1}}
\newcommand{\Rvec}[2]{R^{(#1)}_{#2}}
\newcommand{\momfl}[2]{#1_{#2}}
\newcommand{\momfs}[2]{#1^2_{#2}}
\newcommand{\fpseZ}{A^{^{FP,Z}}}
\newcommand{\fpseA}{A^{^{FP,A}}}
\newcommand{\fptZ}{T^{^{FP,Z}}}
\newcommand{\fptA}{T^{^{FP,A}}}
\newcommand{\dprop}{\overline\Delta}
\newcommand{\dpropi}[1]{d_{#1}}
\newcommand{\dpropic}[1]{d^{c}_{#1}}
\newcommand{\dpropii}[2]{d_{#1}\lpar #2\rpar}
\newcommand{\dpropis}[1]{d^2_{#1}}
\newcommand{\dproppi}[1]{d'_{#1}}
\newcommand{\psf}[4]{P\lpar #1;#2,#3,#4\rpar}
\newcommand{\ssf}[5]{S^{(#1)}\lpar #2;#3,#4,#5\rpar}
\newcommand{\csf}[5]{C_{_S}^{(#1)}\lpar #2;#3,#4,#5\rpar}
%
%
\newcommand{\lvec}{l}
\newcommand{\lvecs}{l^2}
\newcommand{\lveci}[1]{l_{#1}}
\newcommand{\mvec}{m}
\newcommand{\mvecs}{m^2}
\newcommand{\mveci}[1]{m_{#1}}
\newcommand{\nvec}{n}
\newcommand{\nvecs}{n^2}
\newcommand{\nveci}[1]{n_{#1}}
\newcommand{\epi}[1]{\epsilon_{#1}}
\newcommand{\phep}[1]{\ep_{#1}}
\newcommand{\sphep}{\ep}
\newcommand{\vbep}[1]{e_{#1}}
\newcommand{\vbepp}[1]{e^{+}_{#1}}
\newcommand{\vbepm}[1]{e^{-}_{#1}}
\newcommand{\svbep}{e}
%
%
\newcommand{\lpol}{\lambda}
\newcommand{\spol}{\sigma}
\newcommand{\rpol}{\rho  }
\newcommand{\kpol}{\kappa}
\newcommand{\lpols}{\lambda^2}
\newcommand{\spols}{\sigma^2}
\newcommand{\rpols}{\rho^2}
\newcommand{\kpols}{\kappa^2}
\newcommand{\lpoli}[1]{\lambda_{#1}}
\newcommand{\spoli}[1]{\sigma_{#1}}
\newcommand{\rpoli}[1]{\rho_{#1}}
\newcommand{\kpoli}[1]{\kappa_{#1}}
%
%
\newcommand{\uvec}{u}
\newcommand{\uveci}[1]{u_{#1}}
%
%
\newcommand{\imom}{q}
\newcommand{\imomi}[1]{q_{#1}}
\newcommand{\imoms}{q^2}
\newcommand{\pmom}{p}
\newcommand{\pmomp}{p'}
\newcommand{\pmoms}{p^2}
\newcommand{\pmomq}{p^4}
\newcommand{\pmomx}{p^6}
\newcommand{\pmomi}[1]{p_{#1}}
\newcommand{\pmomis}[1]{p^2_{#1}}
\newcommand{\Pmom}{P}
\newcommand{\Pmoms}{P^2}
\newcommand{\Pmomi}[1]{P_{#1}}
\newcommand{\Pmomis}[1]{P^2_{#1}}
\newcommand{\Kmom}{K}
\newcommand{\Kmoms}{K^2}
\newcommand{\Kmomi}[1]{K_{#1}}
\newcommand{\Kmomis}[1]{K^2_{#1}}
\newcommand{\kmom}{k}
\newcommand{\kmoms}{k^2}
\newcommand{\kmomi}[1]{k_{#1}}
\newcommand{\lmom}{l}
\newcommand{\lmoms}{l^2}
\newcommand{\lmomi}[1]{l_{#1}}
\newcommand{\qmom}{q}
\newcommand{\qmoms}{q^2}
\newcommand{\qmomi}[1]{q_{#1}}
\newcommand{\qmomis}[1]{q^2_{#1}}
\newcommand{\smom}{s}
\newcommand{\smoms}{s^2}
\newcommand{\smomi}[1]{s_{#1}}
\newcommand{\tmom}{t}
\newcommand{\tmoms}{t^2}
\newcommand{\tmomi}[1]{t_{#1}}
\newcommand{\Trmom}{Q}
\newcommand{\Prmom}{P}
\newcommand{\gmv}{Q^2}
\newcommand{\Trmoms}{Q^2}
\newcommand{\Prmoms}{P^2}
\newcommand{\Ptmoms}{T^2}
\newcommand{\Pumoms}{U^2}
\newcommand{\Trmomq}{Q^4}
\newcommand{\Prmomq}{P^4}
\newcommand{\Ptmomq}{T^4}
\newcommand{\Pumomq}{U^4}
\newcommand{\Trmomx}{Q^6}
\newcommand{\Trmomi}[1]{Q_{#1}}
\newcommand{\Trmomis}[1]{Q^2_{#1}}
\newcommand{\Prmomi}[1]{P_{#1}}
\newcommand{\pone}{p_1}
\newcommand{\ptwo}{p_2}
\newcommand{\ptre}{p_3}
\newcommand{\pfor}{p_4}
\newcommand{\pones}{p_1^2}
\newcommand{\ptwos}{p_2^2}
\newcommand{\ptres}{p_3^2}
\newcommand{\pfors}{p_4^2}
\newcommand{\poneq}{p_1^4}
\newcommand{\ptwoq}{p_2^4}
\newcommand{\ptreq}{p_3^4}
\newcommand{\pforq}{p_4^4}
\newcommand{\qmomit}[2]{q_{#1#2}}
\newcommand{\modmom}[1]{\mid{\vec{#1}}\mid}
\newcommand{\modmomi}[2]{\mid{\vec{#1}}_{#2}\mid}
\newcommand{\vect}[1]{{\vec{#1}}}
\newcommand{\Energ}{E}
\newcommand{\Energp}{E'}
\newcommand{\Energpp}{E''}
\newcommand{\Energs}{E^2}
\newcommand{\Energc}{E^3}
\newcommand{\Energf}{E^4}
\newcommand{\Energv}{E^5}
\newcommand{\Energx}{E^6}
\newcommand{\Energi}[1]{E_{#1}}
\newcommand{\Energt}[2]{E_{#1}^{#2}}
\newcommand{\Energis}[1]{E^2_{#1}}
\newcommand{\energ}{e}
\newcommand{\energp}{e'}
\newcommand{\energpp}{e''}
\newcommand{\energs}{e^2}
\newcommand{\energi}[1]{e_{#1}}
\newcommand{\energt}[2]{e_{#1}^{#2}}
\newcommand{\energis}[1]{e^2_{#1}}
\newcommand{\wenerg}{w}
\newcommand{\wenergs}{w^2}
\newcommand{\wenergi}[1]{w_{#1}}
\newcommand{\wenergp}{w'}
\newcommand{\wenergpp}{w''}
%
%
\newcommand{\ecut}{e}
\newcommand{\ecuts}{e^2}
\newcommand{\ecuti}[1]{e^{#1}}
\newcommand{\ccut}{c_m}
\newcommand{\ccuti}[1]{c_{#1}}
\newcommand{\ccuts}{c^2_m}
\newcommand{\scuts}{s^2_m}
\newcommand{\ccutis}[1]{c^2_{#1}}
\newcommand{\ccutic}[1]{c^3_{#1}}
\newcommand{\ccutc}{c^3_m}
\newcommand{\rcut}{\varrho}
\newcommand{\rcuts}{\varrho^2}
\newcommand{\rcuti}[1]{\varrho_{#1}}
\newcommand{\rcutu}[1]{\varrho^{#1}}
\newcommand{\Dcut}{\Delta}
%
\newcommand{\dwf}{\delta_{_{WF}}}
\newcommand{\gbar}{\overline g}
\newcommand{\PP}{\mbox{PP}}
\newcommand{\mv}{m_{_V}}
\newcommand{\bGv}{{\overline\Gamma}_{_V}}
\newcommand{\Umuv}{\hat{\mu}_\ssV}
\newcommand{\Svv}{{\Sigma}_\ssV}
\newcommand{\muv}{p_\ssV}
\newcommand{\muvb}{\mu_{\ssV_{0}}}
\newcommand{\URPvv}{{P}_\ssV}
\newcommand{\RPvv}{{P}_\ssV}
\newcommand{\Svvrem}{{\Sigma}_\ssV^{\mathrm{rem}}}
\newcommand{\USvvrem}{\hat{\Sigma}_\ssV^{\mathrm{rem}}}
\newcommand{\Gv}{\Gamma_{_V}}
%
%
\newcommand{\param}{p}
\newcommand{\parami}[1]{p^{#1}}
\newcommand{\paramb}{p_{0}}
\newcommand{\Zcon}{Z}
\newcommand{\Zconi}[1]{Z_{#1}}
\newcommand{\zconi}[1]{z_{#1}}
\newcommand{\Zconim}[1]{{Z^-_{#1}}}
\newcommand{\zconim}[1]{{z^-_{#1}}}
\newcommand{\Zcont}[2]{Z_{#1}^{#2}}
\newcommand{\zcont}[2]{z_{#1}^{#2}}
\newcommand{\zcontm}[2]{z_{#1}^{{#2}-}}
\newcommand{\sZconi}[2]{\sqrt{Z_{#1}}^{\;#2}}
\newcommand{\php}[3]{e^{#1}_{#2}\lpar #3 \rpar}
\newcommand{\gacome}[1]{\lpar #1 - \gfd\rpar}
\newcommand{\sPj}[2]{\Lambda^{#1}_{#2}}
\newcommand{\sPjs}[2]{\Lambda_{#1,#2}}
\newcommand{\amos}{\mbox{$M^2_{_1}$}}
\newcommand{\amts}{\mbox{$M^2_{_2}$}}
\newcommand{\er}{e_{_{R}}}
\newcommand{\epr}{e'_{_{R}}}
\newcommand{\ers}{e^2_{_{R}}}
\newcommand{\erc}{e^3_{_{R}}}
\newcommand{\erq}{e^4_{_{R}}}
\newcommand{\erf}{e^5_{_{R}}}
\newcommand{\sour}{J}
\newcommand{\sourb}{\overline J}
\newcommand{\lrm}{M_{_R}}
%
%
\newcommand{\vlami}[1]{\lambda_{#1}}
\newcommand{\vlamis}[1]{\lambda^2_{#1}}
\newcommand{\Vvert}{V}
\newcommand{\Avert}{A}
\newcommand{\Svert}{S}
\newcommand{\Pvert}{P}
\newcommand{\vvert}{F}
\newcommand{\Cvert}{\cal{V}}
\newcommand{\Bvert}{\cal{B}}
\newcommand{\Vveri}[2]{V_{#1}^{#2}}
\newcommand{\Fveri}[1]{{\cal{F}}^{#1}}
\newcommand{\Cveri}[1]{{\cal{V}}\lpar{#1}\rpar}
\newcommand{\Bveri}[1]{{\cal{B}}\lpar{#1}\rpar}
\newcommand{\Vverti}[3]{V_{#1}^{#2}\lpar{#3}\rpar}
\newcommand{\Averti}[3]{A_{#1}^{#2}\lpar{#3}\rpar}
\newcommand{\Gverti}[3]{G_{#1}^{#2}\lpar{#3}\rpar}
\newcommand{\Zverti}[3]{Z_{#1}^{#2}\lpar{#3}\rpar}
\newcommand{\Hverti}[2]{H^{#1}\lpar{#2}\rpar}
\newcommand{\Wverti}[3]{W_{#1}^{#2}\lpar{#3}\rpar}
\newcommand{\Cverti}[2]{{\cal{V}}_{#1}^{#2}}
\newcommand{\vverti}[3]{F^{#1}_{#2}\lpar{#3}\rpar}
\newcommand{\averti}[3]{{\overline{F}}^{#1}_{#2}\lpar{#3}\rpar}
\newcommand{\fveone}[1]{f_{#1}}
\newcommand{\fvetri}[3]{f^{#1}_{#2}\lpar{#3}\rpar}
\newcommand{\gvetri}[3]{g^{#1}_{#2}\lpar{#3}\rpar}
\newcommand{\cvetri}[3]{{\cal{F}}^{#1}_{#2}\lpar{#3}\rpar}
\newcommand{\hvetri}[3]{{\hat{\cal{F}}}^{#1}_{#2}\lpar{#3}\rpar}
\newcommand{\avetri}[3]{{\overline{\cal{F}}}^{#1}_{#2}\lpar{#3}\rpar}
\newcommand{\fverti}[2]{F^{#1}_{#2}}
\newcommand{\cverti}[2]{{\cal{F}}_{#1}^{#2}}
\newcommand{\fV}{f_{_{\Vvert}}}
\newcommand{\gA}{g_{_{\Avert}}}
\newcommand{\fVi}[1]{f^{#1}_{_{\Vvert}}}
\newcommand{\seai}[1]{a_{#1}}
\newcommand{\seapi}[1]{a'_{#1}}
\newcommand{\seAi}[2]{A_{#1}^{#2}}
\newcommand{\sewi}[1]{w_{#1}}
\newcommand{\seWi}[1]{W_{#1}}
\newcommand{\seWsi}[1]{W^{*}_{#1}}
\newcommand{\seWti}[2]{W_{#1}^{#2}}
\newcommand{\sewti}[2]{w_{#1}^{#2}}
\newcommand{\seSig}[1]{\Sigma_{#1}\lpar\sla{\pmom}\rpar}
\newcommand{\ww}{w}
%
%
\newcommand{\bbff}[1]{{\overline B}_{#1}}
\newcommand{\sW}{p_{_W}}
\newcommand{\sZ}{p_{_Z}}
\newcommand{\ssp}{s_p}
\newcommand{\fW}{f_{_W}}
\newcommand{\fZ}{f_{_Z}}
\newcommand{\tabn}[1]{Tab.(\ref{#1})}
\newcommand{\subMSB}[1]{{#1}_{\mbox{$\overline{\scriptscriptstyle MS}$}}}
\newcommand{\supMSB}[1]{{#1}^{\mbox{$\overline{\scriptscriptstyle MS}$}}}
\newcommand{\redMSB}{{\mbox{$\overline{\scriptscriptstyle MS}$}}}
\newcommand{\gpbb}{g'_{0}}
\newcommand{\Zconip}[1]{Z'_{#1}}
\newcommand{\bpff}[4]{B'_{#1}\lpar #2;#3,#4\rpar}             
\newcommand{\xidf}{\xi^2-1}
\newcommand{\tDdr}{1/{\bar{\varepsilon}}}
\newcommand{\cRz}{{\cal R}_{_Z}}
\newcommand{\cRg}{{\cal R}_{\gamma}}
\newcommand{\Sz}{\Sigma_{_Z}}
\newcommand{\alh}{{\hat\alpha}}
\newcommand{\alhz}{\alpha_{_Z}}
\newcommand{\Phzg}{{\hat\Pi}_{_{\zb\ab}}}
\newcommand{\fvvert}{F^{\rm vert}_{_V}}
\newcommand{\gavert}{G^{\rm vert}_{_A}}
\newcommand{\bmv}{{\overline m}_{_V}}
\newcommand{\Sgn}{\Sigma_{\gamma\hkn}}
\newcommand{\tabns}[2]{Tabs.(\ref{#1}--\ref{#2})}
\newcommand{\rmboxd}{{\rm Box}_d\lpar s,t,u;M_1,M_2,M_3,M_4\rpar}
\newcommand{\rmboxc}{{\rm Box}_c\lpar s,t,u;M_1,M_2,M_3,M_4\rpar}
%
%
\newcommand{\Afaci}[1]{A_{#1}}
\newcommand{\Afacis}[1]{A^2_{#1}}
\newcommand{\upar}[1]{u}
\newcommand{\upari}[1]{u_{#1}}
\newcommand{\vpari}[1]{v_{#1}}
\newcommand{\lpari}[1]{l_{#1}}
\newcommand{\Lpari}[1]{l_{#1}}
\newcommand{\Nff}[2]{N^{(#1)}_{#2}}
\newcommand{\Sff}[2]{S^{(#1)}_{#2}}
\newcommand{\sSff}{S}
\newcommand{\FQED}[2]{F_{#1#2}}
\newcommand{\fbpsif}{{\overline{\psi}_f}}
\newcommand{\fpsif}{\psi_f}
\newcommand{\etafd}[2]{\eta_d\lpar#1,#2\rpar}
\newcommand{\sigdu}[2]{\sigma_{#1#2}}
\newcommand{\scalc}[4]{c_{_0}\lpar #1;#2,#3,#4\rpar}
\newcommand{\scald}[2]{d_{_0}\lpar #1,#2\rpar}
\newcommand{\pir}[1]{\Pi^{\rm ren}\lpar #1\rpar}
\newcommand{\sigh}{\sigma_{\rm had}}
\newcommand{\dah}{\Delta\alpha^{(5)}_{\rm had}}
\newcommand{\dat}{\Delta\alpha_{\rm top}}
\newcommand{\Vqed}[3]{V_1^{\rm sub}\lpar#1;#2,#3\rpar}
\newcommand{\thetah}{{\hat\theta}}
\newcommand{\mtsix}{m^6_t}
\newcommand{\smlon}{\frac{\mlones}{s}}
\newcommand{\lntwo}{\ln 2}
\newcommand{\wmin}{w_{\rm min}}
\newcommand{\kmin}{k_{\rm min}}
\newcommand{\scaldi}[3]{d_{_0}^{#1}\lpar #2,#3\rpar}
\newcommand{\mdls}{\Big|}
\newcommand{\smf}{\frac{\mfs}{s}}
\newcommand{\bint}{\beta_{\rm int}}
\newcommand{\IRv}{V_{_{\rm IR}}}
\newcommand{\IRr}{R_{_{\rm IR}}}
\newcommand{\fssts}{\frac{s^2}{t^2}}
\newcommand{\fssus}{\frac{s^2}{u^2}}
\newcommand{\optM}{1+\frac{t}{M^2}}
\newcommand{\opuM}{1+\frac{u}{M^2}}
\newcommand{\ftM}{\lpar -\frac{t}{M^2}\rpar}
\newcommand{\fuM}{\lpar -\frac{u}{M^2}\rpar}
\newcommand{\omsM}{1-\frac{s}{M^2}}
\newcommand{\xsf}{\sigma_{_{\rm F}}}
\newcommand{\xsb}{\sigma_{_{\rm B}}}
\newcommand{\afb}{A_{_{\rm FB}}}
\newcommand{\rsoft}{\rm soft}
\newcommand{\rms}{\rm s}
\newcommand{\rsmx}{\sqrt{s_{\rm max}}}
\newcommand{\rspm}{\sqrt{s_{\pm}}}
\newcommand{\rsp}{\sqrt{s_{+}}}
\newcommand{\rsm}{\sqrt{s_{-}}}
\newcommand{\sigmx}{\sigma_{\rm max}}
\newcommand{\gG}[2]{G_{#1}^{#2}}
\newcommand{\gacomm}[2]{\lpar #1 - #2\gfd\rpar}
\newcommand{\fcsx}{\frac{1}{\ctwsix}}
\newcommand{\fcq}{\frac{1}{\ctwf}}
\newcommand{\fcs}{\frac{1}{\ctws}}
\newcommand{\affs}[2]{{\cal A}_{#1}\lpar #2\rpar}                   
\newcommand{\stwei}{s_{\theta}^8}
\def\mdan{\vspace{1mm}\mpar{\hfil$\downarrow$new\hfil}\vspace{-1mm}
          \ignorespaces}
\def\muan{\vspace{-1mm}\mpar{\hfil$\uparrow$new\hfil}\vspace{1mm}\ignorespaces}
\def\mlan{\vspace{-1mm}\mpar{\hfil$\rightarrow$new\hfil}\vspace{1mm}\ignorespaces}
\def\mnnew{\mpar{\hfil NEWNEW \hfil}\ignorespaces}
%
%
\newcommand{\boxc}[2]{{\cal{B}}_{#1}^{#2}}
\newcommand{\boxct}[3]{{\cal{B}}_{#1}^{#2}\lpar{#3}\rpar}
\newcommand{\hboxc}[3]{\hat{{\cal{B}}}_{#1}^{#2}\lpar{#3}\rpar}
\newcommand{\vev}{\langle v \rangle}
\newcommand{\vevi}[1]{\langle v_{#1}\rangle}
\newcommand{\vevs}{\langle v^2   \rangle}
\newcommand{\fwfrV}[5]{\Sigma_{_V}\lpar #1,#2,#3;#4,#5 \rpar}
\newcommand{\fwfrS}[7]{\Sigma_{_S}\lpar #1,#2,#3;#4,#5;#6,#7 \rpar}
\newcommand{\fSi}[1]{f^{#1}_{_{\Svert}}}
\newcommand{\fPi}[1]{f^{#1}_{_{\Pvert}}}
\newcommand{\mXs}{m_{_X}}
\newcommand{\mXss}{m^2_{_X}}
\newcommand{\mYs}{M^2_{_Y}}
\newcommand{\xik}{\xi_k}
\newcommand{\xiks}{\xi^2_k}
\newcommand{\mpls}{m^2_+}
\newcommand{\mmis}{m^2_-}
%
\newcommand{\SN}{\Sigma_{_N}}
\newcommand{\SC}{\Sigma_{_C}}
\newcommand{\SPN}{\Sigma'_{_N}}
\newcommand{\PFf}{\Pi^{\fer}_{_F}}
\newcommand{\PFb}{\Pi^{\bos}_{_F}}
\newcommand{\dPZ}{\Delta{\hat\Pi}_{_Z}}
\newcommand{\Sfin}{\Sigma_{_F}}
\newcommand{\Sfir}{\Sigma_{_R}}
\newcommand{\Sfinh}{{\hat\Sigma}_{_F}}
\newcommand{\Sfinf}{\Sigma^{\fer}_{_F}}
\newcommand{\Sfinbh}{\Sigma^{\bos}_{_F}}
\newcommand{\alf}{\alpha^{\fer}}
\newcommand{\alhfz}{\alpha^{\fer}\lpar{\ssZ}\rpar}
\newcommand{\alhfs}{\alpha^{\fer}\lpar{\sman}\rpar}
\newcommand{\gfQ}{g^f_{_{Q}}}
\newcommand{\gfL}{g^f_{_{L}}}
\newcommand{\ccf}{\frac{\gbs}{16\,\pi^2}}
\newcommand{\chq}{{\hat c}^4}
\newcommand{\muuq}{m_{u'}}
\newcommand{\muus}{m^2_{u'}}
\newcommand{\mdd}{m_{d'}}
\newcommand{\clf}[2]{\mathrm{Cli}_{_#1}\lpar\displaystyle{#2}\rpar}
\def\stes{\sin^2\theta}
\def\acal{\cal A}
\def\alr{A_{_{\rm{LR}}}}
\newcommand{\barQ}{\overline Q}
\newcommand{\Sptg}{\Sigma'_{_{3Q}}}
\newcommand{\Sptt}{\Sigma'_{_{33}}}
\newcommand{\Ppgg}{\Pi'_{\ph\ph}}
\newcommand{\Pww}{\Pi_{_{\wb\wb}}}
\newcommand{\capV}[2]{{\cal F}^{#2}_{_{#1}}}
\newcommand{\bt}{\beta_t}
\newcommand{\mhsix}{M^6_{_H}}
\newcommand{\topt}{{\cal T}_{33}}
\newcommand{\topq}{{\cal T}_4}
\newcommand{\Phzgf}{{\hat\Pi}^{\fer}_{_{\zb\ab}}}
\newcommand{\Phzgb}{{\hat\Pi}^{\bos}_{_{\zb\ab}}}
\newcommand{\Sfirh}{{\hat\Sigma}_{_R}}
\newcommand{\Szgh}{{\hat\Sigma}_{_{\zb\ab}}}
\newcommand{\Szghb}{{\hat\Sigma}^{\bos}_{_{\zb\ab}}}
\newcommand{\Szghf}{{\hat\Sigma}^{\fer}_{_{\zb\ab}}}
\newcommand{\Szgb}{\Sigma^{\bos}_{_{\zb\ab}}}
\newcommand{\Szgf}{\Sigma^{\fer}_{_{\zb\ab}}}
\newcommand{\chig}{\chi_{_{\ph}}}
\newcommand{\chiz}{\chi_{_{\zb}}}
\newcommand{\Sfih}{{\hat\Sigma}}
\newcommand{\Szzh}{\hat{\Sigma}_{_{\zb\zb}}}
\newcommand{\dPZf}{\Delta{\hat\Pi}^f_{_{\zb}}}
\newcommand{\khZdf}[1]{{\hat\kappa}^{#1}_{_{\zb}}}
\newcommand{\chf}{{\hat c}^4}
\newcommand{\amp}[2]{{\cal{A}}_{_{#1}}^{\rm{#2}}}
\newcommand{\hatvm}[1]{{\hat v}^-_{#1}}
\newcommand{\hatvp}[1]{{\hat v}^+_{#1}}
\newcommand{\hatvpm}[1]{{\hat v}^{\pm}_{#1}}
\newcommand{\kvz}[1]{\kappa^{\zb #1}_{_V}}
\newcommand{\barp}{\overline p}                
\newcommand{\delw}{\Delta_{_{\wb}}}
\newcommand{\bdelw}{{\bar{\Delta}}_{_{\wb}}}
\newcommand{\bdelf}{{\bar{\Delta}}_{\ff}}
\newcommand{\delz}{\Delta_{_\zb}}
\newcommand{\deli}[1]{\Delta\lpar{#1}\rpar}
\newcommand{\chizb}{\chi_{_\zb}}
\newcommand{\Swwp}{\Sigma'_{_{\wb\wb}}}
\newcommand{\epph}{\varepsilon'/2}
\newcommand{\sbffp}[1]{B'_{#1}}                    
\newcommand{\epss}{\varepsilon^*}
\newcommand{\Ddrhs}{{\ds\frac{1}{\hat{\varepsilon}^2}}}
\newcommand{\lnmsb}{L_{_\wb}}
\newcommand{\lnsmsb}{L^2_{_\wb}}
\newcommand{\tpni}{\lpar 2\pi\rpar^n\ib}
\newcommand{\tpn}{2^n\,\pi^{n-2}}
\newcommand{\cmf}{M_f}
\newcommand{\cmfs}{M^2_f}
\newcommand{\toDdr}{{\ds\frac{2}{{\bar{\varepsilon}}}}}
\newcommand{\troDdr}{{\ds\frac{3}{{\bar{\varepsilon}}}}}
\newcommand{\totDdr}{{\ds\frac{3}{{2\,\bar{\varepsilon}}}}}
\newcommand{\foDdr}{{\ds\frac{4}{{\bar{\varepsilon}}}}}
\newcommand{\smh}{m_{_H}}
\newcommand{\smhs}{m^2_{_H}}
\newcommand{\Ph}{\Pi_{_\hb}}
\newcommand{\Sphh}{\Sigma'_{_{\hb\hb}}}
\newcommand{\bh}{\beta}
\newcommand{\alsn}{\alpha^{(n_f)}_{_S}}
\newcommand{\smq}{m_q}
\newcommand{\smqp}{m_{q'}}
\newcommand{\shb}{h}
\newcommand{\hab}{A}
\newcommand{\hbpm}{H^{\pm}}
\newcommand{\hbp}{H^{+}}
\newcommand{\hbm}{H^{-}}
\newcommand{\msh}{M_h}
\newcommand{\mha}{M_{_A}}
\newcommand{\mhc}{M_{_{H^{\pm}}}}
\newcommand{\mshs}{M^2_h}
\newcommand{\mhas}{M^2_{_A}}
\newcommand{\barfp}{\overline{f'}}                
\newcommand{\chiii}{{\hat c}^3}
\newcommand{\chiv}{{\hat c}^4}
\newcommand{\chv}{{\hat c}^5}
\newcommand{\chvi}{{\hat c}^6}
\newcommand{\alsvi}{\alpha^{6}_{_S}}
\newcommand{\tww}{t_{_W}}
\newcommand{\ti}{t_{_1}}
\newcommand{\tii}{t_{_2}}
\newcommand{\tiii}{t_{_3}}
\newcommand{\tiv}{t_{_4}}
\newcommand{\psla}{\hbox{\rlap/p}}
\newcommand{\qsla}{\hbox{\rlap/q}}
\newcommand{\nsla}{\hbox{\rlap/n}}
\newcommand{\lsla}{\hbox{\rlap/l}}
\newcommand{\msla}{\hbox{\rlap/m}}
\newcommand{\cnsla}{\hbox{\rlap/N}}
\newcommand{\clsla}{\hbox{\rlap/L}}
\newcommand{\cmsla}{\hbox{\rlap/M}}
\newcommand{\blmt}{\lrbr - 3\rrbr}
\newcommand{\blfo}{\lrbr 4 1\rrbr}
\newcommand{\bltp}{\lrbr 2 +\rrbr}
\newcommand{\clitwo}[1]{{\rm{Li}}_{2}\lpar{#1}\rpar}
\newcommand{\clitri}[1]{{\rm{Li}}_{3}\lpar{#1}\rpar}
\newcommand{\xt}{x_{\ft}}
\newcommand{\zt}{z_{\ft}}
\newcommand{\Ht}{h_{\ft}}
\newcommand{\xts}{x^2_{\ft}}
\newcommand{\zts}{z^2_{\ft}}
\newcommand{\Hts}{h^2_{\ft}}
\newcommand{\ztc}{z^3_{\ft}}
\newcommand{\Htc}{h^3_{\ft}}
\newcommand{\ztq}{z^4_{\ft}}
\newcommand{\Htq}{h^4_{\ft}}
\newcommand{\ztv}{z^5_{\ft}}
\newcommand{\Htv}{h^5_{\ft}}
\newcommand{\ztx}{z^6_{\ft}}
\newcommand{\Htx}{h^6_{\ft}}
\newcommand{\ztz}{z^7_{\ft}}
\newcommand{\Htz}{h^7_{\ft}}
\newcommand{\sht}{\sqrt{\Ht}}
\newcommand{\atan}[1]{{\rm{arctan}}\lpar{#1}\rpar}
\newcommand{\dbff}[3]{{\hat{B}}_{_{{#2}{#3}}}\lpar{#1}\rpar}
\newcommand{\ztbs}{{\bar{z}}^{2}_{\ft}}
\newcommand{\ztb}{{\bar{z}}_{\ft}}
\newcommand{\Htbs}{{\bar{h}}^{2}_{\ft}}
\newcommand{\Htb}{{\bar{h}}_{\ft}}
\newcommand{\Hztb}{{\bar{hz}}_{\ft}}
\newcommand{\Ln}[1]{{\rm{Ln}}\lpar{#1}\rpar}
\newcommand{\Lns}[1]{{\rm{Ln}}^2\lpar{#1}\rpar}
\newcommand{\wt}{w_{\ft}}
\newcommand{\wts}{w^2_{\ft}}
\newcommand{\wtb}{\overline{w}}
\newcommand{\fra}{\frac{1}{2}}
\newcommand{\frb}{\frac{1}{4}}
\newcommand{\frc}{\frac{3}{2}}
\newcommand{\frd}{\frac{3}{4}}
\newcommand{\fre}{\frac{9}{2}}
\newcommand{\frf}{\frac{9}{4}}
\newcommand{\frg}{\frac{5}{4}}
\newcommand{\frh}{\frac{5}{2}}
\newcommand{\fri}{\frac{1}{8}}
\newcommand{\frj}{\frac{7}{4}}
\newcommand{\frl}{\frac{7}{8}}
\newcommand{\Spzzh}{\hat{\Sigma}'_{_{\zb\zb}}}
\newcommand{\sss}{s\sqrt{s}}
\newcommand{\sqs}{\sqrt{s}}
\newcommand{\Rtg}{R_{_{3Q}}}
\newcommand{\Rtt}{R_{_{33}}}
\newcommand{\Rww}{R_{_{\wb\wb}}}
\newcommand{\ssZ}{{\scriptscriptstyle{\zb}}}
\newcommand{\ssW}{{\scriptscriptstyle{\wb}}}
\newcommand{\ssH}{{\scriptscriptstyle{\hb}}}
\newcommand{\ssV}{{\scriptscriptstyle{\vb}}}
\newcommand{\ssA}{{\scriptscriptstyle{A}}}
\newcommand{\ssB}{{\scriptscriptstyle{B}}}
\newcommand{\ssC}{{\scriptscriptstyle{C}}}
\newcommand{\ssD}{{\scriptscriptstyle{D}}}
\newcommand{\ssF}{{\scriptscriptstyle{F}}}
\newcommand{\ssG}{{\scriptscriptstyle{G}}}
\newcommand{\ssL}{{\scriptscriptstyle{L}}}
\newcommand{\ssM}{{\scriptscriptstyle{M}}}
\newcommand{\ssN}{{\scriptscriptstyle{N}}}
\newcommand{\ssP}{{\scriptscriptstyle{P}}}
\newcommand{\ssQ}{{\scriptscriptstyle{Q}}}
\newcommand{\ssR}{{\scriptscriptstyle{R}}}
\newcommand{\ssS}{{\scriptscriptstyle{S}}}
\newcommand{\ssT}{{\scriptscriptstyle{T}}}
\newcommand{\ssU}{{\scriptscriptstyle{U}}}
\newcommand{\ssX}{{\scriptscriptstyle{X}}}
\newcommand{\ssY}{{\scriptscriptstyle{Y}}}
\newcommand{\ssWF}{{\scriptscriptstyle{WF}}}
\newcommand{\DiagramFermionToBosonFullWithMomenta}[8][70]{
  \vcenter{\hbox{
  \SetScale{0.8}
  \begin{picture}(#1,50)(15,15)
    \put(27,22){$\nearrow$}      
    \put(27,54){$\searrow$}    
    \put(59,29){$\to$}    
    \ArrowLine(25,25)(50,50)      \Text(34,20)[lc]{#6} \Text(11,20)[lc]{#3}
    \ArrowLine(50,50)(25,75)      \Text(34,60)[lc]{#7} \Text(11,60)[lc]{#4}
    \Photon(50,50)(90,50){2}{8}   \Text(80,40)[lc]{#2} \Text(55,33)[ct]{#8}
    \Vertex(50,50){2,5}          \Text(60,48)[cb]{#5} 
    \Vertex(90,50){2}
  \end{picture}}}
  }
\newcommand{\DiagramFermionToBosonPropagator}[4][85]{
  \vcenter{\hbox{
  \SetScale{0.8}
  \begin{picture}(#1,50)(15,15)
    \ArrowLine(25,25)(50,50)
    \ArrowLine(50,50)(25,75)
    \Photon(50,50)(105,50){2}{8}   \Text(90,40)[lc]{#2}
    \Vertex(50,50){0.5}         \Text(80,48)[cb]{#3}
    \GCirc(82,50){8}{1}            \Text(55,48)[cb]{#4}
    \Vertex(105,50){2}
  \end{picture}}}
  }
\newcommand{\DiagramFermionToBosonEffective}[3][70]{
  \vcenter{\hbox{
  \SetScale{0.8}
  \begin{picture}(#1,50)(15,15)
    \ArrowLine(25,25)(50,50)
    \ArrowLine(50,50)(25,75)
    \Photon(50,50)(90,50){2}{8}   \Text(80,40)[lc]{#2}
    \BBoxc(50,50)(5,5)            \Text(55,48)[cb]{#3}
    \Vertex(90,50){2}
  \end{picture}}}
  }
\newcommand{\DiagramFermionToBosonFull}[3][70]{
  \vcenter{\hbox{
  \SetScale{0.8}
  \begin{picture}(#1,50)(15,15)
    \ArrowLine(25,25)(50,50)
    \ArrowLine(50,50)(25,75)
    \Photon(50,50)(90,50){2}{8}   \Text(80,40)[lc]{#2}
    \Vertex(50,50){2.5}          \Text(60,48)[cb]{#3}
    \Vertex(90,50){2}
  \end{picture}}}
  }
\newcommand{\expgw}{\frac{\gf\mws}{2\srt\,\pi^2}}
\newcommand{\expgz}{\frac{\gf\mzs}{2\srt\,\pi^2}}
\newcommand{\Spww}{\Sigma'_{_{\wb\wb}}}
\newcommand{\shf}{{\hat s}^4}
\newcommand{\acz}{\scff{0}}
\newcommand{\acoo}{\scff{11}}
\newcommand{\acod}{\scff{12}}
\newcommand{\acdo}{\scff{21}}
\newcommand{\acdd}{\scff{22}}
\newcommand{\acdt}{\scff{23}}
\newcommand{\acdq}{\scff{24}}
\newcommand{\acto}{\scff{31}}
\newcommand{\actd}{\scff{32}}
\newcommand{\actt}{\scff{33}}
\newcommand{\actq}{\scff{34}}
\newcommand{\actc}{\scff{35}}
\newcommand{\acts}{\scff{36}}
\newcommand{\acoA}{\scff{1A}}
\newcommand{\acdA}{\scff{2A}}
\newcommand{\acdB}{\scff{2B}}
\newcommand{\acdC}{\scff{2C}}
\newcommand{\acdD}{\scff{2D}}
\newcommand{\actA}{\scff{3A}}
\newcommand{\actB}{\scff{3B}}
\newcommand{\actC}{\scff{3C}}
\newcommand{\ada}{\sdff{0}}
\newcommand{\adb}{\sdff{11}}
\newcommand{\adc}{\sdff{12}}
\newcommand{\add}{\sdff{13}}
\newcommand{\ade}{\sdff{21}}
\newcommand{\adf}{\sdff{22}}
\newcommand{\adg}{\sdff{23}}
\newcommand{\adh}{\sdff{24}}
\newcommand{\adi}{\sdff{25}}
\newcommand{\adj}{\sdff{26}}
\newcommand{\adl}{\sdff{27}}
\newcommand{\adm}{\sdff{31}}
\newcommand{\adn}{\sdff{32}}
\newcommand{\ado}{\sdff{33}}
\newcommand{\adp}{\sdff{34}}
\newcommand{\adq}{\sdff{35}}
\newcommand{\adr}{\sdff{36}}
\newcommand{\ads}{\sdff{37}}
\newcommand{\adt}{\sdff{38}}
\newcommand{\adu}{\sdff{39}}
\newcommand{\adw}{\sdff{310}}
\newcommand{\adv}{\sdff{311}}
\newcommand{\ady}{\sdff{312}}
\newcommand{\adz}{\sdff{313}}
\newcommand{\admt}{\frac{\tman}{\sman}}
\newcommand{\admu}{\frac{\uman}{\sman}}
\newcommand{\frm}{\frac{3}{8}}
\newcommand{\frn}{\frac{5}{8}}
\newcommand{\fro}{\frac{15}{8}}
\newcommand{\frp}{\frac{3}{16}}
\newcommand{\frq}{\frac{5}{16}}
\newcommand{\frr}{\frac{1}{16}}
\newcommand{\frs}{\frac{7}{2}}
\newcommand{\frt}{\frac{7}{16}}
\newcommand{\fru}{\frac{1}{3}}
\newcommand{\frw}{\frac{2}{3}}
\newcommand{\frz}{\frac{4}{3}}
\newcommand{\fry}{\frac{13}{3}}
\newcommand{\fraa}{\frac{11}{4}}
\newcommand{\bee}{\beta_{e}}
\newcommand{\beW}{\beta_{_\wb}}
\newcommand{\toDdrh}{{\ds\frac{2}{{\hat{\varepsilon}}}}}
\newcommand{\bqas}{\begin{eqnarray*}}
\newcommand{\eqas}{\end{eqnarray*}}
\newcommand{\mhcub}{M^3_{_H}}
\newcommand{\adComA}{\sdff{A}}
\newcommand{\adComB}{\sdff{B}}
\newcommand{\adComC}{\sdff{C}}
\newcommand{\adComD}{\sdff{D}}
\newcommand{\adComE}{\sdff{E}}
\newcommand{\adComF}{\sdff{F}}
\newcommand{\adComG}{\sdff{G}}
\newcommand{\adComH}{\sdff{H}}
\newcommand{\adComI}{\sdff{I}}
\newcommand{\adComJ}{\sdff{J}}
\newcommand{\adComL}{\sdff{L}}
\newcommand{\adComM}{\sdff{M}}
\newcommand{\adComN}{\sdff{N}}
\newcommand{\adComO}{\sdff{O}}
\newcommand{\adComP}{\sdff{P}}
\newcommand{\adComQ}{\sdff{Q}}
\newcommand{\adComR}{\sdff{R}}
\newcommand{\adComS}{\sdff{S}}
\newcommand{\adComT}{\sdff{T}}
\newcommand{\adComU}{\sdff{U}}
\newcommand{\adComAc}{\sdff{A}^c}
\newcommand{\adComBc}{\sdff{B}^c}
\newcommand{\adComCc}{\sdff{C}^c}
\newcommand{\adComDc}{\sdff{D}^c}
\newcommand{\adComEc}{\sdff{E}^c}
\newcommand{\adComFc}{\sdff{F}^c}
\newcommand{\adComGc}{\sdff{G}^c}
\newcommand{\adComHc}{\sdff{H}^c}
\newcommand{\adComIc}{\sdff{I}^c}
\newcommand{\adComJc}{\sdff{J}^c}
\newcommand{\adComLc}{\sdff{L}^c}
\newcommand{\adComMc}{\sdff{M}^c}
\newcommand{\adComNc}{\sdff{N}^c}
\newcommand{\adComOc}{\sdff{O}^c}
\newcommand{\adComPc}{\sdff{P}^c}
\newcommand{\adComQc}{\sdff{Q}^c}
\newcommand{\adComRc}{\sdff{R}^c}
\newcommand{\adComSc}{\sdff{S}^c}
\newcommand{\adComTc}{\sdff{T}^c}
\newcommand{\adComUc}{\sdff{U}^c}
\newcommand{\adComAf}{\sdff{A}^f}
\newcommand{\adComBf}{\sdff{B}^f}
\newcommand{\adComCf}{\sdff{F}^f}
\newcommand{\adComDf}{\sdff{D}^f}
\newcommand{\adComEf}{\sdff{E}^f}
\newcommand{\adComFf}{\sdff{F}^f}
\newcommand{\adComGf}{\sdff{G}^f}
\newcommand{\adComHf}{\sdff{H}^f}
\newcommand{\adComIf}{\sdff{I}^f}
\newcommand{\adComJf}{\sdff{J}^f}
\newcommand{\adComLf}{\sdff{L}^f}
\newcommand{\adComMf}{\sdff{M}^f}
\newcommand{\adComNf}{\sdff{N}^f}
\newcommand{\adComOf}{\sdff{O}^f}
\newcommand{\adComPf}{\sdff{P}^f}
\newcommand{\adComQf}{\sdff{Q}^f}
\newcommand{\adComRf}{\sdff{R}^f}
\newcommand{\adComSf}{\sdff{S}^f}
\newcommand{\adComTf}{\sdff{T}^f}
\newcommand{\adComUf}{\sdff{U}^f}
\newcommand{\adComBfc}{\sdff{B}^{fc}} 
\newcommand{\adComCfco}{\sdff{C}^{fc1}}
\newcommand{\adComCfcd}{\sdff{C}^{fc2}} 
\newcommand{\adComCfct}{\sdff{C}^{fc3}} 
\newcommand{\adComDfc}{\sdff{D}^{fc}}
\newcommand{\adComEfc}{\sdff{E}^{fc}}
\newcommand{\adComFfc}{\sdff{F}^{fc}}
\newcommand{\adComGfc}{\sdff{G}^{fc}}
\newcommand{\adComHfc}{\sdff{H}^{fc}}
\newcommand{\adComLfc}{\sdff{L}^{fc}}
\newcommand{\afba}[1]{A^{#1}_{_{\rm FB}}}
\newcommand{\alra}[1]{A^{#1}_{_{\rm LR}}}
\newcommand{\adComAt}{\sdff{A}^t}
\newcommand{\adComBt}{\sdff{B}^t}
\newcommand{\adComCt}{\sdff{T}^t}
\newcommand{\adComDt}{\sdff{D}^t}
\newcommand{\adComEt}{\sdff{E}^t}
\newcommand{\adComFt}{\sdff{T}^t}
\newcommand{\adComGt}{\sdff{G}^t}
\newcommand{\adComHt}{\sdff{H}^t}
\newcommand{\adComIt}{\sdff{I}^t}
\newcommand{\adComJt}{\sdff{J}^t}
\newcommand{\adComLt}{\sdff{L}^t}
\newcommand{\adComMt}{\sdff{M}^t}
\newcommand{\adComNt}{\sdff{N}^t}
\newcommand{\adComOt}{\sdff{O}^t}
\newcommand{\adComPt}{\sdff{P}^t}
\newcommand{\adComQt}{\sdff{Q}^t}
\newcommand{\adComRt}{\sdff{R}^t}
\newcommand{\adComSt}{\sdff{S}^t}
\newcommand{\adComTt}{\sdff{T}^t}
\newcommand{\adComUt}{\sdff{U}^t}
\newcommand{\adComAtt}{\sdff{A}^{\tau}}
\newcommand{\adComBtt}{\sdff{B}^{\tau}}
\newcommand{\adComCtt}{\sdff{T}^{\tau}}
\newcommand{\adComDtt}{\sdff{D}^{\tau}}
\newcommand{\adComEtt}{\sdff{E}^{\tau}}
\newcommand{\adComFtt}{\sdff{T}^{\tau}}
\newcommand{\adComGtt}{\sdff{G}^{\tau}}
\newcommand{\adComHtt}{\sdff{H}^{\tau}}
\newcommand{\adComItt}{\sdff{I}^{\tau}}
\newcommand{\adComJtt}{\sdff{J}^{\tau}}
\newcommand{\adComLtt}{\sdff{L}^{\tau}}
\newcommand{\adComMtt}{\sdff{M}^{\tau}}
\newcommand{\adComNtt}{\sdff{N}^{\tau}}
\newcommand{\adComOtt}{\sdff{O}^{\tau}}
\newcommand{\adComPtt}{\sdff{P}^{\tau}}
\newcommand{\adComQtt}{\sdff{Q}^{\tau}}
\newcommand{\adComRtt}{\sdff{R}^{\tau}}
\newcommand{\adComStt}{\sdff{S}^{\tau}}
\newcommand{\adComTtt}{\sdff{T}^{\tau}}
\newcommand{\adComUtt}{\sdff{U}^{\tau}}
\newcommand{\etavz}[1]{\eta^{\zb #1}_{_V}}
\newcommand{\phanst}{$\hphantom{\sigma^{s+t}\ }$}
\newcommand{\phanat}{$\hphantom{A_{FB}^{s+t}\ }$}
\newcommand{\phanss}{$\hphantom{\sigma^{s}\ }$}
\newcommand{\phanas}{$\hphantom{A_{FB}^{s}\ }$} 
\newcommand{\pbb}{\,\mbox{\bf pb}}
\newcommand{\pe}{\,\%\:}
\newcommand{\pc}{\,\%}
\newcommand{\temiv}{10^{-4}}
\newcommand{\temv}{10^{-5}}
\newcommand{\temvi}{10^{-6}}
\newcommand{\di}[1]{d_{#1}}
\newcommand{\delip}[1]{\Delta_+\lpar{#1}\rpar}
\newcommand{\propbb}[5]{{{#1}\over {\lpar #2^2 + #3 - \ib\varepsilon\rpar
\lpar\lpar #4\rpar^2 + #5 -\ib\varepsilon\rpar}}}
\newcommand{\cfft}[5]{C_{#1}\lpar #2;#3,#4,#5\rpar}    
\newcommand{\ppl}[1]{p_{+{#1}}}
\newcommand{\pmi}[1]{p_{-{#1}}}
\newcommand{\bpox}{\beta^2_{\xi}}
\newcommand{\bffdiff}[5]{B_{\rm d}\lpar #1;#2,#3;#4,#5\rpar}             
\newcommand{\cffdiff}[7]{C_{\rm d}\lpar #1;#2,#3,#4;#5,#6,#7\rpar}    
\newcommand{\affdiff}[2]{A_{\rm d}\lpar #1;#2\rpar}             
\newcommand{\Dqf}{\Delta\qf}
\newcommand{\bposx}{\beta^4_{\xi}}
\newcommand{\svverti}[3]{f^{#1}_{#2}\lpar{#3}\rpar}
\newcommand{\Mods}{\mbox{$M^2_{12}$}}
\newcommand{\Mots}{\mbox{$M^2_{13}$}}
\newcommand{\Motq}{\mbox{$M^4_{13}$}}
\newcommand{\Mdts}{\mbox{$M^2_{23}$}}
\newcommand{\Mdos}{\mbox{$M^2_{21}$}}
\newcommand{\Mtds}{\mbox{$M^2_{32}$}}
\newcommand{\dffpt}[3]{D_{#1}\lpar #2,#3;}           
\newcommand{\quu}{Q_{uu}}
\newcommand{\qdd}{Q_{dd}}
\newcommand{\qud}{Q_{ud}}
\newcommand{\qdu}{Q_{du}}
\newcommand{\msPj}[6]{\Lambda^{#1#2#3}_{#4#5#6}}
\newcommand{\bdiff}[4]{B_{\rm d}\lpar #1,#2;#3,#4\rpar}             
\newcommand{\bdifff}[7]{B_{\rm d}\lpar #1;#2;#3;#4,#5;#6,#7\rpar}             
\newcommand{\adiff}[3]{A_{\rm d}\lpar #1;#2;#3\rpar}  
\newcommand{\aw}{a_{_\wb}}
\newcommand{\az}{a_{_\zb}}
\newcommand{\sct}[1]{sect.~\ref{#1}}
\newcommand{\dreim}[1]{\varepsilon^{\rm M}_{#1}}
\newcommand{\drem}{\varepsilon^{\rm M}}
\newcommand{\hcapV}[2]{{\hat{\cal F}}^{#2}_{_{#1}}}
\newcommand{\swww}{{\scriptscriptstyle \wb\wb\wb}}
\newcommand{\szhz}{{\scriptscriptstyle \zb\hb\zb}}
\newcommand{\shzh}{{\scriptscriptstyle \hb\zb\hb}}
\newcommand{\bwith}[3]{\beta^{#3}_{#1}\lpar #2\rpar}
\newcommand{\Shhh}{{\hat\Sigma}_{_{\hb\hb}}}
\newcommand{\Sphhh}{{\hat\Sigma}'_{_{\hb\hb}}}
\newcommand{\seWilc}[1]{w_{#1}}
\newcommand{\seWtilc}[2]{w_{#1}^{#2}}
\newcommand{\eilc}{\gamma}
\newcommand{\eilcs}{\gamma^2}
\newcommand{\eilcc}{\gamma^3}
\newcommand{\eilcb}{{\overline{\gamma}}}
\newcommand{\eilcbs}{{\overline{\gamma}^2}}
\newcommand{\Sttww}{\Sigma_{_{33;\wb\wb}}}
\newcommand{\bSttww}{{\overline\Sigma}_{_{33;\wb\wb}}}
\newcommand{\Pggtg}{\Pi_{\ph\ph;3Q}}
\newcommand{\bDelta}{{\bar\Delta}}
\newcommand{\tDelta}{{\tilde\Delta}}
\newcommand{\tcft}[1]{C_{#1}\lpar t\rpar}
\newcommand{\tcftt}[1]{C_{#1}\lpar tt\rpar}
\newcommand{\tcfp}[1]{C^+_{#1}}
\newcommand{\tcfm}[1]{C^-_{#1}}

%
%
\title{Unstable Particles and Non-Conserved Currents:\\
A Generalization of the Fermion-Loop Scheme}

\author{
Giampiero Passarino  \\
{\em Dipartimento di Fisica Teorica, Universit\`a di Torino, Italy} \\
{\em INFN, Sezione di Torino, Italy} \\ \\ \\
}

\date{}
\maketitle

\begin{abstract}
  \normalsize \noindent 
The incorporation of finite-width effects in the theoretical predictions
for tree-level processes $e^+e^- \to n\,$fermions requires that gauge 
invariance must not be violated.
Among various schemes proposed in the literature, the most satisfactory,
from the point of view of field theory is the so-called Fermion-Loop scheme.
It consists in the re-summation of the fermionic one-loop corrections to the 
vector-boson propagators and the inclusion of all remaining fermionic one-loop 
corrections, in particular those to the Yang--Mills vertices. 
In the original formulation, the Fermion-Loop scheme requires that vector
bosons couple to conserved currents, i.e., that the masses of all external
fermions be neglected. There are several examples where fermion masses must
be kept to obtain a reliable prediction. The most famous one is the so-called
single-$\wb$ production mechanism, the process $e^+e^- \to e^- \barnu_e f_1
\barf_2$ where the outgoing electron is collinear, within a small cone, with
the incoming electron. Therefore, $\me$ cannot be neglected.
Furthermore, among the $20$ Feynman diagrams that contribute (for $e\barnu_e 
u\bard$ final states, up to $56$ for $e^+e^-\nu_e\barnu_e$) there are 
multi-peripheral ones that require a non-vanishing mass also for the other
fermions. A generalization of the Fermion-Loop scheme is introduced to
account for external, non-conserved, currents. Dyson re-summed transitions
are introduced without neglecting the $p_{\mu}p_{\nu}$-terms and including
the contributions from the Higgs-Kibble ghosts in the 't Hooft-Feynman gauge.
Running vector boson masses are introduced and their relation with the 
corresponding complex poles are investigated. It is shown that any 
$\Smat$-matrix element takes a very simple form when written in terms of
these running masses. A special example of Ward identity, the U(1) Ward
identity for single-$\wb$, is derived in a situation where all currents
are non-conserved and where the top quark mass is not neglected inside loops.

\end{abstract}

\clearpage

%
%

\section{Introduction.}
The incorporation of finite-width effects in the theoretical predictions for
LEP2 processes and beyond necessitates a careful treatment. Independently of 
how finite widths of propagating particles are introduced, this requires 
a re-summation of the vacuum-polarization effects. Furthermore,
the principle of gauge invariance must not be violated, i.e.,
the Ward identities, have to be preserved.

In a series of two papers, \cite{kn:bhf1} and \cite{kn:bhf2}, one can find a 
complete description of several schemes that allow the incorporation of 
finite-width effects in tree-level amplitudes without spoiling gauge 
invariance. 

In~\cite{kn:bhf1} it was argued that the preferable (Fermion-Loop) scheme 
consists in the re-summation of the fermionic one-loop corrections to the 
vector-boson propagators and the inclusion of all remaining fermionic one-loop 
corrections, in particular those to the Yang--Mills vertices. 
This re-summation of one-particle-irreducible (1PI) fermionic $\ord{\alpha}$ 
corrections involves the closed set of all $\ord{[N^f_c \alpha/\pi]^i}$ 
(leading color-factor) corrections, and is as such
manifestly gauge-invariant. These corrections constitute the bulk of the
width effects for gauge bosons and an important part of the complete set of 
weak corrections. 

In Ref.~\cite{kn:bhf1} the main incentive was the discussion of the process
$e^+e^- \to e^-\barnu_e u\bard$ at small scattering angles and LEP2 
energies. 
Naive inclusion of the finite $\wb$-boson width breaks U(1) electromagnetic
gauge invariance and leads to a totally wrong cross-section 
in the collinear limit, as e.g.\ discussed in Ref.~\cite{kn:Kurihara}. 
By taking into account in addition the imaginary
parts arising from cutting the massless fermion loops in the
triple-gauge-boson vertex, U(1) gauge invariance is restored and a sensible
cross-section is obtained.

In Ref.~\cite{kn:bhf2} the authors presented the details of the full-fledged 
Fermion-Loop scheme, taking into account the complete fermionic one-loop
corrections including all real and imaginary parts, and all contributions of 
the massive top quark. 
A proper treatment of the neutral gauge-boson propagators is performed
by solving the Dyson equations for the photon, $\zb$-boson, 
and mixed photon--$\zb$ propagators.
This is necessary to guarantee the unitarity cancellations at high energies. 
The top-quark contributions are particularly important for 
delayed-unitarity effects. In this respect also terms involving the
totally-antisymmetric $\varepsilon$-tensor (originating from vertex
corrections) are relevant.
While such terms are absent for complete generations of massless fermions
owing to the anomaly cancellations, they show up  for finite fermion
masses. As the $\varepsilon$-dependent terms satisfy the Ward identities by 
themselves, they can be left out in more minimal treatments like the one 
used in Ref.~\cite{kn:bhf1}.

In Ref.~\cite{kn:bhf2} a renormalization of the fermion-loop corrections is 
formulated that uses the language of running couplings.
One rewrites bare amplitudes in terms of these renormalized
couplings and demonstrates that the resulting renormalized amplitudes 
respect gauge invariance, i.e., that they fulfill the relevant Ward identities.

In both papers the external fermionic currents are assumed to be conserved,
i.e. one neglects the masses of the external fermions.
The effect of including masses is notoriously very small, except in collinear 
regions like, for instance, the one of single-$\wb$ production.

In other words, one should remember that the current attached to the photon
propagator is strictly conserved, while the ones attached to the (massive)
vector boson propagators are not. The missing terms are of order $\mfs$ and,
therefore, they are negligible if we can show that collinear limits and 
high-energy limit are not upset by ignoring these terms.
Although no formal proof of the extension of the Fermion-Loop scheme is 
found in the literature, some partial considerations and a numerical 
analysis are given in Ref.~\cite{kn:thosetwo}.

In Ref.~\cite{kn:bhf1}, the CC20 process, $e^+e^- \to e^-\barnu_e u\bard$
has been analyzed in the so-called $\wb\wb$ configuration, where we
require that the outgoing electron is away from the collinear region,
$\theta_e > \theta_c$, although numerical results were presented with
$\theta_c$ as low as $0.1^\circ$. However, the opposite limit, 
$\theta_e < \theta_c$,
is also of theoretical and experimental importance, defining the so-called
single-$\wb$ production cross-section. A complete description of all
aspects present in single-$\wb$ production can be found in Ref~\cite{kn:swc}.
Complementary aspects can also be found in Ref.~\cite{kn:bd}.

This process has been extensively analyzed in the literature.
It is a sensitive probe of anomalous electromagnetic couplings of the $\wb$ 
boson and represents a background to searches for new physics beyond the 
standard model.

As we have discussed above, the issue of gauge invariance in the CC20 family 
has been solved by the introduction of the Fermion-Loop scheme but several 
subtleties remain, connected with the region of vanishing scattering angle of 
the electron and with the limit of massless final state fermions in a fully 
extrapolated setup. A satisfactory solution to compute the total 
cross-section is, therefore, given by the extension of the Fermion-Loop scheme
to the case of external, non-conserved currents.

First of all, for single$\wb$ production one cannot neglect the electron mass,
nor in the matrix element, neither in the kinematics of the process.
However, keeping a finite electron mass through the calculation is not enough.
Massless quarks in the final state induce a singularity, even for finite 
$\me$, if a cut is not imposed on the invariant mass $M(u\bard)$.

With a cut on $M(\bard u)$ the singularity at zero momentum transfer
($Q^2 = 0$) is avoided but we still have additional singularities.
Indeed, there are two multi-peripheral diagrams contributing to the CC20 
process $e^+e^- \to e^- \barnu_e f_1 \barf_2$, see \fig{cc20gz}.
When $Q^2 = 0$, i.e. the electron is lost in the beam pipe, and the
(massless) $f_1(f_2)$-fermion is emitted parallel to the (quasi-real) photon
then the internal fermion propagator will produce an enhancement in the cross 
section. Taking into account a $\ln \mes$ from the photon flux-function, 
three options follow:

\begin{enumerate}
\item to consider massive ($f_1/f_2$) fermions, giving a result that is
proportional to $\ln\mes\ln\mfs$,
\item to use massless fermions, giving instead $\ln^2\mes$,
\item to introduce an angular cut on the outgoing $f_1$ and $\barf_2$
fermions with respect to the beam axis, $\theta(f_1,{\bar f}_2) \ge 
\theta_{\rm cut}$, giving $\ln\mes\ln\theta_{\rm cut}$.
\end{enumerate}

The first option is clean but ambiguous when the final state fermions are 
light quarks, what to use for $m_u, m_d$?
The second one presents no problems for a fully leptonic CC20 final state
but completely fails to describe quarks, as it can be shown by discussing QCD 
corrections~\cite{kn:swc}.
The last option is also theoretically clean and can be used to give 
differential distributions for the final state jets. It is, however,
disliked by the experimentalists when computing the total sample of events: 
hadronized jets are seen and not isolated quarks. Even if the quark is 
parallel to the beam axis the jet could be broad enough and the event selected.
These events are also interesting since they correspond to a situation where
the electron and one of the quarks are lost in the beam pipe, while
the other quark is recoiling against the neutrino, i.e. one has a totally 
unbalanced mono-jet structure, background to new particle 
searches\footnote{M.~Gr\"unewald, private communication.}.

The singularity induced by massless quarks in $e^+e^- \to e^- \barnu_e u \bard$
can only be treated within the context of QCD final state corrections and of 
the photon hadronic structure function (PHSF) scenario. 
To discuss QCD corrections in the low ($u\bard$) invariant mass region
is not the purpose of this work, therefore we assume that the light quarks
have a mass and that no kinematical cuts are imposed on the process. Our
goal will be to formulate a consistent scheme that takes into account all 
fermion masses.

For many purposes it is enough to introduce a fixed width for the $\wb$, both 
for the $s$-channel and the $t$-channel, i.e. the fixed-width scheme. 
However, there is another important point in favor of adopting the full
Fermion-Loop scheme, since it guarantees automatically the correct choice
of scale for the running of $\alpha_{\rm QED}$. The latter is particularly
relevant in a process that is dominated by a small momentum transfer.

In dealing with theoretical predictions we must distinguish between an
Input-Parameter-Set (IPS) and a Renormalization-Scheme (RS). IPS one always 
has, RS comes only when one starts including loop corrections. The IPS can 
be made equal in all calculations, RS is the author's choice, very much as the 
choice of gauge.

Apart from some recent development, each calculation aimed to provide some 
estimate for $e^+e^- \to 4\,$f production is, at least nominally, a tree level 
calculation. Among other things it will require the choice of some IPS
and of certain relations among the parameters. In the literature, although 
improperly, this is usually referred to as the choice of the Renormalization 
Scheme.

Typically we have at our disposal four experimental data point 
(plus $\alpha_s$), i.e. the measured vector boson masses $\mz, \mw$ and the
coupling constants, $\gf$ and $\alpha$. However we only 
have three bare parameters at our disposal, the charged vector boson mass, 
the $SU(2)$ coupling constant and the sinus of the weak mixing angle. While 
the inclusion of one loop corrections would allow us to fix at least the 
value of the top quark mass from a consistency relation, this cannot be done 
at the tree level. Thus, different choices of the basic relations among the 
input parameters can lead to different results with deviations which, in some 
case, can be sizeable.

For instance, a possible choice is to fix the coupling constant $g$ as
\bq
g^2 = {{4\pi\alpha}\over {s_{_W}^2}}, \quad
s_{_W}^2 = {{\pi\alpha}\over {{\sqrt 2}\gf\mw^2}},
\eq
where $\gf$ is the Fermi coupling constant. Another possibility would be to use
\bq
g^2 = 4{\sqrt 2}\gf\mw^2, 
\eq
but, in both cases, we miss the correct running of the coupling. Ad hoc 
solutions should be avoided, and the running of the parameters must always 
follow from a fully consistent scheme.
Therefore, the only satisfactory solution is in the extension of the full 
Fermion-Loop to having non zero external masses, or non-conserved currents.
Unless, of course, one can compute the full set of corrections.

We will term massive-massless the version of the Fermion-Loop scheme 
developed in~\cite{kn:bhf2}. Our generalization will be denoted as the
massive-massive version of the Fermion-Loop scheme.
Work is in progress towards the implementation of the massive-massive 
Fermion-Loop scheme in the Fortran code {\tt WTO}~\cite{kn:wto}. 
The implementation of the Imaginary--Fermion-Loop scheme (see 
Ref.~\cite{kn:bhf2} for a definition) has been done in the Fortran code 
{\tt WPHACT}~\cite{kn:prep}.

For a review on the status of single-$\wb$ we refer to the work of 
Ref.~\cite{kn:balle} and to some recent activity within the LEP2/MC 
Workshop\footnote{see http://www.to.infn.it/$\tilde{}\,$giampier/lep2.html} 
with comparisons among different groups~\cite{kn:groups}. Experimental
findings are reported in~\cite{kn:expf}.

The outline of the paper will be as follows. In Sect. 2 we recall the building 
blocks for the construction of the massive-massless Fermion-Loop scheme.
In Sect. 3 we give the explicit construction of all transitions in the
charged sector of the theory. The running $\wb$ mass is introduced in
Sect. 4. The single-$\wb$ Ward identity is proved in Sect. 5. With
Sect. 6 we give a detailed discussion of the numerically relevant 
approximations in the massive-massive Fermion-Loop scheme. 
Cancellation of ultraviolet divergences, within the Fermion-Loop scheme, is 
examined in Sect. 7. Re-summation in the neutral sector of the theory is 
discussed in Sect. 8, where we also introduce the notion of $\zb$ running mass 
and describe its relation with the other running parameters of the scheme.
In Sect. 9 the imaginary parts of all corrections are explicitly computed.

\section{The Fermion-Loop scheme}

In this Section we briefly discuss the main ingredients entering into the
Fermion-Loop scheme~\cite{kn:bhf2}. Here, we assume that all external currents 
are conserved, i.e. we assume that the external world is massless.

In the 't Hooft--Feynman gauge, the $\drii{\mu}{\nu}$ part of the 
vector--vector transitions can be cast in the following form~\cite{kn:book},
where $\stw(\ctw)$ is the sine(cosine) of the weak mixing angle:
\bqa
S_{\ph\ph}  &=& \frac{\gbs\stws}{16\,\pi^2}\,\Pgg(\pmoms)\,\pmoms, 
\quad
S_{\ssZ\ssZ} = \frac{\gbs}{16\pi^2\ctws}\,\Szz(\pmoms),
\nl
S_{\ssZ\ph} &=& \frac{\gbs\stw}{16\pi^2\ctw}\,\Sigma_{_{\ssZ\ph}}(\pmoms),
\quad
S_{\ssW\ssW} = \frac{\gbs}{16\pi^2}\,\Sww(\pmoms).
\label{S_self}
\eqa
Next we have can transform to the $(3,Q)$ basis, where:
\bqa
\Szz(\pmoms) &=& \Stt(\pmoms)-2\stws\Stg(\pmoms)+\stwf\Pgg(\pmoms)\,\pmoms,
\nl
\Sigma_{_{\ssZ\ph}}(\pmoms) &=& \Stg(\pmoms)-\stws\Pgg(\pmoms)\,\pmoms.
\label{Sig_self}
\eqa
To recall the derivation of the Fermion-Loop scheme, we start with some another 
drastic approximation; namely, we put to zero also all fermions masses in 
loops, i.e. we assume a massless internal world.
Under this assumption we will introduce the following quantities:
\bqa
\sbff{n}\equiv\bff{n}{\pmoms}{0}{0} &=& 2\,\bff{21}{\pmoms}{0}{0} 
- \bff{0}{\pmoms}{0}{0},  
\nl
\sbff{c}\equiv\bff{c}{\pmoms}{0}{0} &=& \bff{21}{\pmoms}{0}{0} 
+ \bff{0}{\pmoms}{0}{0} = \frac{1}{2}\,\sbff{n}.
\eqa
Here the $\bff{ij}{p^2}{0}{0}$ are scalar one-loop integrals~\cite{kn:pv}.
Furthermore, we will use the fact that a $\pmoms$ can be factorized:
\bq
\Stg(\pmoms) = \Ptg(\pmoms)\,\pmoms, \qquad
\Stt(\pmoms) = \Ptt(\pmoms)\,\pmoms.
\eq
As a result, all the vector boson--vector boson transitions simplify 
drastically:
\bqa
\Pgg(\sman) &=& \frac{32}{3}\asums{g}\,\sbff{n},  
\quad
\Ptg(\sman) = 4\,\asums{g}\,\sbff{n} = \frac{3}{8}\,\Pgg(\sman),  
\nl
\Stt(\sman) &=& - 4\sman\,\asums{g}\,\sbff{n} = - \sman\Ptg(\sman),  
\nl
\Sww(\sman) &=& - 8\sman\,\asums{g}\,\sbff{c} = - 4\sman\,\asums{g}\,
\sbff{n} = -\sman\Ptt(\sman),
\eqa
where $\pmoms=-\sman$ and where the sum is over the fermion 
generations.

The resulting expressions retain some simplicity even if we do not ignore 
the top quark mass. Self-energies may still be written in compact form:
\bqa
\Pgg(\sman) &=& \frac{32}{3}\,\asums{g}\,\sbff{n} + \frac{16}{3}\,\bbff{n}
,  
\quad
\Ptg(\sman) = 4\,\asums{g}\,\sbff{n} + 2\,\bbff{n},  
\nl
\Stt(\sman) &=& -\sman\,\Ptg(\sman) + \frac{1}{2}\sman\,\bbff{n} 
- \frac{3}{2}\,\mts\bff{0}{-\sman}{\mt}{\mt},  
\nl
\Sww(\sman) &=& - s\,\lrbr 4\,\asums{g}\,\sbff{n} + 6\,\bbff{c}\rrbr 
-3\,\mts\,\sbff{mc},
\eqa
where we have introduced new auxiliary functions, defined by
\bqa
\bbff{n} &=& \bff{n}{-\sman}{\mt}{\mt} - \bff{n}{-\sman}{0}{0},  
\nl
\bbff{c} &=& \bff{c}{-\sman}{\mt}{0} - \bff{c}{-\sman}{0}{0},  
\nl
\sbff{mc}&=& \bff{1}{-\sman}{\mt}{0}  + \bff{0}{-\sman}{\mt}{0}.
\eqa
A useful way of presenting these results will be to split the universal part,
proportional to $\Ptg(\sman)$, from the remainder:
\bq
\Stt(\sman) = -\sman\,\Ptg(\sman) + \fZ(\sman),  
\quad
\Sww(\sman) = -\sman\,\Ptg(\sman) + \fW(\sman),
\eq
where the two $f$-functions are expressible as
\bqa
\fW(\sman) &=& -2\,\sman\,
\Bigl[ 3\bff{21}{-\sman}{\mt}{0} - 2\bff{21}{-\sman}{\mt}{\mt} 
      - \bff{21}{-\sman}{0}{0} 
\nl &&
+ 3\bff{1}{-\sman}{\mt}{0} 
- 3\bff{1}{-\sman}{0}{0} 
+ \bff{0}{-\sman}{\mt}{\mt} 
- \bff{0}{-\sman}{0}{0}
\Bigr]
\nl &&
- 3\,\mts\,\Big[\bff{1}{-\sman}{\mt}{0} + \bff{0}{-\sman}{\mt}{0}\Big],  
\nl
\fZ(\sman) &=& \frac{1}{2}\,\sman\,
\Bigl[2\bff{21}{-\sman}{\mt}{\mt} - 2\bff{21}{-\sman}{0}{0} 
\nl &&
- \bff{0}{-\sman}{\mt}{\mt} 
+ \bff{0}{-\sman}{0}{0} 
\Bigr] 
- \frac{3}{2}\,\mts\,\bff{0}{-\sman}{\mt}{\mt}.
\eqa
\subsection{Running couplings}

We now consider three parameters, the e.m. coupling constant $\ec$,
the $SU(2)$ coupling constant $\gb$ and the sine of the weak mixing
angle $\stw$. At the tree level they are not independent, but rather
they satisfy the relation $\gbs\stws = \ecs$.
The running of the e.m. coupling constant is easily derived and gives
\bq
\frac{1}{\ecs(\sman)} = \frac{1}{\gbs\stws}-\frac{1}{16\,\pi^2}\,\Pgg(\sman).
\eq
However, we have a natural scale to use since at $\sman = 0$ we have 
the fine structure constant at our disposal. 
Therefore, the running of $\ecs(\sman)$ is
completely specified in terms of $\alpha$ by
\bq
\frac{1}{\ecs(\sman)} = \frac{1}{4\,\pi\alpha}\,
\lrbr 1 - \frac{\alpha}{4\,\pi}\,\Pi(\sman)\rrbr,  
\qquad \mbox{with} \quad
\Pi(\sman) = \Pgg(\sman) - \Pgg(0).
\eq
For the running of $\gbs$ we derive a similar equation:
\bq
\frac{1}{\gbs(\sman)} = \frac{1}{\gbs} - \frac{1}{16\,\pi^2}\,\Ptg(\sman).
\eq
The running of the third parameter, $\stws(\sman)$, is now fixed by
\bq
\stws(\sman) = \frac{\ecs(\sman)}{\gbs(\sman)}.
\eq

\subsection{Propagator functions.}

The re-summed propagators for the vector bosons are:
\bqa
G_{\ph}(\pmoms) &=& \biggl\{\pmoms - S_{\ph\ph}(\pmoms) 
-{{\lrbr S_{\ssZ\ph}(\pmoms)\rrbr^2}\over
{\pmoms + \bzms - S_{\ssZ\ssZ}(\pmoms)}}\biggr\}^{-1},  
\nl
G_{\ssZ\ph}(\pmoms) &=& {{S_{\ssZ\ph}(\pmoms)}\over
{\bigl[\pmoms - S_{\ph\ph}(\pmoms)\bigr]\,
 \bigl[\pmoms + \bzms - S_{\ssZ\ssZ}(\pmoms)\bigr] 
-\bigl[S_{\ssZ\ph}(\pmoms)\bigr]^2}},  
\nl
G_{\ssZ}(\pmoms) &=& \biggl\{\pmoms + \bzms - S_{\ssZ\ssZ} - 
{{\lrbr S_{\ssZ\ph}(\pmoms)\rrbr^2}\over 
{\pmoms - S_{\ph\ph}(\pmoms)}}\biggr\}^{-1},  
\nl
G_{\ssW}(\pmoms) &=& \Bigl[ \pmoms + \LMs - S_{_{\ssW\ssW}}(\pmoms)\Bigr]^{-1}.
\label{defprops}
\eqa
The quantity $\bzm = M/\ctw$ is the bare $\zb$ mass.
An essential ingredient of the construction is represented by the location
of the complex poles; they are determined by the following two equations:
\bqa
\sW &=& \LMs - S_{_{\ssW\ssW}}(\sW),  
\nl
\sZ &=& \bzms - Z(\sZ),\qquad \mbox{where} \quad Z(\sman) =  
S_{\ssZ\ssZ}(\sman) 
-{{\bigl[S_{\ssZ\ph}(\sman)\bigr]^2}\over{\sman + S_{\ph\ph}(\sman)}}.
\label{poleq}
\eqa
Substituting \eqn{poleq} into the expressions for the propagators, 
\eqn{defprops}, we see that all ultraviolet divergences not proportional
to $\pmoms$ cancel. We obtain
\bqa
G_{\ssZ}(\sman) &=& \Bigl[-\sman + \sZ - Z(\sman) + Z(\sZ)\Bigr]^{-1},  
\nl
G_{\ssW}(\sman) &=& \Bigl[-\sman + \sW - S_{_{\ssW\ssW}}(\sman) 
+ S_{_{\ssW\ssW}}(\sW)\Bigr]^{-1},  
\nl
G_{\ssZ\ph}(\sman) &=& -\,{{S_{\ssZ\ph}(\sman)}\over{\sman + S_{\ph\ph}(\sman)}}\,
G_{\ssZ}(\sman),  
\nl
G_{\ph}(\sman) &=& -\,{1\over{\sman + S_{\ph\ph}(\sman)}} 
+\lrbr{{S_{\ssZ\ph}(\sman)}\over{\sman + S_{\ph\ph}(\sman)}}\rrbr^2\,G_{\ssZ}(\sman).
\eqa
Using the self-energies and the running parameters we can write
\bqa
1 + \frac{Z(\sman)}{\sman} &=& \frac{\gbs}{\ctws}\,
\lrbr \frac{\canys(\sman)}{\gbs(\sman)} 
+ \frac{1}{16\,\pi^2}\,\frac{f_{\ssZ}(\sman)}{\sman}\rrbr,  
\nl
1 + \frac{S_{_{\ssW\ssW}}(\sman)}{\sman} &=& \gbs\,\lrbr \frac{1}{\gbs(\sman)} 
+ \frac{1}{16\,\pi^2}\,\frac{f_{\ssW}(\sman)}{\sman}\rrbr,  
\eqa
As a result, the vector boson propagators are now expressed as
\bqa
G_{\ssW}(\sman)&=&-\frac{\gbs(\sman)}{\gbs}\,
\frac{\omega_{\ssW}(\sman)}{\sman},  
\qquad
G_{\ssZ}(\sman) = -\frac{\ctws}{\gbs}\,\frac{\gbs(\sman)}{\canys(\sman)}\,
\frac{\omega_{\ssZ}(\sman)}{\sman},  
\nl
G_{\ssZ\ph}(\sman)&=&\frac{\stw}{\ctw}\lrbr 1 - \frac{\sanys(\sman)}{\stws}
\rrbr\,G_{\ssZ}(\sman),  
\nl
G_{\ph}(\sman) &=& \frac{\ecs(\sman)}{\ecs} + \frac{\stws}{\ctws}
\lrbr 1 - \frac{\sanys(\sman)}{\stws}\rrbr^2\,G_{\ssZ}(\sman),
\label{defG}
\eqa
where the propagation functions are
\bqa
\omega^{-1}_{\ssW}(\sman) &=& 1 - \frac{\gbs(\sman)}{\sman}\,
\lcbr
\frac{\sW}{\gbs(\sW)} - \frac{1}{16\,\pi^2}\,
\Bigl[f_{\ssW}(\sman) - f_{\ssW}(\sW)\Bigr]
\rcbr,  
\nl
\omega^{-1}_{\ssZ}(\sman) &=& 1 - \frac{\gbs(\sman)}{\canys(\sman)\sman}\,
\lcbr
\frac{\canys(\sZ)}{\gbs(\sZ)}\,\sZ - \frac{1}{16\,\pi^2}\,
\Bigl[f_{\ssZ}(\sman) - f_{\ssZ}(\sW)\Bigr]
\rcbr.
\eqa
The $\omega$-functions are ultraviolet finite since the ultraviolet
poles in $f_{\ssZ}$ and $f_{\ssW}$ do not depend on the scale. 
The result of including vector boson transitions is illustrated schematically
by the following diagrams (\eqn{folldia}):
\bqa
\quad
\DiagramFermionToBosonFull{}{$\ph$} & = &
             \DiagramFermionToBosonPropagator{}{$\ph$}{$\ph$}
           + \DiagramFermionToBosonPropagator[75]{}{$\ph$}{$\zb$},
\nl
\quad
\DiagramFermionToBosonFull{}{$\zb$} & = &
             \DiagramFermionToBosonPropagator{}{$\zb$}{$\ph$}
           + \DiagramFermionToBosonPropagator[75]{}{$\zb$}{$\zb$}.
\label{folldia}
\eqa
Here the open circles denote re-summed propagators and the dot
a vertex. The dot on the right-hand side of the diagrams indicates
that the corresponding leg is not amputated, i.e. that the propagator is 
included. 

\subsection{A simple recipe for implementing the Fermion-Loop scheme.}

\begin{figure}[h]
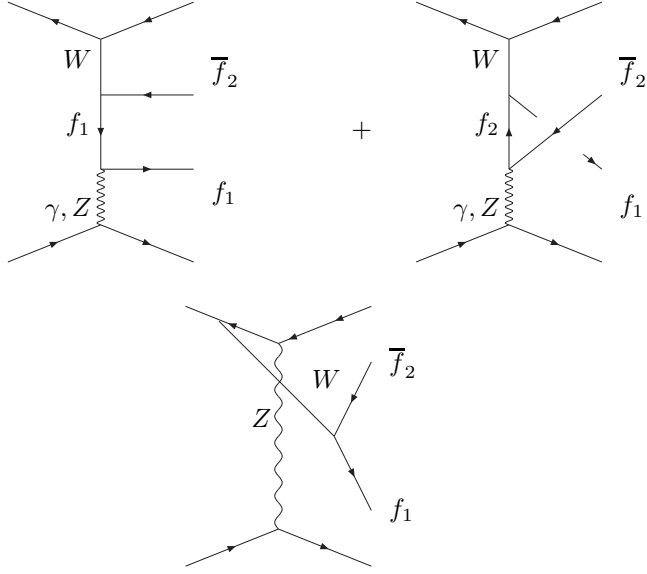

\caption[]{The CC20$_{\ph \oplus \zb}$ family of diagrams.}
\label{cc20gz}
\end{figure}

There is a simple recipe for implementing the Fermion-Loop scheme: take any
process where the external sources are physical and on-shell, eg fermionic
currents, then the complete procedure for the re-summation of self-energies 
and transitions, for the inclusion of running couplings and of re-summed 
propagators amounts to rewrite the corresponding Born amplitude in terms of 
re-summed, running, quantities without the inclusion of transitions like the 
$\ph-\zb$ one. Let us consider one example in more detail.
First, we split the 20 Feynman diagrams of the CC20 family into the nine 
diagrams of \fig{cc20gz}, characterized by the presence of a $t$-channel 
photon or $\zb$-boson, and the rest
\bq
\mbox{CC20} = \mbox{CC20}_{\gamma \oplus \zb} + \mbox{CC20}_{\rm R}.
\label{splitCC20}
\eq
\vspace{0.2cm}
\bqas
\ba{ccc}
\vcenter{\hbox{
  \SetScale{0.7}
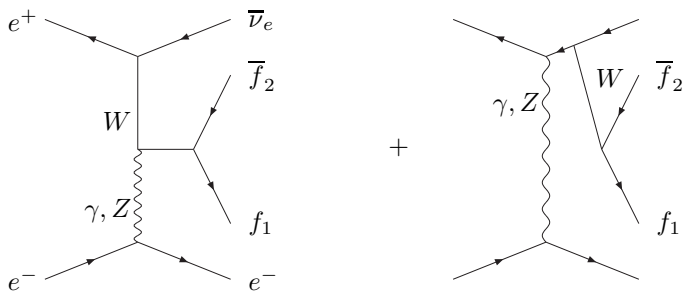
  \begin{picture}(110,100)(0,0)
  \ArrowLine(50,120)(0,140)
  \ArrowLine(100,140)(50,120)
  \ArrowLine(0,0)(50,20)
  \ArrowLine(50,20)(100,0)
  \ArrowLine(100,110)(80,70)
  \ArrowLine(80,70)(100,30)
  \Photon(50,20)(50,70){2}{7}
  \Line(50,70)(50,120)
  \Line(50,70)(80,70)
  \Text(-14,98)[lc]{$e^+$}
  \Text(77,98)[lc]{$\barnu_e$}
  \Text(-14,0)[lc]{$e^-$}
  \Text(77,0)[lc]{$e^-$}
  \Text(77,77)[lc]{$\barf_2$}
  \Text(77,21)[lc]{$f_1$}
  \Text(22,60)[lc]{$\wb$}
  \Text(15,26)[lc]{$\gamma,\zb$}
  \end{picture}}}
&\quad+&
\vcenter{\hbox{
  \SetScale{0.7}
  \begin{picture}(110,100)(0,0)
  \ArrowLine(50,120)(0,140)
  \ArrowLine(65,126)(50,120)
  \ArrowLine(100,140)(65,126)
  \ArrowLine(0,0)(50,20)
  \ArrowLine(50,20)(100,0)
  \ArrowLine(100,110)(80,70)
  \ArrowLine(80,70)(100,30)
  \Photon(50,20)(50,120){2}{7}
  \Line(65,126)(80,70)
  \Text(77,77)[lc]{$\barf_2$}
  \Text(77,21)[lc]{$f_1$}
  \Text(54,76)[lc]{$\wb$}
  \Text(15,66)[lc]{$\gamma,\zb$}
  \end{picture}}}
\ea
\eqas
\bqas
\ba{ccc}
\vcenter{\hbox{
  \SetScale{0.7}
  \begin{picture}(110,100)(0,0)
  \ArrowLine(50,120)(0,140)
  \ArrowLine(100,140)(50,120)
  \ArrowLine(0,0)(50,20)
  \ArrowLine(50,20)(100,0)
  \ArrowLine(100,90)(50,90)
  \ArrowLine(50,90)(50,50)
  \ArrowLine(50,50)(100,50)
  \Photon(50,20)(50,50){2}{7}
  \Line(50,90)(50,120)
  \Text(77,72)[lc]{$\barf_2$}
  \Text(77,26)[lc]{$f_1$}
  \Text(22,52)[lc]{$f_1$}
  \Text(14,20)[lc]{$\gamma,\zb$}
  \Text(21,77)[lc]{$\wb$}
  \end{picture}}}
&\quad+&
\vcenter{\hbox{
  \SetScale{0.7}
  \begin{picture}(110,100)(0,0)
  \ArrowLine(50,120)(0,140)
  \ArrowLine(100,140)(50,120)
  \ArrowLine(0,0)(50,20)
  \ArrowLine(50,20)(100,0)
  \ArrowLine(100,90)(50,50)
  \Line(50,90)(65,78)
  \ArrowLine(90,58)(100,50)
  \ArrowLine(50,50)(50,90)
  \Photon(50,20)(50,50){2}{7}
  \Line(50,90)(50,120)
  \Text(77,72)[lc]{$\barf_2$}
  \Text(77,21)[lc]{$f_1$}
  \Text(22,52)[lc]{$f_2$}
  \Text(14,20)[lc]{$\gamma,\zb$}
  \Text(21,77)[lc]{$\wb$}
  \end{picture}}}
\ea
\eqas
\bqas
\ba{ccc}
\vcenter{\hbox{
  \SetScale{0.7}
  \begin{picture}(110,100)(0,0)
  \ArrowLine(50,120)(0,140)
  \ArrowLine(65,126)(50,120)
  \ArrowLine(100,140)(65,126)
  \ArrowLine(0,0)(50,20)
  \ArrowLine(50,20)(100,0)
  \ArrowLine(100,110)(80,70)
  \ArrowLine(80,70)(100,30)
  \Photon(50,20)(50,120){2}{7}
  \Line(18,132)(80,70)
  \Text(77,77)[lc]{$\barf_2$}
  \Text(77,21)[lc]{$f_1$}
  \Text(48,71)[lc]{$\wb$}
  \Text(25,56)[lc]{$\zb$}
  \end{picture}}}
&{}&{}
\ea
\eqas
\vspace{-2mm}
\vskip 20pt

Momenta are assigned as follows:
\bq
e^+(p_+)\, e^-(p_-) \to e^-(q_-)\, \barnu_e(q_+)\, u(k)\, \bard(\kbar),
\quad Q_{\pm} = p_{\pm}-q_{\pm}.
\eq
First we consider the diagram with the non-abelian coupling and generalize it
to have re-summed $\wb$-boson propagators and $\ph-\ph, \zb-\ph$ (1a) 
transitions and $\zb-\zb, \ph-\zb$ (1b) transitions.
The corresponding amplitudes will then become
\bq
M_1 = {\rm M}_{1{\rm a}\mu} \otimes \lpar -i g\stw\rpar \gapu{\mu} +
{\rm M}_{1{\rm b}\mu} \otimes \frac{i g}{2\,\ctw}\gapu{\mu}\,\lpar 2\,\stws - 
\frac{1}{2}\gdp\rpar,
\eq
where $\gdpm = 1 \pm \gfd$ and the sub-amplitudes are
\bqa
{\rm M}_{1{\rm a}\mu} &=& \frac{1}{8}\,\frac{G_sG_t}{g^3\stw}\,
\frac{\omega^s_{\ssW}\omega^t_{\ssW}}{(p_--q_+)^2(k+\kbar)^2Q^2_-}\,
\lrbr e^2(T) - \frac{G_{\ssT}}{c^2(T)}\,\omega^{\ssT}_{\ssZ}\Delta_{\ph}\rrbr 
\nl
{}&\times& \gapu{\alpha}\gdp \otimes \gapu{\beta}\gdp\,V^0_{\mu\alpha\beta},
\nl\nl
{\rm M}_{1{\rm b}\mu} &=& -\frac{1}{8}\,\frac{G_sG_t}{g^3}\,
\frac{\omega^s_{\ssW}\omega^t_{\ssW}}{(p_--q_+)^2(k+\kbar)^2Q^2_-}\,
G_{\ssT}\omega^{\ssT}_{\ssZ}  \nl
{}&\times& \gapu{\alpha}\gdp \otimes \gapu{\beta}\gdp\,V^0_{\mu\alpha\beta},
\eqa
and where $V^0_{\mu\alpha\beta}$ is the tree-level non-abelian coupling. 
Furthermore,
\bq
\Delta_{\ph} = \stwf - 2\,s^2(T)\stws + s^4(T) + \stws\ctws - s^2(T)\ctws.
\eq
In the previous equations we have introduced the following quantities:
\bqa
T &=& -(p_- - q_-)^2,  \nl
G_{\ssT} &=& g^2(T), \quad G_t = g^2((p_+-q_+)^2), \quad G_s = g^2((k+\kbar)^2).
\eqa
Our sign convention is such that
\bq
g^2(s) \equiv g^2(p^2)\mid_{p^2 = -s}.
\eq
In the same way we have introduced $\omega_{\ssW}^s$, the propagation function 
for a $\wb$-boson with $p^2 = -(k+\kbar)^2$. When we combine the two results we 
obtain
\bqa
M &=& -\frac{i}{8}\,\frac{G_sG_t}{g^2}
\frac{\omega^s_{\ssW}\omega^t_{\ssW}}{(p_--q_+)^2(k+\kbar)^2Q^2_-}\,
\lcbr e^2(T)\gapu{\mu} \right.  \nl
{}&+&\left. \frac{1}{4}G_{\ssT}\omega^{\ssT}_{\ssZ}\gapu{\mu}\,
\lrbr 4\,s^2(T) - \gdp\rrbr\rcbr\,
\otimes \gapu{\alpha}\gdp \otimes \gapu{\beta}\gdp\,V^0_{\mu\alpha\beta}.
\label{g2fact}
\eqa
The only remaining bare quantity is an overall $1/g^2$ factor which, as 
discussed in Sect. 7, is essential in performing the renormalization of the 
amplitude. The rest is exactly the sum of two Born-like diagrams with
$\ph$ and $\zb$ exchange and with running parameters instead of bare ones.
In the following we will denote by $Q_f$ the charge of the fermion.
Similarly, for the other diagrams we obtain
\bqa
M_2 &=& {\rm M}_{2{\rm a}\mu} \otimes \lpar -i g\stw\rpar \gapu{\mu} +
{\rm M}_{2{\rm b}\mu} \otimes \frac{i g}{2\,\ctw}\gapu{\mu}\,\lpar 2\,\stws - 
\frac{1}{2}\gdp\rpar  \nl
{}&=& \frac{i}{8}\,\frac{G_se^2(T)}{(p_-+Q_-)^2(k+\kbar)^2Q^2_-}\,
\omega^s_{\ssW}\,\gapu{\mu}\lpar \sla{p_-}+\sla{Q_-}\rpar \otimes
\gapu{\alpha}\gdp \otimes \gapu{\alpha}\gdp \otimes \gapu{\mu}  \nl
{}&-& \frac{i}{8}\,\frac{G_sG_{\ssT}}{(p_-+Q_-)^2(k+\kbar)^2Q^2_-}\,
\omega^s_{\ssW}\omega^{\ssT}_{\ssZ}\,\frac{G_t}{16\,c^2(T)}  \nl
{}&\times& \gapu{\mu}\lpar 4\,s^2(T) - \gdp\rpar \lpar \sla{p_-}+\sla{Q_-}\rpar 
\gapu{\alpha}\gdp \otimes \gapu{\alpha}\gdp \otimes \gapu{\mu}\,\lpar
4\,s^2(T) - \gdp\rpar, 
\nl\nl
M_3 &=& {\rm M}_{3{\rm a}\mu} \otimes \lpar -i g\stw\rpar \gapu{\mu} +
{\rm M}_{3{\rm b}\mu} \otimes \frac{i g}{2\,\ctw}\gapu{\mu}\,\lpar 2\,\stws - 
\frac{1}{2}\gdp\rpar  \nl
{}&=& \frac{i}{8}\,\frac{G_te^2(T)}{Q^2_+Q^2_-}\,Q_u\,
\omega^t_{\ssW}\,\gapu{\alpha}\gdp \otimes \gapu{\mu}\frac{\sla{Q_-}-
\sla{k} - i\,\muq}{(Q_--k)^2+\mus}\,\gapu{\alpha}\gdp \otimes \gapu{\mu}  \nl
{}&+& \frac{i}{8}\,\frac{G_tG_{\ssT}}{16\,c^2(T)\,Q^2_+Q^2_-}\,
\omega^t_{\ssW}\omega^{\ssT}_{\ssZ}\,\gapu{\alpha}\gdp  \nl
{}&\otimes& \lpar \gdp - 4\,Q_us^2(T)\rpar
\frac{\sla{Q_-} - \sla{k} - i\,\muq}{(Q_--k)^2+\mus}\,\gapu{\alpha}\gdp\,
\otimes \gapu{\mu}\,\lpar 4\,s^2(T) - \gdp\rpar,
\nl\nl
M_4 &=& {\rm M}_{4{\rm a}\mu} \otimes \lpar -i g\stw\rpar \gapu{\mu} +
{\rm M}_{4{\rm b}\mu} \otimes \frac{i g}{2\,\ctw}\gapu{\mu}\,\lpar 2\,\stws - 
\frac{1}{2}\gdp\rpar  \nl
{}&=& \frac{i}{8}\,\frac{G_te^2(T)}{Q^2_+Q^2_-}\,Q_d\,
\omega^t_{\ssW}\,\gapu{\alpha}\gdp \otimes \gapu{\mu}\frac{\sla{Q_-}-
\sla{\kbar} + i\,\md}{(Q_--\kbar)^2+\mds}\,\gapu{\alpha}\gdp \otimes 
\gapu{\mu}  \nl
{}&+& \frac{i}{8}\,\frac{G_tG_{\ssT}}{16\,c^2(T)\,Q^2_+Q^2_-}\,
\omega^t_{\ssW}\omega^{\ssT}_{\ssZ}\,\gapu{\alpha}\gdp  \nl
{}&\otimes& \lpar -\gdp - 4\,Q_ds^2(T)\rpar
\frac{\sla{Q_-} - \sla{\kbar} + i\,\md}{(Q_--\kbar)^2+\mds}\,\gapu{\alpha}\gdp\,
\otimes \gapu{\mu}\,\lpar 4\,s^2(T) - \gdp\rpar,
\nl\nl
M_5 &=& {\rm M}_{5{\rm a}\mu} \otimes \lpar -i g\stw\rpar \gapu{\mu} +
{\rm M}_{5{\rm b}\mu} \otimes \frac{i g}{2\,\ctw}\gapu{\mu}\,\lpar 2\,\stws - 
\frac{1}{2}\gdp\rpar  \nl
{}&=& -\frac{i}{8}\,\frac{G_sG_t}{16\,c^2(T)\,(k+\kbar)^2(Q_--q_+)^2
Q^2_-}\,\omega^s_{\ssW}\omega^{\ssT}_{\ssZ}  \nl
{}&\times& \gapu{\alpha}\gdp\lpar \sla{Q_-}-\sla{q_+}\rpar\gapu{\mu}\gdp 
\otimes \gapu{\mu}\,\lpar 4\,s^2(T) - \gdp\rpar \otimes \gapu{\alpha}\gdp.
\eqa
This set of equations proves the assertion that Fermion-Loop is exactly
an improved, gauge preserving, Born approximation.

To complete the construction of the Fermion-Loop scheme one must include the 
one-loop fermionic vertices. At the Born level the $\ph\wbp\wbm$ and the 
$\zb\wbp\wbm$ vertices are the same, once we have factorized $\stw$ and $\ctw$ 
in front of them. For one-loop corrected vertices this is no longer true and 
one may wonder whether this fact spoils the transition from bare quantities to
re-summed ones. 

The lowest order interaction for $\vb(\Pmom) \to \wbp(\qmomi{+})
\wbm(\qmomi{-})$ is specified by the tensor
\bqa
V^0_{\mu\alpha\beta}\lpar\Pmom;\qmomi{+},\qmomi{-}\rpar &=&
 \drii{\mu}{\beta}   \lpar\Pmom    -\qmomi{-}\rpar_{\alpha} 
+\drii{\alpha}{\beta}\lpar\qmomi{-}-\qmomi{+}\rpar_{\mu} 
+\drii{\mu}{\alpha}  \lpar\qmomi{+}-\Pmom\rpar_{\beta}.
\eqa
At the one-loop level we need seven independent form-factors, if the external
sources are physical; they are as follows:
\bqa
V^1_{\mu\alpha\beta} &=& \frac{g^2\stw}{16\,\pi^2}\,\sqrt{s}\,
\asums{i=1,7}\,I_i\,W^i_{\mu\alpha\beta},  \nl
W^1_{\mu\alpha\beta} &=& \frac{4}{\sqrt{s}}\,\lrbr \drii{\alpha}{\beta}
\qmomit{-}{\mu} + \drii{\mu}{\beta}\Pmom_{\alpha} +
\drii{\mu}{\alpha}\qmomit{+}{\beta}\rrbr,  \nl
W^{2,3}_{\mu\alpha\beta} &=& \frac{2}{\sqrt{s}}\,\lrbr \drii{\mu}{\beta}
\Pmom_{\alpha} \pm \drii{\mu}{\alpha}\qmomit{+}{\beta}\rrbr,  \nl
W^4_{\mu\alpha\beta} &=& \frac{2}{s\sqrt{s}}\,
\qmomit{-}{\mu}\Pmom_{\alpha}\qmomit{+}{\beta},  \nl
W^{5,6}_{\mu\alpha\beta} &=& \frac{1}{\sqrt{s}}\,\lrbr \varepsilon\lpar
\qmomi{+},\mu,\alpha,\beta\rpar \pm \varepsilon\lpar \qmomi{-},\mu,\alpha,\beta
\rpar\rrbr,  \nl
W^7_{\mu\alpha\beta} &=& \frac{1}{s\sqrt{s}}\, \varepsilon\lpar
\qmomi{-},\qmomi{+},\alpha,\beta\rpar\,\qmomit{+}{\mu},
\label{fvphysical}
\eqa
where we have introduced a special notation,
\bq
\varepsilon(a,b,c,d) = \varepsilon^{\mu\nu\alpha\beta}\,a_{\mu}b_{\nu}
c_{\alpha}d_{\beta}.
\eq
For a massless fermion generation there is no difference between the $\zb$
and $\ph$ coefficients once we have factorized $\ctw (\stw)$ in front
of the full vertex. For the third generation we have the same for all loops
except the $(\mt,0,\mt)$ one, where the difference between $\zb$ and $\ph$
is given by the following relations:
\bq
I^{\ssZ\ssW\ssW}_i = I^{\ph\ssW\ssW}_i + \frac{1}{\ctws}\,\Delta I_i,
\qquad i=1,\dots,7
\label{extrat}
\eq
where the explicit expression for the extra term, $\Delta I_i$, is of no
concern here. The amplitude $M_1$, containing the non-abelian coupling,
retains its structure when we substitute $V^0_{\ph\ssW\ssW}$ with 
$V^1_{\ph\ssW\ssW}$ and we have additional contributions
\bqa
\delta_i &=& \frac{g^3\Delta I_i}{\ctws}\,\ctw\,G_{\ssZ}\,\lrbr \frac{\stw}
{\ctw}\,\lpar 1 - \frac{s^2(T)}
{\stws}\rpar\,\lpar -i g\stw\rpar \otimes \gapu{\mu} \right.  \nl
{}&+&\left. \frac{i g}{2\,\ctw} \otimes \gapu{\mu}\,
\lpar 2\,\stws - \frac{1}{2}\,\gdp\rpar\rrbr  \nl
{}&=& + \frac{i}{4} g^2\,\Delta_i\,\frac{G_{\ssT}}{c^2(T)}
\frac{\omega^{\ssT}_{\ssZ}}
{Q^2_-} \otimes \gapu{\mu}\,\lpar 4\,s^2(T) - \gdp\rpar.
\eqa
Therefore, also for one-loop vertices we can write a Born-like
amplitude and promote all bare quantities to running ones, i.e. we take
the full one-loop corrected vertex of \eqn{fvphysical} with factors
$\stw(\ctw)$ factorized and also replace $1/\ctws$ with $1/c^2(T)$ in
the extra term of \eqn{extrat}.

\section{Re-summed propagators in the charged sector.}

We now proceed to the construction of the Fermion-Loop scheme for non-conserved
currents. The first step consists in re-deriving the propagators in
a situation where all external fermion masses are kept. We work in the
't Hooft-Feynman gauge and compute the following transitions:

\begin{itemize}

\item The $\wb-\wb$ transition, $S^{\mu\nu}_{\ssW}$ with
\bq
S^{\mu\nu}_{\ssW} = \frac{g^2}{16\,\pi^2}\,\Sigma^{\mu\nu}_{\ssW},  \quad
\Sigma^{\mu\nu}_{\ssW} = \Sigma^0_{\ssW}\,\delta^{\mu\nu} + 
\Sigma^1_{\ssW}\,p^{\mu}p^{\nu}.
\eq
We also introduce a special notation,
\bq
\Sigma^T_{\ssW} = \Sigma^0_{\ssW} + p^2\,\Sigma^1_{\ssW}.
\eq
\item The $\phi-\phi$ transition, $S_{\phi}$ with
\bq
S_{\phi} = \frac{g^2}{16\,\pi^2}\,\Sigma_{\phi}.
\eq
\item The $\wb-\phi$ and $\phi-\wb$ transitions,
\bq
S^{\mu}_{\ssW\phi} = +\frac{g^2}{16\,\pi^2}\,\Sigma_{\ssW\phi}\,ip^{\mu},  \quad
S^{\mu}_{\phi\ssW} = -\frac{g^2}{16\,\pi^2}\,\Sigma_{\ssW\phi}\,ip^{\mu}.
\eq
\end{itemize}
Let us introduce indices $a,b,\dots = 1,\dots,5$; the re-summation amounts to
write the following equation:
\bqa
{\bar\Delta}_{ab} &=& \Delta_{ab} + \delta_{ac}\,S_{cd}\,\Delta_{db} + \dots
\nl
{}&=& \delta_{ac}\,\lpar \delta_{cb} + S_{cd}\,\Delta_{db} + \dots\rpar =
\Delta_{ac}\,X_{cb},  \nl
X_{ab} &=& \lpar 1 - S\,\Delta\rpar^{-1}_{ab}.
\label{dysonc}
\eqa
Here we have made use of Born propagators given, in the 't Hooft-Feynman gauge,
by
\bq
\Delta^{\mu\nu}_{\ssW\ssW} = \frac{\delta_{\mu\nu}}{p^2+M^2}, \qquad
\Delta_{\phi\phi} = \frac{1}{p^2+M^2}.
\eq
Examples of Dyson re-summation are as follows:
\bqa
{\bar\Delta}^{\mu\nu}_{\ssW\ssW} &=& \Delta^{\mu\nu}_{\ssW\ssW} + 
\Delta^{\mu\alpha}_{\ssW\ssW}\,
S^{\alpha\beta}_{\ssW}\,\delta^{\beta\nu}_{\ssW\ssW} + \dots,  \nl
{\bar\Delta}_{\phi\phi} &=& \Delta_{\phi\phi} + \Delta_{\phi\phi}\,
S_{\phi}\,\Delta_{\phi\phi} + \dots,  \nl
{\bar\Delta}^{\mu}_{\ssW\phi} &=& \Delta^{\mu\alpha}_{\ssW\ssW}\,
S^{\alpha}_{\ssW\phi}\,\Delta_{\phi\phi} + \dots
\eqa.
After performing the inversion of the matrix in \eqn{dysonc}, we obtain
\bq
{\bar\Delta}^{\mu\nu}_{\ssW\ssW} = \frac{1}{p^2+M^2-S^0_{\ssW}}\,\bigl[
\delta^{\mu\nu} + {{S^1_{\ssW} + \lpar S_{\ssW\phi}\rpar^2/\lpar 
p^2+M^2-S_{\phi}\rpar}\over {p^2+M^2 - S^T_{\ssW} - p^2\lpar S_{\ssW\phi}\rpar^2/
\lpar p^2+M^2-S_{\phi}\rpar}}\,p^{\mu}p^{\nu}\Big]
\eq
The previous result can be cast into a simpler form when we use some
important relation originating from Ward identities applied to two-point
functions. The Ward identities for transitions in the charged sector are 
shown in \fig{fig:wised} where we used the symbol $=$, attached to a vector 
boson line, to indicate multiplication by $\ib\,\pmomi{\mu}$. 
\begin{figure}[th]
\vspace*{-9mm}
\begin{eqnarray*}
\vcenter{\hbox{
  \begin{picture}(60,50)(0,10)
  \Photon(0,35)(20,35){2}{5}
  \GCirc(25,35){5}{0.5}
  \Photon(30,35)(50,35){2}{5}
  \Text(-2,32)[cb]{$=$}
  \Text(52,32)[cb]{$=$}
  \Text(5,45)[cb]{$\pmom$}
  \Text(5,20)[cb]{$\wb$}
  \Text(45,20)[cb]{$\wb$}
  \end{picture}}}
&+\;\;&
\vcenter{\hbox{
  \begin{picture}(60,50)(0,10)
  \Photon(0,35)(20,35){2}{5}
  \GCirc(25,35){5}{0.5}
  \DashLine(30,35)(50,35){3}
  \Text(-2,32)[cb]{$=$}
  \Text(45,40)[cb]{$\LM$}
  \Text(5,20)[cb]{$\wb$}
  \Text(45,20)[cb]{$\hkg$}
  \end{picture}}}
+\;\;
\vcenter{\hbox{
  \begin{picture}(60,50)(0,10)
  \DashLine(0,35)(20,35){3}
  \GCirc(25,35){5}{0.5}
  \Photon(30,35)(50,35){2}{5}
  \Text(5,40)[cb]{$\LM$}
  \Text(52,32)[cb]{$=$}
  \end{picture}}}
+\;\;
\vcenter{\hbox{
  \begin{picture}(60,50)(0,10)
  \DashLine(0,35)(20,35){3}
  \GCirc(25,35){5}{0.5}
  \DashLine(30,35)(50,35){3}
  \Text(5,40)[cb]{$\LM$}
  \Text(45,40)[cb]{$\LM$}
  \end{picture}}}
=\;\;0
\end{eqnarray*}
\vspace{-0.5cm}
\caption[]{Example of Ward identities for transitions in the charged sector.
\label{fig:wised}}
\end{figure}
\vskip 5pt
\noindent
From the explicit expressions of these Ward identities we derive the following 
results:
\bqa
S_{\ssW\phi} &=& \frac{M}{p^2}\,S_{\phi}, \qquad S^T_{\ssW} = \frac{M^2}{p^2}\,
S_{\phi},  \nl
S^1_{\ssW} &=& \frac{1}{p^2}\,\lpar \frac{M^2}{p^2}\,S_{\phi} - S^0_{\ssW}\rpar.
\eqa
With their help the re-summed $\wb$ propagator becomes
\bqa
{\bar\Delta}^{\mu\nu}_{\ssW\ssW} &=& \frac{1}{p^2+M^2-S^0_{\ssW}}\,\lpar
\delta^{\mu\nu} + \Delta_L\,p^{\mu}p^{\nu}\rpar,  \nl
\Delta_L &=& \frac{1}{p^2}\,\Big[ M^2\,S_{\phi} - p^2\,S^0_{\ssW} +
{{M^2\lpar S_{\phi}\rpar^2}\over {p^2+M^2-S_{\phi}}}\Big]  \nl
{}&\times&
\Big[ p^2\,\lpar p^2+M^2\rpar - M^2\,S_{\phi} - {{M^2\,\lpar S_{\phi}\rpar^2}
\over {p^2+M^2-S_{\phi}}}\Big]^{-1}.
\eqa
The re-summed $\wb$ propagator satisfies the following identity:
\bq
p_{\mu}\,{\bar\Delta}^{\mu\nu}_{\ssW\ssW} = {{p^{\nu}}\over
{p^2+M^2-\frac{M^2}{p^2}\,S_{\phi}\,\Big[ 1 + S_{\phi}/\lpar p^2+M^2-S_{\phi}
\rpar\Big]}}.
\eq
Similarly, we obtain the $\wb-\phi$ re-summed transition,
\bq
{\bar\Delta}^{\mu}_{\ssW\phi} = i\,\frac{Mp^{\mu}}{p^2}\,
{{S_{\phi}}\over {p^2+M^2-S_{\phi}}}\,
{1\over {p^2+M^2-\frac{M^2}{p^2}\,S_{\phi}\,\Big[ 1 + S_{\phi}/
\lpar p^2+M^2-S_{\phi}\rpar\Big]}},
\eq
and the $\phi-\phi$ re-summed transition,
\bq
{\bar\Delta}_{\phi\phi} = {{p^2+M^2-M^2/p^2\,S_{\phi}}\over
{\lpar p^2+M^2-S_{\phi}\rpar\,\lpar p^2+M^2-M^2/p^2\,S_{\phi}\rpar - 
M^2/p^2\,\lpar S_{\phi}\rpar^2}}.
\eq
Before continuing we define some auxiliary quantities:
\bqa
{\bar\Delta}^{\mu\nu}_{\ssW\ssW} &=& \frac{1}{p^2+M^2-S^0_{\ssW}}\,
\lpar \delta^{\mu\nu} + \frac{N_v}{p^4\,D}\,p^{\mu}p^{\nu}\rpar,  \nl
{\bar\Delta}^{\mu}_{\ssW\phi} &=& i\,\frac{Mp^{\mu}}{p^2}\,
{{S_{\phi}}\over {\lpar p^2+M^2-S_{\phi}\rpar\,D}},  \nl
{\bar\Delta}_{\phi\phi} &=& \frac{1}{p^2+M^2-S_{\phi}}\,\frac{N_s}{D},
\eqa
where we have introduced
\bqa
D &=& p^2+M^2-\frac{M^2}{p^2}\,S_{\phi}\,\Big[ 1 + \frac{S_{\phi}}
{p^2+M^2-S_{\phi}}\Big],  \nl
N_v &=& M^2\,S_{\phi} - p^2\,S^0_{\ssW} + {{M^2\lpar S_{\phi}\rpar^2}\over
{p^2+M^2-S_{\phi}}},  \nl
N_s &=& p^2+M^2 - \frac{M^2}{p^2}\,S_{\phi}.
\eqa
We have seen that, for zero external masses, the re-summation of transitions
has a very simple effect on the Born amplitude, it promotes all quantities to
running ones and all bare couplings disappear from the amplitude itself. Is
it possible to have a generalization of this phenomenon that accounts for non
zero external masses? The answer to this question will be the subject of the
next section.

\section{The running $\wb$-boson mass.}

Running couplings appear naturally in the one-loop corrected amplitude
when we consider an $\Smat$-element, i.e. a transition between physical
sources. Therefore, we start with some simple example to illustrate the
generalization of the Fermion-Loop scheme. 
Consider the process $\nu_{\mu} \mu \to \barnu_e e$, as given in \fig{subp1}.
\vspace{0.2cm}
\bqas
\ba{ccc}
\vcenter{\hbox{
  \SetScale{0.7}
  \begin{picture}(110,50)(0,0)
  \ArrowLine(50,0)(0,50)
  \ArrowLine(0,-50)(50,0)
  \Line(50,0)(75,0)
  \GCirc(80,0){5}{1}
  \Line(85,0)(110,0)
  \ArrowLine(110,0)(160,-50)
  \ArrowLine(160,50)(110,0)
  \Text(-14,-40)[lc]{$\mu^-$}
  \Text(-14,45)[lc]{$\nu_{\mu}$}
  \Text(114,-40)[lc]{$e^-$}
  \Text(114,45)[lc]{$\barnu_e$}
  \Text(35,10)[lc]{$\wb$}
  \Text(65,10)[lc]{$\wb$}
  \end{picture}}}
&\quad{}&
\vcenter{\hbox{
  \SetScale{0.7}
  \begin{picture}(110,50)(0,0)
  \ArrowLine(50,0)(0,50)
  \ArrowLine(0,-50)(50,0)
  \Line(50,0)(75,0)
  \GCirc(80,0){5}{1}
  \DashLine(85,0)(110,0){3}
  \ArrowLine(110,0)(160,-50)
  \ArrowLine(160,50)(110,0)
  \Text(35,10)[lc]{$\wb$}
  \Text(65,10)[lc]{$\phi$}
  \end{picture}}}
\ea
\eqas
\vskip 0.8cm
\bqas
\ba{ccc}
\vcenter{\hbox{
  \SetScale{0.7}
  \begin{picture}(110,50)(0,0)
  \ArrowLine(50,0)(0,50)
  \ArrowLine(0,-50)(50,0)
  \DashLine(50,0)(75,0){3}
  \GCirc(80,0){5}{1}
  \Line(85,0)(110,0)
  \ArrowLine(110,0)(160,-50)
  \ArrowLine(160,50)(110,0)
  \Text(35,10)[lc]{$\phi$}
  \Text(65,10)[lc]{$\wb$}
  \end{picture}}}
&\quad{}&
\vcenter{\hbox{
  \SetScale{0.7}
  \begin{picture}(110,50)(0,0)
  \ArrowLine(50,0)(0,50)
  \ArrowLine(0,-50)(50,0)
  \DashLine(50,0)(75,0){3}
  \GCirc(80,0){5}{1}
  \DashLine(85,0)(110,0){3}
  \ArrowLine(110,0)(160,-50)
  \ArrowLine(160,50)(110,0)
  \Text(35,10)[lc]{$\phi$}
  \Text(65,10)[lc]{$\phi$}
  \end{picture}}}
\ea
\eqas
\vspace{2cm}
\begin{figure}[h]
\caption[]{The process $\mu \nu_{\mu} \to e \barnu_e$.}
\label{subp1}
\end{figure}
\vskip 20pt

For our purposes only four diagrams are relevant. They correspond to 
the inclusion of re-summed $\wb-\wb, \wb-\phi, \phi-\wb$ and $\phi-\phi$ 
transitions. We obtain
\bqa
M &=& \lpar \frac{i g}{2\srt}\rpar^2\,{\cal M},  \nl
{\cal M} &=& \gapu{\mu}\gdp \otimes \gapu{\nu}\gdp\,
{\bar\Delta}^{\mu\nu}_{\ssW\ssW} + \frac{\me}{M}\,\gapu{\mu}\gdp \otimes
\gdp\,{\bar\Delta}^{\mu}_{\ssW\phi}  \nl
{}&-& \frac{\mm}{M}\, \gdm \otimes \gapu{\mu}\gdp\,
{\bar\Delta}^{\mu}_{\phi\ssW} - \frac {\me\mm}{M^2}\, \gdm \otimes \gdp\,
{\bar\Delta}_{\phi\phi}.
\eqa
The amplitude can we written as the sum of two terms, a familiar one where
$\me = \mm = 0$ and an {\em extra} contribution given by:
\bq
{\cal M} = \frac{1}{p^2+M^2-S^0_{\ssW}} \gapu{\mu}\gdp \otimes \gapu{\mu}\gdp
+ {\cal M}_{\rm extra}.
\label{extt}
\eq
After some lengthy but straightforward algebra, making use of the Dirac 
equation and of the relation,
\bq
D\,\lpar p^2+M^2-S_{\phi}\rpar = \lpar p^2+M^2\rpar^2\,\lpar 1 - 
\frac{S_{\phi}}{p^2}\rpar,
\eq
the second term in \eqn{extt} can be written as follows:
\bq
{\cal M}_{\rm extra} = \me\mm\,\gdm \otimes \gdp\,\frac{1}{p^2+M^2-S^0_{\ssW}}\,
{{S^0_{\ssW} - p^2 - \frac{M^2}{p^2}\,S_{\phi}}\over {M^2\,\lpar p^2-S_{\phi}
\rpar}}.
\eq
There are two ingredients that we need to continue our construction of the
Fermion-Loop scheme. First the complex $\wb$-pole. We start with the
relation
\bq
-g^2\Delta_{\ssW} = - g^2\,\Big[ p^2+M^2-S^0_{\ssW}(p^2)\Big]^{-1} =
\Big[\frac{s-M^2}{g^2} + \frac{1}{16\,\pi^2}\,\Sigma^0_{\ssW}
\Big]^{-1},
\eq
and define the complex pole as a solution of
\bq
\frac{\sW-M^2}{g^2} + \frac{1}{16\,\pi^2}\,\Sigma^0_{\ssW}\lpar\sW\rpar = 0.
\eq
Therefore, for the $\wb$ propagator, we obtain
\bq
-g^2\,\Delta_{\ssW} = \Big\{ \frac{s-\sW}{g^2(s)} +\frac{\sW}{16\,\pi^2}\,
\Big[ \Ptg(\sW) - \Ptg(s)\Big] + \frac{1}{16\,\pi^2}\,\Big[ f_{\ssW}(s) -
f_{\ssW}(\sW)\Big]\Big\}^{-1}.
\eq
As expected, the position of the complex $\wb$-boson pole is solely fixed
by the $S^0_{\ssW}$-component of its self-energy.
Next, we need a second ingredient, the running $\wb$-boson mass.

\newtheorem{guess}{Definition}
\begin{guess}
The $\wb$-boson running mass is defined by the following equation:
\bq
\frac{1}{M^2(p^2)} = \frac{1}{M^2}\,{{p^2-S^0_{\ssW}+\frac{M^2}{p^2}\,S_{\phi}}
\over {p^2-S_{\phi}}}.
\eq
\end{guess}
It is not an independent quantity but, instead, it is related to the complex
pole. To give a simple illustration of this relation, let us consider the case 
of massless fermions in the one-loop transitions. This means, in particular, 
$S_{\phi} = 0$. Therefore, we have
\bq 
\frac{1}{M^2(p^2)} = \lpar 1 - \frac{S^0_{\ssW}}{p^2}\rpar\,\frac{1}{M^2}.
\eq
Using the fact that the bare mass and the complex pole are related by
\bq
M^2 = \sW + S^0_{\ssW}(\sW),
\eq
and combining it with the following two relations,
\bq
S^0_{\ssW}(p^2) = p^2\,\frac{g^2}{16\,\pi^2}\,\Ptg(p^2), \qquad
\frac{1}{g^2(p^2)} = \frac{1}{g^2} - \frac{1}{16\,\pi^2}\,\Ptg(p^2),
\eq
for a massless internal world we obtain
\bq
M^2(p^2) = \frac{g^2(p^2)}{g^2(\sW)}\,\sW, \qquad M^2(\sW) = \sW.
\eq
Equipped with this result, we can write ${\cal M}_{\rm extra}$ as
\bq
{\cal M}_{\rm extra} = \gapu{\mu}\gdp \otimes \gapu{\nu}\gdp\,
\frac{1}{p^2+M^2-S^0_{\ssW}}\, \frac{p^{\mu}p^{\nu}}{M^2(p^2)}.
\eq
To summarize our findings, the complete one-loop re-summation in the
't Hooft-Feynman gauge is equivalent to some {\em effective} unitary-gauge
$\wb$-propagator. The whole amplitude can be written in terms of a $\wb$-boson 
exchange diagram, if we make use of the following effective propagator:
\bq
\Delta^{\mu\nu}_{\rm eff} = \frac{1}{p^2+M^2-S^0_{\ssW}}\,\Big[
\delta^{\mu\nu} + \frac{p^{\mu}p^{\nu}}{M^2(p^2)}\Big].
\eq
we obtain a similar, although a little more complicated, result also when the 
top quark mass is not neglected in loop corrections. We start with
\bq
S^0_{\ssW} = \frac{g^2}{16\,\pi^2}\,\Big[ p^2\,\Ptg(p^2) + f_{\ssW}(p^2)\Big],
\eq
and derive
\bq
S_{\phi} = \frac{g^2}{16\,\pi^2}\,\frac{\mts}{M^2}\,f_{\phi}(p^2).
\eq
With $f_i(p^2) = p^2\sigma_i(p^2)$, we end up with the following result for 
the running mass:
\bqa
\frac{\sW}{M^2(p^2)} &=& \frac{g^2(\sW)}{g^2(p^2)}\,\Big\{ 
1 - \frac{g^2(p^2)}{16\,\pi^2}\,\Big[ \sigma_{\ssW}(p^2) - \frac{\mts}{p^2}\,
\sigma_{\phi}(p^2)\Big]\Big\}  \nl
{}&\times& 
\Big\{1 - \frac{g^2(\sW)}{16\,\pi^2}\,\Big[ \sigma_{\ssW}(\sW) + 
\frac{\mts}{\sW}\,\sigma_{\phi}(p^2)\Big]\Big\}^{-1}.
\label{rmassM}
\eqa
The above result is ultraviolet finite. Indeed we obtain
\bq
f_{\ssW}(p^2)\mid_{_{\rm UV}} = - \frac{3}{2}\,\mts\,\Ddr,
\eq
where $\Ddrd$ is the ultraviolet regulator,
\bq
\Ddr = \frac{2}{\dre} - \eilc - \lpi,
\eq 
and $\eilc=0.577216$ is the Euler constant. From the explicit expression 
for $S_{\phi}$, i.e.
\bq
S_{\phi}(p^2) = - \frac{3}{2}\,\Big[ \aff{0}{\mt} + \lpar p^2+\mts\rpar \,
\bff{0}{p^2}{\mt}{0}\Big],
\eq
we get the ultraviolet part of $f_{\phi}$, 
\bq
f_{\phi}(p^2)\mid_{_{\rm UV}} = - \frac{3}{2}\,p^2\,\Ddr,
\eq
giving a cancellation of the ultraviolet divergent term, $\tDdr$, inside
\eqn{rmassM}.

There are several examples where one can show that external fermion masses
can be easily included in the Fermion-Loop scheme. We simply promote all 
quantities to be running ones and use the unitary-gauge expression for the 
$\wb$-boson propagator, but with a running mass. Some of these examples are 
very instructive, since they clearly show how the strategy works only for 
$\Smat$-matrix elements, i.e. for amputated Green's function with on-shell and 
properly renormalized external sources. 

Consider $\wbp(q)+\ph(k) \to u \bard$, a component of the single-$\wb$ process.
We have an amplitude that can be written as follows:
\bq
M_{\mu\alpha} = V^0_{\mu\alpha\beta}\,{{\gapu{\beta}\gdp}\over 
{p^2+M^2-S^0_{\ssW}}} + i\,{\cal M}_{\mu\alpha}\,\lpar \muq\gdp - \md\gdm\rpar.
\eq
The first result that we obtain is for a situation where the incoming
particles are physical, i.e.
\bq
k^2= 0, \quad q^2 = - M^2, \qquad \spro{k}{e(k)}= 0, \quad 
\spro{q}{\varepsilon(q)} = 0,
\eq
where $e_{\mu}(k)$ and $\varepsilon_{\alpha}(q)$ are the photon and the $\wb$
polarization vectors. In this case we find
\bq
{\cal M}_{\mu\alpha} = {{V^0_{\mu\alpha\beta}}\over {p^2+M^2-S^0_{\ssW}}}\,
\frac{p^{\beta}}{M^2(p^2)},
\eq
where $p = q+k$. To go further, we only require conservation of the e.m. 
current, $\spro{k}{e(k)} = 0$, but not the mass-shell condition $k^2 = 0$.
In this case, even for $\me = 0$, a residual term remains:
\bq
\delta_{\mu\alpha}\,p^2\,{{M^2+q^2-k^2}\over {M^2\,\lpar p^2+M^2\rpar\,
\lpar p^2-S_{\phi}\rpar}}.
\eq
To understand the mechanism of cancellation we embed the sub-process (four 
diagrams) corresponding to the annihilation $\wbp \ph \to u \bard$ into
$e^+ \ph \to \barnu_e u \bard$. We obtain the following result:
\vspace{0.2cm}
\bqas
\ba{ccc}
\vcenter{\hbox{
  \SetScale{0.7}
  \begin{picture}(110,50)(0,0)
  \Line(50,0)(0,50)
  \Photon(0,-50)(50,0){2}{7}
  \Line(50,0)(75,0)
  \GCirc(80,0){5}{1}
  \Line(85,0)(110,0)
  \ArrowLine(110,0)(160,-50)
  \ArrowLine(160,50)(110,0)
  \ArrowLine(0,50)(-50,50)
  \ArrowLine(50,75)(0,50)
  \Text(50,-65)[lc]{\bf annihilation}
  \Text(0,15)[lc]{$\wb$}
  \Text(36,10)[lc]{$\wb/\phi$}
  \Text(62,10)[lc]{$\wb/\phi$}
  \end{picture}}}
&\quad\qquad&
\vcenter{\hbox{
  \SetScale{0.7}
  \begin{picture}(110,50)(0,0)
  \ArrowLine(0,50)(-50,50)
  \ArrowLine(50,75)(0,50)
  \Photon(0,50)(0,-50){2}{7}
  \Line(50,75)(110,0)
  \GCirc(80,38){5}{1}
  \ArrowLine(110,0)(160,-50)
  \ArrowLine(160,50)(110,0)
  \ArrowLine(75,90)(50,75)
  \Text(30,-65)[lc]{\bf bremsstrahlung}
  \Text(45,48)[lc]{$\wb$}
  \Text(66,20)[lc]{$\wb/\phi$}
  \end{picture}}}
\ea
\eqas
\vspace{1.4cm}
\begin{figure}[h]
\caption[]{Different topologies for the process $e^+ \ph \to \barnu_e u \bard$.}
\label{subp2}
\end{figure}
\vskip 10pt
\bqa
M^{\rm ann}_{\mu} &=& {{V^0_{\mu\alpha\beta}}\over {p^2+M^2-S^0_{\ssW}}}\,\Big[
\delta^{\beta\lambda} + {{p^{\beta}p^{\lambda}}\over {M^2(p^2)}}\,
\gapu{\alpha}\gdp \otimes \gapu{\lambda}\gdp  \nl
{}&+& \delta_{\mu\alpha}\,\frac{p^2}{M^2}\,{{M^2+q^2-k^2}\over 
{\lpar p^2+M^2\rpar\,\lpar p^2-S_{\phi}\rpar}}\,\gapu{\alpha}\gdp
\otimes \sla{p}\gdp.
\label{extra1}
\eqa
Some unwanted term remains.
However, there is another contribution, where the $u\bard$ pair is emitted by 
a $\wb$-boson that, in turn, is coming from the splitting $e \to \nu_e \wb$,
see \fig{subp2}.
Here, in the limit $\me = 0$, two diagrams contribute, a $\wb-\wb$ propagator
and a $\wb-\phi$ propagator. We find
\bqa
M^{\rm brem}_{\mu} &=& 
\frac{1}{p^2+M^2-S^0_{\ssW}}\,\Big[ \delta_{\alpha\beta} + 
{{p_{\alpha}p_{\beta}}\over {M^2(p^2)}}\Big]\,
\gapu{\mu}\,{{i\lpar \sla{p_+}+\sla{k}\rpar}\over
{\lpar p_++k\rpar^2}}\,\gapu{\alpha}\gdp \otimes \gapu{\beta}\gdp  \nl
{}&-& \frac{p^2}{M^2}\, {1\over {\lpar p^2+M^2\rpar\,\lpar p^2-S_{\phi}\rpar}}
\,\gapu{\mu}\,{{i\lpar \sla{p_+}+\sla{k}\rpar}\over
{\lpar p_++k\rpar^2}}\,\lpar \sla{p_+}+\sla{k}\rpar\,\gdp \otimes
\sla{p}\gdp.
\label{extra2}
\eqa
The two extra terms coming from \eqns{extra1}{extra2} add up to something
proportional to $k^2$, therefore vanishing for $k^2= 0$. This line of
arguments is correct only as long as we neglect re-summation in the $t$-channel
$\wb$-propagator.

\section{Ward identities for single-$\wb$.}

Before proving the relevant U(1) Ward identity for the single-$\wb$
amplitude we recall that the total of $20$ Feynman diagrams is, first of all,
split into a $10$ $s$-channel part and a $10$ $t$-channel part. In the latter
part, one diagram has a $\wb$-exchange, five a $\zb$-exchange and four a 
$\ph$-exchange. As we have shown, this picture is not changed by the inclusion
of one-loop fermionic corrections: when all transitions are properly taken 
into account, we still end up with the $1-5-4$ subdivision
described above, as long as the fermion--anti-fermion-vector--boson couplings
are described in terms of the re-summed expressions.

This $s\,\oplus\,t$ splitting is a gauge invariant one, although we 
can further restrict the number of diagrams. The argument is as follows:
Take $e^+ \mu^- \to \barnu_e \mu^- u \bard$. Only the CC20 $t$-channel 
diagrams contribute and, since these are all diagrams that we need, this set 
is gauge invariant.
Next, take $e^+ \nu_{\mu} \to \barnu_e \nu_{\mu} u \bard$. Its is, again,
only $t$-channel but the photon does not contribute, so that we have
$10-4 = 6$ diagrams. Moreover, if one writes any $\zb$ current as 
$J = J_{\ssQ} + J_{\ssL}$ (with $J_{\ssL}$ proportional to $\gdp$), only 
$J_{\ssL}$ contributes here, because of the neutrinos, so that we have five 
$J^{\ssZ}_{\ssL}$ diagrams plus one $\wb$ diagram that form a gauge 
invariant set.
Since the whole $t$-channel is gauge invariant, the eight diagrams, four
with photons and four with $J^Z_Q$, must form a gauge invariant sub-set.
This remains true for one-loop corrections when writing everything in terms 
of running objects and including vertices. 

The gauge invariance property that we are referring to is the full SU(2) one.
However, the first step is to prove the simpler U(1) gauge invariance or,
stated differently, we write the amplitude squared for the four 
photon-mediated $t$-channel diagrams as the product of a leptonic tensor 
$L_{\mu\nu}$ and of a single-$\wb$ tensor, $W_{\mu\nu}$. Successively, we must 
be able to prove that $Q^{-\mu}W_{\mu\nu} = Q^{-\nu}W_{\mu\nu} = 0$ (where 
$Q_-$ 
is the photon momentum), so that the only terms that survive in the product
$L_{\mu\nu}W_{\mu\nu}$ are those proportional to $Q^2_-$ or to $\mes$, giving 
rise
to the familiar contributions to the cross-section. Either logarithmically 
enhanced, $1/Q^2_-$, or the sub-leading, constant one, $\mes/Q^4_-$.

To be more specific, we give the explicit expression for $L_{\mu\nu}$.
With momenta assignment $e^+(p_+) e^-(p_-) \to e^-(q_-) \barnu_e(q_+) u(k) 
\bard(\kbar)$, we have
\bq
L_{\mu\nu} = 2\,Q^2_-\delta_{\mu\nu} + 4\,\lpar p_{-\mu}q_{-\nu} + 
q_{-\mu}p_{-\nu}\rpar.
\eq
U(1) gauge invariance, yet to be proven, requires that the following 
decomposition holds for $W_{\mu\nu}$,
\bqa
W_{\mu\nu} &=& W_1\, \Big[ - \delta_{\mu\nu}+\frac{Q_{-\mu}Q_{\nu}}{Q^2_-}\Big] -
             W_2\,\frac{Q^2_-}{\lpar\spro{p_+}{Q_-}\rpar^2}\,
\lpar p_{+\mu}-\frac{\spro{p_+}{Q_-}}{Q^2_-}\,Q_{-\mu} \rpar  \nl
{}&\times&
\lpar p_{+\nu}-\frac{\spro{p_+}{Q_-}}{Q^2_-}\,Q_{-\nu} \rpar.
\label{decom}
\eqa
After contraction, and for $X \to 0$, we obtain
\bq
\frac{1}{4}\,L_{\mu\nu}W^{\mu\nu} = 
W_1\,\Big[ 2\,\frac{\mes}{\lpar Xys\rpar^2} - \frac{1}{Xys}\Big] +
2\,W_2\,\frac{y-1}{Xy^3s} + \ord{X},
\eq
where we have introduced the variable $y$, equivalent to the fraction of the 
electron energy carried by the photon, and
\bqa
\spro{p_+}{Q_-} &=& \frac{1}{2}\,\Big[ \mes - \lpar X + 1\rpar\,ys\Big], \nl
Q^2_- &=& X\,ys,  \qquad \lpar p_+ + p_-\rpar^2 = - ys.
\eqa
The decomposition of \eqn{decom} requires that we can prove a U(1) Ward
identity. This we will do without performing any approximation, within the
framework of the Fermion-Loop scheme. One has to be fully aware that the seven 
tensor structures, introduced in \eqn{fvphysical}, are not enough to describe 
the situation, simply because currents are not-conserved. 

Before giving the most general structure for the vertex, we will introduce the 
invariants needed to describe the sub-process, $e^+(p_+) \ph(Q_-) \to 
\barnu_e(q_+) u(k) \bard(\kbar)$.
They are specified in the following list:
\bqa
\spro{p_+}{Q_-} &=& \frac{1}{2}\,\lpar \mes-Q^2_--y\,s\rpar,  \nl
\spro{k}{\kbar} &=& \frac{1}{2}\,\lpar -s'+\mus+\mds\rpar,  \nl
\spro{Q_-}{k} &=& \frac{1}{2}\,\lpar Q^2_--\mus+t\rpar,  \nl
\spro{p_+}{\kbar} &=& \frac{1}{2}\,\lpar -\mes-\mds+t'\rpar,  \nl
\spro{Q_-}{\kbar} &=& \frac{1}{2}\,\lpar Q^2_--\mds+u\rpar,  \nl
\spro{p_+}{k} &=& \frac{1}{2}\,\lpar -\mes-\mus+u'\rpar,  \nl
\spro{p_+}{q_+} &=& \frac{1}{2}\,\lpar \kappa_+-\mes\rpar,  \nl
\spro{Q_-}{q_+} &=& \frac{1}{2}\,\lpar Q^2_-+\kappa_-\rpar,  \nl
\spro{\kbar}{q_+} &=& \frac{1}{2}\,\lpar -\mds+\zeta_+\rpar,  \nl
\spro{k}{q_+} &=& \frac{1}{2}\,\lpar \mus+\zeta_-\rpar,  
\eqa
The linearly independent invariants are:
\bqa
t &=& \tau ys,  \quad s' = x_2 y s,  \quad \kappa_- = zys,  \nl
\zeta_- &=& (x_2-x_1)\,ys,  \quad Q^2_- = Xys,  
\eqa
with the following solution for the remaining ones:
\bqa
\kappa_+ &=& \mes + y\,s (  - 1 - X + x_2 - z ),  \nl
u &=& \mes + \mus + \mds + y\,s (  - 1 - \tau - 2 X - z ),  \nl
u' &=& \mes + 2\,\mus + \mds + y\,s (  - \tau - X - x_1 ),  \nl
t' &=& - \mus + y\,s ( \tau + X + x_1 - x_2 + z ),  \nl
\zeta_+ &=& - \mus + \mds + y\,s (  - 1 + x_1 ).
\label{refkp}
\eqa
We introduce another auxiliary quantity
\bq
Q^2_+ = -\,Yys,
\eq
and also the vector $Q= Q_+ + Q_- = k + \kbar$.
In the one-loop corrected $\ph\wbp\wbm$ vertex of \fig{fig:threepoint} all 
vector-boson lines are off mass-shell and non-conserved. 
\begin{figure}[h]
\vspace*{-8mm}
\[
  \vcenter{\hbox{
  \begin{picture}(140,130)(-15,-15)
    \ArrowLine(30,50)(73,75)        \Text(50,72)[cb]{$m$}
    \ArrowLine(73,75)(73,25)        \Text(50,25)[cb]{$m$}
    \ArrowLine(73,25)(30,50)        \Text(82,50)[cb]{$m'$}
    \Photon(0,50)(30,50){2}{7}    
    \Line(100,100)(73,75)   
    \Line(100,0)(73,25)   
    \LongArrow(10,57)(20,57)   \Text(13,65)[cb]{$\mu,\,Q_-$}
    \LongArrow(89,98)(82,91)   \Text(79,103)[cb]{$\beta,\,Q_+$}
    \LongArrow(89,2)(82,9)     \Text(79,-10)[cb]{$\alpha,\,-\,Q$}
  \end{picture}}}
\qquad
  \vcenter{\hbox{
  \begin{picture}(140,130)(-15,-15)
    \ArrowLine(73,75)(30,50)        \Text(50,72)[cb]{$m'$}
    \ArrowLine(73,25)(73,75)        \Text(50,25)[cb]{$m'$}
    \ArrowLine(30,50)(73,25)        \Text(82,50)[cb]{$m$}
    \Photon(0,50)(30,50){2}{7}    
    \Line(100,100)(73,75)   
    \Line(100,0)(73,25)   
    \LongArrow(10,57)(20,57)   \Text(13,65)[cb]{$\mu,\,Q_-$}
    \LongArrow(89,98)(82,91)   \Text(79,103)[cb]{$\beta,\,Q_+$}
    \LongArrow(89,2)(82,9)     \Text(79,-10)[cb]{$\alpha,\,-\,Q$}
  \end{picture}}}
\]
\vspace{-0.75cm}
\caption[]{The $\ph\wbp\wbm$ vertex.\label{fig:threepoint}}
\end{figure}
\vskip 5pt
For the e.m. current this is a consequence of the fact that we are computing 
a Ward identity and not an amplitude.
As a consequence, several more structures are needed and we list them below:
\bqa
W^1_{\mu\alpha\beta} &=& 4\,{\cal N}\,\Big[
 \drii{\alpha}{\beta}Q_{+\mu} + \drii{\mu}{\beta}Q_{-\alpha} -
\drii{\mu}{\alpha}Q_{\beta}\Big],  
\nl
W^{2,3}_{\mu\alpha\beta} &=& 2\,{\cal N}\,
\Big[ \drii{\mu}{\beta}Q_{_\alpha} \mp \drii{\mu}{\alpha}Q_{\beta}\Big],
\nl
W^4_{\mu\alpha\beta} &=& 2\,{\cal N}\,
\drii{\alpha}{\beta}Q_{-\mu},
\nl
W^5_{\mu\alpha\beta} &=& 2\,{\cal N}\,
\drii{\mu}{\alpha}Q_{+\beta},
\nl
W^6_{\mu\alpha\beta} &=& 2\,{\cal N}\,
\drii{\mu}{\beta}Q_{+\alpha},
\nl
W^7_{\mu\alpha\beta} &=& -2\,{\cal N}\,
\drii{\mu}{\beta}Q_{\alpha},
\nl
W^8_{\mu\alpha\beta} &=& -2\,{\cal N}\,
\drii{\alpha}{\beta}Q_{\mu},
\nl\nl
W^9_{\mu\alpha\beta} &=& 2\,{\cal N}^3
Q_{+\mu}Q_{-\alpha}Q_{\beta},
\nl
W^{10}_{\mu\alpha\beta} &=& 2\,{\cal N}^3\,
Q_{-\mu}Q_{-\alpha}Q_{\beta},
\nl
W^{11}_{\mu\alpha\beta} &=& -2\,{\cal N}^3\,
Q_{+\mu}Q_{+\alpha}Q_{+\beta},
\nl
W^{12}_{\mu\alpha\beta} &=& 2\,{\cal N}^3\,
Q_{+\mu}Q_{+\alpha}Q_{\beta},
\nl
W^{13}_{\mu\alpha\beta} &=& 2\,{\cal N}^3\,
Q_{+\mu}Q_{\alpha}Q_{+\beta},
\nl
W^{14}_{\mu\alpha\beta} &=& -2\,{\cal N}^3\,
Q_{+\mu}Q_{\alpha}Q_{\beta},
\nl
W^{15}_{\mu\alpha\beta} &=& 2\,{\cal N}^3\,
Q_{\mu}Q_{+\alpha}Q_{+\beta},
\nl
W^{16}_{\mu\alpha\beta} &=& -2\,{\cal N}^3\,
Q_{\mu}Q_{+\alpha}Q_{\beta},
\nl
W^{17}_{\mu\alpha\beta} &=& -2\,{\cal N}^3\,
Q_{\mu}Q_{\alpha}Q_{+\beta},
\nl
W^{18}_{\mu\alpha\beta} &=& 2\,{\cal N}^3\,
Q_{\mu}Q_{\alpha}Q_{\beta},
\nl\nl
W^{19}_{\mu\alpha\beta} &=& -2\,{\cal N}\, \varepsilon(Q_-,\mu,\alpha,\beta),
\nl
W^{20}_{\mu\alpha\beta} &=& -2\,{\cal N}\, 
\Big[ \varepsilon(Q_-,\mu,\alpha,\beta) + 2\,\varepsilon(Q_+,\mu,\alpha,\beta)
\Big],
\nl\nl
W^{21}_{\mu\alpha\beta} &=& -{\cal N}^3\,\varepsilon(Q_+,Q_-,\alpha,\beta)
Q_{\mu},  
\nl 
W^{22}_{\mu\alpha\beta} &=& {\cal N}^3\,
\varepsilon(Q_+,Q_-,\alpha,\beta)Q_{+\mu},
\nl
W^{23}_{\mu\alpha\beta} &=& {\cal N}^3\,\Big[
\varepsilon(Q_+,Q_-,\mu,\alpha)Q_{+\beta}-\varepsilon(Q_+,Q_-,\mu,\beta)
Q_{+\alpha}\Big],
\nl
W^{24}_{\mu\alpha\beta} &=& -\,{\cal N}^3\,\varepsilon(Q_+,Q_-,\mu,\beta)
Q_{\alpha}.
\label{operators}
\eqa
The normalization factor is ${\cal N} = (Xys)^{-1/2}$.
In presenting the vertices we face the usual problem of introducing a full
{\em scalarization} of the result; that is, all results should be presented 
in terms of one-loop scalar-integral coefficient functions. This procedure is 
mandatory since the  symmetry of the vertices has to be verified.
However, there is no need to show the scalarized version of the full vertices,
only the corresponding contraction entering into the Ward identity is needed 
and, as a matter of fact, the latter is considerably simpler and much
shorter. To summarize, we will not present explicitly the full vertex
\bq
V^1_{\mu\alpha\beta} = \frac{g^3\stw}{16\,\pi^2}\,(Xys)^{1/2}\,
\asums{i=1,24}\,I_i\,W^i_{\mu\alpha\beta},
\eq
but only the contraction. Actually, we cannot limit our analysis to the
$\ph\wbp\wbm$ vertex but we must include also the $\ph\wbp\hkm, \ph\hkp\wbm$
and $\ph\hkp\hkm$ vertices. 
For $\ph\wbp\hkm$ the operators are as follows:
\bqa
W^1_{\mu\beta} &=& \drii{\mu\beta},
\nl
W^2_{\mu\beta} &=& {\cal N}^2\,Q_{+\mu}Q_{\beta},
\nl
W^3_{\mu\beta} &=& -{\cal N}^2\,Q_{+\mu}Q_{+\beta},
\nl
W^4_{\mu\beta} &=& {\cal N}^2\,Q_{\mu}Q_{+\beta},
\nl
W^5_{\mu\beta} &=& -{\cal N}^2\,Q_{\mu}Q_{\beta},
\nl
W^6_{\mu\beta} &=& {\cal N}^2\,\varepsilon(Q_+,Q_-,\mu,\beta).
\eqa
For $\ph\hkp\wbm$ the operators are as follows:
\bqa
W^1_{\mu\alpha} &=& \drii{\mu}{\alpha},
\nl
W^2_{\mu\alpha} &=& {\cal N}^2\,Q_{+\mu}Q_{\alpha},
\nl
W^3_{\mu\alpha} &=& -{\cal N}^2\,Q_{+\mu}Q_{+\alpha},
\nl
W^4_{\mu\alpha} &=& {\cal N}^2\,Q_{+\mu}Q_{\alpha},
\nl
W^5_{\mu\alpha} &=& -{\cal N}^2\,Q_{\mu}Q_{\alpha},
\nl
W^6_{\mu\alpha} &=& {\cal N}^2\,\varepsilon(Q_+,Q_-,\mu,\alpha).
\eqa
Finally for $\ph\hkp\hkm$ we have
\bq
W^1_{\mu} = Q_{+\mu}, \qquad W^2_{\mu} = Q_{\mu}.
\eq
For $\ph\wbp\wbm$ we will, therefore, present the contraction
\bqa
V^{\ssW\ssW}_{\alpha\beta} &=& Q^{\mu}_- V^1_{\mu\alpha\beta} =
\frac{g^3\stw}{16\,\pi^2}\,\Big[ W^{\ssW\ssW}_0\,\drii{\alpha}{\beta} 
+ W^{\ssW\ssW}_1\,\lpar
Q_{+\alpha}Q_{-\beta} + Q_{+\beta}Q_{-\alpha}\right.
\nl 
{}&+&\left. Q_{-\alpha}Q_{-\beta}\rpar + W^{\ssW\ssW}_2\,Q_{+\alpha}Q_{+\beta}
\Big].
\label{cont1}
\eqa
This, is all what we need for a massless internal world, i.e. $\mt = 0$.
Otherwise, additional contractions are needed. They are:
\bqa
V^{\ssW\phi}_{\beta} &=& \frac{ig^3\stw}{16\,\pi^2 M}\,\Big[
W^{\ssW\phi}_1\,Q_{+\beta} + W^{\ssW\phi}_2\,Q_{-\beta}\big],
\nl 
V^{\phi\ssW}_{\alpha} &=& \frac{ig^3\stw}{16\,\pi^2 M}\,\Big[
W^{\phi\ssW}_1\,Q_{+\alpha} + W^{\phi\ssW}_2\,Q_{-\alpha}\Big],
\nl 
V^{\phi\phi} &=& \frac{g^3\stw}{16\,\pi^2 M^2}\,W^{\phi\phi}.
\label{cont2}
\eqa
The Ward identity that we want to prove can be written as follows:
\bq 
{\rm WI} = Q^{\mu}_-\,\asums{i=1,4}\,D^i_{\mu} + D^1_V,
\label{defWI}
\eq
where the $D^i$ are the four diagrams of $t$-channel with a photon line
and the electron line removed and $D^1_V$ is $D^1$ with the inclusion
of the all one-loop vertices and with the saturation already performed.

The sum of all diagrams, needed for our Ward identity, is as follows:
\bqa
D^1_{\mu} &=& \lpar\frac{i g}{2\,\srt}\rpar^2 g \stw \Big[
\iap{p_+} S^V_{l\beta}(1) \oap{q_+} \op{k} S^V_{q\alpha}(1) \oap{\kbar} 
V^0_{\wb\wb\mu\lambda\rho} \bDelta^{\alpha\lambda}_{\ssW\ssW}(s)
\bDelta^{\beta\rho}_{\ssW\ssW}(t) \nl
{}&+&
\iap{p_+} S^V_{l\beta}(1) \oap{q_+} \op{k} S^S_q(1) \oap{\kbar} 
V^0_{\wb\wb\mu\lambda\rho} \bDelta^{\lambda}_{\ssW\phi}(s) 
\bDelta^{\beta\rho}_{\ssW\ssW}(t) \nl
{}&+&
\iap{p_+} S^V_{l\beta}(1) \oap{q_+} \op{k} S^V_{q\alpha}(1) \oap{\kbar} 
V^0_{\wb\phi\mu\rho} \bDelta^{\alpha}_{\phi\ssW}(s) 
\bDelta^{\beta\rho}_{\ssW\ssW}(t) \nl
{}&+&
\iap{p_+} S^V_{l\beta}(1) \oap{q_+} \op{k} S^S_q(1) \oap{\kbar} 
V^0_{\wb\phi\mu\rho} \bDelta_{\phi\phi}(s) \bDelta^{\beta\rho}_{\ssW\ssW}(t) \nl
{}&+&
\iap{p_+} S^V_{l\beta}(1) \oap{q_+} \op{k} S^V_{q\alpha}(1) \oap{\kbar} 
V^0_{\phi\wb\mu\lambda} \bDelta^{\alpha\lambda}_{\ssW\ssW}(s) 
\bDelta^{\beta}_{\ssW\phi}(t) \nl
{}&+&
\iap{p_+} S^V_{l\beta}(1) \oap{q_+} \op{k} S^S_q(1) \oap{\kbar} 
V^0_{\phi\wb\mu\lambda} \bDelta^{\lambda}_{\ssW\phi}(s) 
\bDelta^{\beta}_{\ssW\phi}(t) \nl
{}&+&
\iap{p_+} S^V_{l\beta}(1) \oap{q_+} \op{k} S^V_{q\alpha}(1) \oap{\kbar} 
V^0_{\phi\phi\mu} \bDelta^{\alpha}_{\phi\ssW}(s) \bDelta^{\beta}_{\ssW\phi}(t) \nl
{}&+&
\iap{p_+} S^V_{l\beta}(1) \oap{q_+} \op{k} S^S_q(1) \oap{\kbar} 
V^0_{\phi\phi\mu} \bDelta_{\phi\phi}(s) \bDelta^{\beta}_{\ssW\phi}(t) \nl
{}&+&
\iap{p_+} S^S_l(1) \oap{q_+} \op{k} S^V_{q\alpha}(1) \oap{\kbar} 
V^0_{\wb\wb\mu\lambda\rho} \bDelta^{\alpha\lambda}_{\ssW\ssW}(s) 
\bDelta^{\rho}_{\phi\ssW}(t) \nl
{}&+&
\iap{p_+} S^S_l(1) \oap{q_+} \op{k} S^S_q(1) \oap{\kbar} 
V^0_{\wb\wb\mu\lambda\rho} \bDelta^{\lambda}_{\ssW\phi}(s) 
\bDelta^{\rho}_{\phi\ssW}(t) \nl
{}&+&
\iap{p_+} S^S_l(1) \oap{q_+} \op{k} S^V_{q\alpha}(1) \oap{\kbar} 
V^0_{\wb\phi\mu\rho} \bDelta^{\alpha}_{\phi\ssW}(s) \bDelta^{\rho}_{\phi\ssW}(t)
\nl
{}&+&
\iap{p_+} S^S_l(1) \oap{q_+} \op{k} S^S_q(1) \oap{\kbar} 
V^0_{\wb\phi\mu\rho} \bDelta_{\phi\phi}(s) \bDelta^{\rho}_{\phi\ssW}(t) \nl
{}&+&
\iap{p_+} S^S_l(1) \oap{q_+} \op{k} S^V_{q\alpha}(1) \oap{\kbar} 
V^0_{\phi\wb\mu\lambda} \bDelta^{\alpha\lambda}_{\ssW\ssW}(s) 
\bDelta_{\phi\phi}(t) \nl
{}&+&
\iap{p_+} S^S_l(1) \oap{q_+} \op{k} S^S_q(1) \oap{\kbar} 
V^0_{\phi\wb\mu\lambda} \bDelta^{\lambda}_{\ssW\phi}(s) 
\bDelta_{\phi\phi}(t) \nl
{}&+&
\iap{p_+} S^S_l(1) \oap{q_+} \op{k} S^V_{q\alpha}(1) \oap{\kbar} 
V^0_{\phi\phi\mu} \bDelta^{\alpha}_{\phi\ssW}(s) \bDelta_{\phi\phi}(t) \nl
{}&+&
\iap{p_+} S^S_l(1) \oap{q_+} \op{k} S^S_q(1) \oap{\kbar} 
V^0_{\phi\phi\mu} \bDelta_{\phi\phi}(s) \bDelta_{\phi\phi}(t)\Big]
\label{defD1}
\eqa
\bqa
D^2_{\mu} &=& \lpar\frac{i g}{2\,\srt}\rpar^2 \lpar -i g \stw\rpar\,\Big[
\iap{p_+} S^V_{l\mu\beta}(2) \oap{q_+} \op{k} S^V_{q\alpha}(2) \oap{\kbar} 
       \bDelta^{\beta\alpha}_{\ssW\ssW}(s) \nl
{}&+&
\iap{p_+} S^V_{l\mu\beta}(2) \oap{q_+} \op{k} S^S_q(2) \oap{\kbar} 
       \bDelta^{\beta}_{\ssW\phi}(s) \nl
{}&+&
\iap{p_+} S^S_{l\mu}(s) \oap{q_+} \op{k} S^V_{q\alpha}(2) \oap{\kbar} 
       \bDelta^{\alpha}_{\phi\ssW}(s) \nl
{}&+&
\iap{p_+} S^S_{l\mu}(s) \oap{q_+} \op{k} S^S_q(2) \oap{\kbar} 
       \bDelta_{\phi\phi}(s)\Big]
\eqa
\bqa
D^3_{\mu} &=& \lpar\frac{i g}{2\,\srt}\rpar^2 \lpar i g Q_u \stw\rpar\,\Big[
\iap{p_+} S^V_{l\beta}(3) \oap{q_+} \op{k} S^V_{q\mu\alpha}(3) \oap{\kbar} 
       \bDelta^{\beta\alpha}_{\ssW\ssW}(t) \nl
{}&+& 
\iap{p_+} S^V_{l\beta}(3) \oap{q_+} \op{k} S^S_{q\mu}(3) \oap{\kbar} 
       \bDelta^{\beta}_{\ssW\phi}(t) \nl
{}&+& 
\iap{p_+} S^S_l(3) \oap{q_+} \op{k} S^V_{q\mu\alpha}(3) \oap{\kbar} 
       \bDelta^{\alpha}_{\phi\ssW}(t) \nl
{}&+& 
\iap{p_+} S^S_l(3) \oap{q_+} \op{k} S^S_{q\mu}(3) \oap{\kbar} 
       \bDelta_{\phi\phi}(t)\Big] 
\eqa
\bqa 
D^4_{\mu} &=& \lpar\frac{i g}{2\,\srt}\rpar^2 \lpar i g Q_d \stw\rpar\,\Big[
\iap{p_+} S^V_{l\beta}(4) \oap{q_+} \op{k} S^V_{q\mu\alpha}(4) \oap{\kbar} 
       \bDelta^{\beta\alpha}_{\ssW\ssW}(t) \nl
{}&+& 
\iap{p_+} S^V_{l\beta}(4) \oap{q_+} \op{k} S^S_{q\mu}(4) \oap{\kbar} 
       \bDelta^{\beta}_{\ssW\phi}(t) \nl
{}&+& 
\iap{p_+} S^s_l(4) \oap{q_+} \op{k} S^V_{q\mu\alpha}(4) \oap{\kbar} 
       \bDelta^{\alpha}_{\phi\ssW}(t) \nl
{}&+& 
\iap{p_+} S^s_l(4) \oap{q_+} \op{k} S^S_{q\mu}(4) \oap{\kbar} 
       \bDelta_{\phi\phi}(t)\Big] 
\eqa
Here, the argument $s$ or $t$ for propagators denotes the variable
$s'$ and $\kappa_+$, of \eqn{refkp}, respectively. Furthermore,
\bq
Q_{\pm} = p_{\pm} - q_{\pm}, \qquad Q = Q_+ + Q_-.
\eq
The strings of gamma-matrices are given by
\bqa
S^V_{l\mu}(1) &=& \gadu{\mu} \gdp, \quad
S^S_l(1) = \frac{\me}{M} \gdp,  \quad
S^V_{q\mu}(1) = \gadu{\mu} \gdp, \nl
S^S_q(1) &=& \frac{1}{2} \Big[\frac{\muq-\md}{M} (\gdp+\gdm)+\frac{\muq+\md}{M}
(\gdp-\gdm)\Big].
\eqa
\bqa 
S^V_{l\mu\beta}(2) &=& \gadu{\mu} \frac{i (\sla{q_+}+\sla{Q})+\me}
{(q_++Q)^2+\mes} \gadu{\beta} \gdp,  \nl
S^S_{l\mu}(2) &=& \gadu{\mu} \frac{i (\sla{q_+}+\sla{Q})+\me}
{(q_++Q)^2+\mes} \frac{\me}{M} \gdp, \quad
S^V_{q\mu}(2) = \gadu{\mu} \gdp, \nl
S^S_q(2) &=& \frac{1}{2} \Big[\frac{\muq-\md}{M} (\gdp+\gdm)+\frac{\muq+\md}{M} 
(\gdp-\gdm)\Big].
\eqa
\bqa
S^V_{l\mu}(3) &=& \gadu{\mu} \gdp, \quad
S^S_l(3) = \frac{\me}{M} \gdp,  \nl
S^V_{q\mu\alpha}(3) &=& \gadu{\mu} \frac{i (\sla{Q_-}-\sla{k})+\muq}
{(Q_--k)^2+\mus} \gadu{\alpha} \gdp,  \nl
S^S_{q\mu}(3) &=& \gadu{\mu} \frac{i (\sla{Q_-}-\sla{k})+\muq}
{(Q_--k)^2+\mus} (\frac{\muq}{M} \gdp-\frac{\md}{M} \gdm).
\eqa
\bqa 
S^V_{l\mu}(4) &=& \gadu{\mu} \gdp, \quad
S^S_l(4) = \frac{\me}{M} \gdp,  \nl
S^V_{q\mu\alpha}(4) &=& \gadu{\alpha} \gdp 
\frac{-i (\sla{Q_-}-\sla{\kbar})+\md}{(Q_--\kbar)^2+\mds} \gadu{\mu}, \nl
S^S_{q\mu}(4) &=& (\frac{\muq}{M} \gdp-\frac{\md}{M} \gdm) 
\frac{-i (\sla{Q_-}-\sla{\kbar})+\md}{(Q_--\kbar)^2+\mds} \gadu{\mu}.
\eqa
The remaining tree-level vertices, with $g\stw$ factorized out, are given by
\bq
V^0_{\wb\phi\mu\alpha} = i M \drii{\mu}{\alpha}, \quad
V^0_{\phi\wb\mu\alpha} = -i M \drii{\mu}{\alpha}, \quad
V^0_{\phi\phi\mu} = Q_{+\mu}+Q_{\mu}.
\eq
The $D^1_V$ one-loop contraction of \eqn{defWI} is obtained from $D^1_{\mu}$
of \eqn{defD1} after multiplication by $Q_{-\mu}$ and after replacing the
lowest order triple vertices ($\ph\wbp\wbm$, $\ph\wbp\hkm, \ph\hkp\wbm$
and $\ph\hkp\hkm$) with the one-loop corrected ones. The corresponding 
contractions are given in \eqns{cont1}{cont2}.

\subsection{The massless internal world.}

To prove the Ward identity is particularly simple if we neglect all fermion 
masses in loops. To make the final result more compact we introduce the
following notations:
\bqa
\Gamma^{\pm}_l(\mu) &=& \iap{p_+}\gadu{\mu}\gdpm\oap{q_+},
\nl
\Gamma^{\pm}_q(\mu) &=& \op{k}\gadu{\mu}\gdpm\oap{\kbar},
\nl
\Gamma^{\pm}_l &=& \iap{p_+}\gdpm\oap{q_+},
\nl
\Gamma^{\pm}_q &=& \op{k}\gdpm\oap{\kbar},
\eqa
and also,
\bqa
\Gamma\lpar Q_-,+,1,\pm\rpar &=& \muq\,\Gamma^+_l(Q_-)\,\Gamma^+_q -
\md\,\Gamma^+_l(Q_-)\,\Gamma^-_q,
\nl
\Gamma\lpar 1,+,1,\pm\rpar &=& \muq\,\Gamma^+_l\,\Gamma^+_q -
\md\,\Gamma^+_l\,\Gamma^-_q.
\eqa
For the non-vertex part of the Ward identity, after using the Dirac equation,
we get the following result:
\bq
\frac{1}{g\stw}\,Q^{\mu}_-\,\asums{i=1,4}\,D^i_{\mu} = {\rm WI_{nv}},
\eq
where, with $\mu^2_y = M^2/ys$, the contractions is
\bqa
{\rm WI_{nv}} &=& \frac{i\,g^2}{16}\,\Gamma\lpar Q_-,+,1,\pm\rpar\,
                  {{R_s}\over {x_2-\mu^2_y}}\,{1\over P_s}\,
                  \lpar \frac{1}{ys} + \frac{Y-x_2}{P_t}\rpar,  
\nl
{}&+&   \frac{g^2\me}{16}\,\Gamma\lpar 1,+,1,\pm\rpar\,
        \frac{R_sR_t}{P_sP_t}\,\Big[ \frac{ys}{M^2}\lpar x_2-Y\rpar +
        \frac{1}{2}\,{{x_2+X-Y}\over {x_2-\mu^2_y}} 
\nl
{}&+&
        \frac{1}{2}\,{{x_2-X-Y}\over {Y-\mu^2_y}}\Big],
\nl
{}&+&   \frac{g^2\me}{16\,M^2}\,\Gamma\lpar 1,+,1,\pm\rpar\,
        \frac{R_s}{P_s}\,\Big[ \frac{ys}{P_t}\lpar x_2-Y\rpar - 1\Big],
\nl
{}&+&   \frac{g^2\me}{16\,M^2}\,\Gamma\lpar 1,+,1,\pm\rpar\,
        \frac{R_t}{P_t}\,\Big[ \frac{ys}{P_s}\lpar x_2-Y\rpar + 1\Big],
\nl
{}&+&   \frac{g^2\me}{16\,M^2}\,\Gamma\lpar 1,+,1,\pm\rpar\,
        \Big[ \frac{x_2-Y}{P_sP_t} + \frac{1}{P_t} - \frac{1}{P_s}\Big]
\nl
{}&+&   \frac{i\,g^2\me}{16}\,\Gamma\lpar 1,+,Q_-,+\rpar\,
        {{R_t}\over {Y-\mu^2_y}}\,\frac{1}{P_t}\,
        \Big[ \frac{Y-x_2}{P_s} - 1\Big],
\nl
{}&+&   \frac{g^2}{16}\,\Gamma^+_l(\mu)\,\Gamma^+_q(\mu)\,
        \Big[ \frac{ys}{P_sP_t}\lpar x_2-Y\rpar + \frac{1}{P_t} -
        \frac{1}{P_s}\Big].
\eqa
The propagators are expressed as
\bq
P_s = \Big[ -s'+M^2-S^0_{\ssW}(s')\Big]^{-1}, \qquad
P_t = \Big[ -\kappa_++M^2-S^0_{\ssW}(\kappa_+)\Big]^{-1}.
\eq
Moreover we have ratios of bare to running masses,
\bq
R(p^2) = \frac{M^2}{M^2(p^2)} - 1, \qquad R_s = R(s'), \quad R_t = R(\kappa_+).
\eq
This case is relatively simple and, therefore, we give the derivation in some 
details. The expression that we have obtained is still written in terms of
bare quantities and renormalization can be achieved by using the following
set of relations:
\bqa
S_{\phi}(p^2) &=& 0, \qquad S^0_{\ssW}(p^2) = p^2\,\Big[ 1 - 
\frac{g^2}{g^2(p^2)}\Big],
\nl
M^2 &=& M^2(p^2)\,\frac{g^2}{g^2(p^2)} = \frac{g^2}{g^2(\sW)}\,\sW, \qquad
R(p^2) = \frac{g^2}{g^2(p^2)} - 1,
\nl
p^2 + M^2 &=& \frac{g^2}{g^2(\sW)}\,\sW\,\Big\{\Big[ 1 + 
\frac{g^2(\sW)}{16\,\pi^2}\,\Ptg(\sW)\Big]\,\frac{p^2}{\sW} + 1\Big\}
\label{renorm1}
\eqa
Furthermore, the propagators are rewritten as
\bqa 
P^{-1}_s &=& -\frac{G_s}{g^2}\,\frac{1}{p_s}, \quad
p_s = x_2ys - \sW + \lpar 1 - \frac{G_s}{G_{\ssW}}\rpar\,\sW,
\nl
P^{-1}_t &=& -\frac{G_t}{g^2}\,\frac{1}{p_t}, \quad
p_t = Yys - \sW + \lpar 1 - \frac{G_t}{G_{\ssW}}\rpar\,\sW,
\label{renorm2}
\eqa
where $G_{\ssW} = g^2(\sW)$. We introduce an auxiliary function
\bq
\Phi_{\ssW}(x) = \Big[\lpar 1 + \frac{G_{\ssW}}{16\,\pi^2}\,\Pi_{\ssW}\rpar\,
\frac{xys}{\sW} - 1 \Big]^{-1}, \qquad \Pi_{\ssW} = \Ptg(\sW).
\eq 
In this way, the non-vertex part of the Ward identity will become
\bqa
\frac{p_sp_t}{G_sG_t}\,{\rm WI_{nv}} &=& 
\frac{i}{16}\,\Gamma\lpar Q_-,+,1,\pm\rpar\,\frac{\Pi_s}{16\,\pi^2}  
\nl
{}&\times&
\Big[ 1 - \frac{Yys}{\sW}\,\frac{\Pi_t}{16\,\pi^2}\,G_{\ssW}\,
\Phi_{\ssW}(x_2)\Big]
\nl\nl
{}&+&
\frac{i\me}{16}\,\Gamma\lpar 1,+,1,\pm\rpar\,\Big\{
{{\Pi_s-\Pi_t}\over {16\,\pi^2}} + G_{\ssW}\,{{\Pi_s\Pi_t}\over {256\,\pi^4}}\,
\frac{ys}{2\,\sW} 
\nl
{}&\times&
\Big[ \lpar x_2+X-y\rpar\,\Phi_{\ssW}(x_2) + \lpar x_2-X-Y\rpar\,\Phi_{\ssW}(Y)
\Big]\Big\}
\nl\nl
{}&+&
\frac{i\me}{16}\,\Gamma^+_l\,\Gamma^+_q(Q_-)\,\frac{\Pi_t}{16\,\pi^2}\,
\Big[ \frac{x_2ys}{\sW}\,\frac{\Pi_s}{16\,\pi^2}\,
\Phi_{\ssW}(Y) - 1\Big]
\nl\nl
{}&+&
\Gamma^+_l(\mu)\,\Gamma^+_q(\mu)\,{{x_2\Pi_s- Y\Pi_t}\over {16\,\pi^2}}.
\eqa
As usual, we have introduced subscripts $s,t$ to denote $\Pi_s = \Ptg(s')$
and $\Pi_t = \Ptg(\kappa_+)$.
Also for the vertices we get a contribution to the Ward identity that we 
write as
\bq
\frac{1}{g\stw}\,D^1_V = {\rm WI_v}.
\eq
Two additional quantities are needed, 
\bq
\mu^2_y = \frac{M^2}{ys}, \qquad \Phi(x) = (x - \mu^2_y)^{-1}.
\eq
The function $\Phi$ is related to $\Phi_{\ssW}$ by
\bqa
\Phi(x) &=& \frac{G_{\ssW}}{g^2}\,\frac{ys}{\sW}\,\Phi_{\ssW}(x) =
\lpar \Phi_{\ssW} + 1\rpar\,\frac{\Phi_{\ssW}(x)}{\Phi_{\ssW}},  \nl
\Phi_{\ssW} &\equiv& \Phi_{\ssW}(1).
\eqa
The explicit expression for this part of the Ward identity is
\bq
16\,\pi^2\,\frac{p_sp_t}{G_sG_t}\,{\rm WI_v} = {\rm WI'_v},
\eq
\bqa
{\rm Wi'_v} &=&
\frac{i}{16}\,\Gamma\lpar Q_-,+,1,\pm\rpar 
\Big[ R_s\,\Phi(x_2)\, \lpar \frac{1}{ys}W^{\ssW\ssW}_0 -
x_2\,W^{\ssW\ssW}_1\rpar - W^{\ssW\ssW}_1\Big]
\nl\nl
{}&+&
\frac{\me}{32}\,\Gamma\lpar 1,+,1,\pm\rpar\,\Big\{
R_sR_t\,\Phi(x_2)\,\Phi(Y)
\Big[ \frac{2\,x_2+X}{ys}\,W^{\ssW\ssW}_0 - x_2\lpar X+2\,Y\rpar\,
W^{\ssW\ssW}_2\Big]
\nl\nl
{}&+&
R_sR_t\,\Phi(x_2)
\Big[\frac{1}{ys}\,W^{\ssW\ssW}_0 +\lpar x_2+X+Y\rpar\,W^{\ssW\ssW}_1 -
\lpar X+Y\rpar\,W^{\ssW\ssW}_2\Big]
\nl\nl
{}&+&
R_sR_t\,\Phi(Y)
\Big[ - \frac{1}{ys}\,W^{\ssW\ssW}_0 - \lpar x_2+X+Y\rpar\,W^{\ssW\ssW}_1 +
\lpar X+Y\rpar\,W^{\ssW\ssW}_2\Big]
\nl\nl
{}&+&
R_s\,\Phi(x_2)
\Big[ \frac{2}{ys}\,W^{\ssW\ssW}_0 +\lpar -x_2+X+Y\rpar\,W^{\ssW\ssW}_1 -
\lpar x_2+X+Y\rpar\,W^{\ssW\ssW}_2\Big]
\nl\nl
{}&+&
R_t\,\Phi(Y)
\Big[\frac{2}{ys}\,W^{\ssW\ssW}_0 +\lpar -x_2-X+Y\rpar\,W^{\ssW\ssW}_1 -
2\,Y\,W^{\ssW\ssW}_2\Big] - 2\,W^{\ssW\ssW}_2\Big\}
\nl\nl
{}&+&
\frac{i\me}{16}\,\Gamma^+_l\,\Gamma^+_q(Q_-)\,\Big\{R_t\,\Phi(Y)
\Big[\frac{1}{ys}\,W^{\ssW\ssW}_0 + Y\,W^{\ssW\ssW}_1 - Y\,W^{\ssW\ssW}_2\Big]
\nl\nl
{}&+&
W^{\ssW\ssW}_1 - W^{\ssW\ssW}_2 \Big\}
- \frac{1}{16}\,\Gamma^+_l(\mu)\,\Gamma^+_q(\mu)\,W^{\ssW\ssW}_0
\eqa
To go one step further, to this expression we apply the renormalization 
procedure of \eqns{renorm1}{renorm2}. By inspection, its is seen that the Ward 
identity
\bq
{\rm WI_{nv}} + {\rm WI_v} = 0,
\label{wieq}
\eq
is satisfied if the following conditions hold:
\bqa
W^{\ssW\ssW}_0 &=& \lpar x_2\,\Pi_s - Y\,\Pi_t\rpar\,ys,  \nl
W^{\ssW\ssW}_1 &=& \Pi_s,  \nl
W^{\ssW\ssW}_2 &=& \Pi_s - \Pi_t.
\eqa
An explicit calculation shows that
\bqa
\Pi_s &=& \frac{4}{9} - \frac{4}{3}\,\bff{0}{-s'}{0}{0},  \nl
\Pi_t &=& \frac{4}{9} - \frac{4}{3}\,\bff{0}{-\kappa_+}{0}{0},  \nl
W^{\ssW\ssW}_0 &=& \Big\{ \frac{4}{9}\,\lpar x_2-Y\rpar + \frac{4}{3}\,
\Big[ Y\,\bff{0}{-\kappa_+}{0}{0} - x_2\,\bff{0}{-s'}{0}{0}\Big]\Big\},
\nl
W^{\ssW\ssW}_1 &=& \frac{4}{9} - \frac{4}{3}\,\bff{0}{-s'}{0}{0},  \nl
W^{\ssW\ssW}_2 &=& \frac{4}{3}\,\Big[ \bff{0}{-\kappa_+}{0}{0} -
\bff{0}{-s'}{0}{0}\Big],
\eqa
which represents the solution of \eqn{wieq}.

\subsection{The case of non-zero $\mt$.}

When the internal world is not massless the derivation of the Ward identity,
although straightforward, is considerably lengthy. First, we have to change
the normalization equations, \eqns{renorm1}{renorm2}. Now they can be written 
as
\bqa 
P^{-1}_s &=& -\frac{G_s}{g^2}\,\frac{1}{p_s}, \quad
p_s = x_2ys - \sW + \lpar 1 - \frac{G_s}{G_{\ssW}}\rpar\,\sW + G_s\,\big[
f_{\ssW}(s') - f_{\ssW}(\sW)\big],
\nl
P^{-1}_t &=& -\frac{G_t}{g^2}\,\frac{1}{p_t}, \quad
p_t = Yys - \sW + \lpar 1 - \frac{G_t}{G_{\ssW}}\rpar\,\sW + G_t\,\big[
f_{\ssW}(\kappa_+) - f_{\ssW}(\sW)\big],
\nl
\label{renormM}
\eqa
Furthermore, the relations between bare and running quantities is now
modified into the following form:
\bqa
{\sW\over {M^2(p^2)}} &=& \frac{g^2(\sW)}{g^2(p^2)}\,\eta(p^2),
\nl
\eta(p^2) &=& \Big\{ 
1 - \frac{g^2(p^2)}{16\,\pi^2}\,\Big[ \sigma_{\ssW}(p^2) - \frac{\mts}{p^2}\,
\sigma_{\phi}(p^2)\Big]\Big\}  \nl
{}&\times& 
\Big\{1 - \frac{g^2(\sW)}{16\,\pi^2}\,\Big[ \sigma_{\ssW}(\sW) + 
\frac{\mts}{\sW}\,\sigma_{\phi}(p^2)\Big]\Big\}^{-1},
\nl\nl
M^2 &=& M^2(p^2)\,\frac{g^2}{g^2(p^2)}\,\eta(p^2) + \frac{g^2}{16\,\pi^2}\,
f_{\ssW}(\sW),
\nl
R(p^2) &=& \frac{g^2}{g^2(p^2)}\,\eta(p^2)\,\Big[
1 - \frac{g^2(\sW)}{16\,\pi^2}\,\sigma_{\ssW}(\sW)\Big] - 1.
\eqa
Moreover, all vertices contribute. We have created a FORM program that computes
the massive-massive Ward identity~\footnote{available from 
http://www.to.infn.it/$\tilde{}$giampier/form/mflwi.f} and have found that the latter 
is satisfied if
\bqa
W^{\ssW\ssW}_0 &=& x_2ys\,\lpar \Pi_s + \sigma_{\ssW}^s \rpar - 
               Yys  \,\lpar \Pi_t + \sigma_{\ssW}^t \rpar,
\nl
W^{\ssW\ssW}_1 &=& \Pi_s + \sigma^{\wb}_s + \frac{\mts}{x_2ys}\,\sigma^{\phi}_s,
\nl
W^{\ssW\ssW}_2 &=& \Pi_s - \Pi_t + \sigma_{\ssW}^s - \sigma_{\ssW}^t +
               \frac{\mts}{x_2ys}\,\sigma_{\phi}^s -
               \frac{\mts}{Yys}\,\sigma_{\phi}^t,
\nl
W^{\ssW\phi}_1 &=&  \mts\,\lpar  \sigma_{\phi}^t - \sigma_{\phi}^s \rpar,
\nl
W^{\ssW\phi}_2 &=&  - \mts\,\sigma_{\phi}^s,
\nl
W^{\phi\phi} &=& \mts\,\lpar x_2 \sigma_{\phi}^s - Y \sigma_{\phi}^t\rpar\,ys.
\label{wiMM}
\eqa
We have also written a FORM program for the evaluation of 
vertices~\footnote{available from 
http://www.to.infn.it/$\tilde{}$giampier/form/vertsw.f} derived from a more 
general one that computes all one-loop diagrams in the standard model. From it 
we derive
\bq
W^{\phi\ssW}_i = - W^{\ssW\phi}_i,
\eq
and, with $N_g$ generations of fermions, the result is as follows:
\bqa
W^{\ssW\ssW}_0 &=& N_g y s \Big\{ \frac{4}{9} (x_2-Y) + 
        \frac{4}{3} \Big[Y \bff{0}{-\kappa_+}{0}{0} -
        x_2 \bff{0}{-s'}{0}{0}\Big]\Big\}
\nl
{}&+&
        \frac{1}{3} y s  ( x_2 - Y )
       -\frac{1}{2} \frac{\mtq}{Yys} \bff{0}{-\kappa_+}{0}{\mt} 
       + Yys \bff{0}{-\kappa_+}{0}{\mt}    
\nl
{}&-&
       \frac{1}{2} \mts \bff{0}{-\kappa_+}{0}{\mt}   
       + \frac{1}{2} \frac{\mtq}{x_2ys}   \bff{0}{-s'}{0}{\mt} 
\nl
{}&-&
         x_2ys \bff{0}{-s'}{0}{\mt}    
       + \frac{1}{2} \mts \bff{0}{-s'}{0}{\mt} 
       + \frac{1}{2}( \frac{1}{x_2} - \frac{1}{Y} ) 
       \frac{\mts}{ys} \aff{0}{\mt}
\nl
{}&-&
        \frac{1}{3} y s   ( x_2 - Y )
       - Yys \bff{0}{-\kappa_+}{0}{0} 
       + x_2ys \bff{0}{-s'}{0}{0},
\nl\nl
W^{\ssW\ssW}_1 &=& N_g \Big[ \frac{4}{9} - \frac{4}{3} \bff{0}{-s'}{0}{0}\Big]
       - \frac{\mts}{x_2ys}   
       + 2 \lpar\frac{\mts}{x_2ys}\rpar^2   \bff{0}{-s'}{0}{\mt}
\nl
{}&-& 
         \bff{0}{-s'}{0}{\mt} \frac{\mts}{x_2ys}  
       - \bff{0}{-s'}{0}{\mt}
       + 2 \frac{\mts}{\lpar x_2ys\rpar^2} \aff{0}{\mt}  
\nl
{}&-&
         \frac{1}{x_2ys} \aff{0}{\mt}
       + \bff{0}{-s'}{0}{0},
\nl\nl
W^{\ssW\ssW}_2 &=&  \frac{4}{3} N_g \Big[ \bff{0}{-\kappa_+}{0}{0} -
          \bff{0}{-s'}{0}{0}\Big]
\nl
{}&+&
         \frac{\mts}{ys}   (  - \frac{1}{x_2} + \frac{1}{Y} )
       - 2  \lpar\frac{\mts}{Yys}\rpar^2 \bff{0}{-\kappa_+}{0}{\mt}  
\nl
{}&+&
         \frac{\mts}{Yys} \bff{0}{-\kappa_+}{0}{\mt}  
       +  \bff{0}{-\kappa_+}{0}{\mt}
       + 2  \lpar\frac{\mts}{x_2ys}\rpar^2 \bff{0}{-s'}{0}{\mt}  
\nl
{}&-&
         \frac{\mts}{x_2ys} \bff{0}{-s'}{0}{\mt}  
       - \bff{0}{-s'}{0}{\mt} 
       +  2 \frac{\mts}{\lpar ys\rpar^2}   
       ( \frac{1}{x_2^2} - \frac{1}{Y^2} ) \bff{0}{-s'}{0}{\mt}
\nl
{}&+&
        (  - \frac{1}{x_2ys} + \frac{1}{Yys} ) \aff{0}{\mt} 
       - \bff{0}{-\kappa_+}{0}{0}
       + \bff{0}{-s'}{0}{0},
\nl\nl 
W^{\ssW\phi}_1 &=& \frac{3}{2}  \frac{\mtq}{Yys}  \bff{0}{-\kappa_+}{0}{\mt}
       - \frac{3}{2} \mts \bff{0}{-\kappa_+}{0}{\mt} 
\nl
{}&-&
         \frac{3}{2}  \frac{\mtq}{x_2ys}   \bff{0}{-s'}{0}{\mt}
       + \bff{0}{-s'}{0}{\mt} \mts   ( \frac{3}{2} )
       - \frac{3}{2} \frac{\mts}{ys}   
       (   \frac{1}{x_2} - \frac{1}{Y} ) \aff{0}{\mt},
\nl\nl
W^{\ssW\phi}_2 &=& - \frac{3}{2}  \frac{\mtq}{x_2ys}   \bff{0}{-s'}{0}{\mt}
\nl
{}&+&
         \frac{3}{2} \mts \bff{0}{-s'}{0}{\mt} 
       -  \frac{3}{2} \frac{\mts}{x_2ys}   \aff{0}{\mt},
\nl\nl 
W^{\phi\phi} &=& \frac{3}{2} Yys\,\mts \bff{0}{-\kappa_+}{0}{\mt} 
       - \frac{3}{2} \mtq \bff{0}{-\kappa_+}{0}{\mt} 
\nl
{}&-&
         \frac{3}{2} x_2ys\,\mts \bff{0}{-s'}{0}{\mt} 
       + \frac{3}{2} \mtq \bff{0}{-s'}{0}{\mt} 
\label{sol1}
\eqa
The remaining quantities have already been given, but we repeat them, for
completeness, in their scalarized form:
\bqa
f_{\ssW}(s') &=& -\frac{1}{3} \mts
       -\frac{4}{3} \bff{0}{-s'}{\mt}{\mt} \mts
\nl
{}&-&
       \frac{2}{3} x_2 y s \bff{0}{-s'}{\mt}{\mt}
       -\frac{1}{2} \bff{0}{-s'}{0}{\mt} \mts
       -\frac{1}{2} \frac{\mtq}{x_2 y s} \bff{0}{-s'}{0}{\mt} 
\nl
{}&+&
       x_2 y s \bff{0}{-s'}{0}{\mt}
       -\frac{1}{3} x_2 y s \bff{0}{-s'}{0}{0}
       -\frac{1}{2} \frac{\mts}{x_2 y s} \aff{0}{\mt}  
       -\frac{1}{3} \aff{0}{\mt},
\nl\nl
f_{\ssW}(\kappa_+) &=& -\frac{1}{3} \mts
       -\frac{4}{3} \bff{0}{-\kappa_+}{\mt}{\mt} \mts
\nl
{}&-&
       \frac{2}{3} Y y s \bff{0}{-\kappa_+}{\mt}{\mt}
       -\frac{1}{2} \bff{0}{-\kappa_+}{0}{\mt} \mts
       -\frac{1}{2} \frac{\mtq}{Y y s} \bff{0}{-\kappa_+}{0}{\mt} 
\nl
{}&+&
       Y y s \bff{0}{-\kappa_+}{0}{\mt}
       -\frac{1}{3} Y y s \bff{0}{-\kappa_+}{0}{0}
       -\frac{1}{2} \frac{\mts}{Y y s} \aff{0}{\mt}  
       -\frac{1}{3} \aff{0}{\mt}.
\nl
\label{sol2}
\eqa
Furthermore, we have
\bqa
\sigma_{\phi}^s &=& \frac{3}{2} \frac{1}{x_2ys}\,
\Big[\aff{0}{\mt} + \lpar \mts - x_2ys\rpar \bff{0}{-s'}{0}{\mt}\Big],
\nl
\sigma_{\phi}^t &=& \frac{3}{2} \frac{1}{Yys}\,
\Big[\aff{0}{\mt} + \lpar \mts - Yys\rpar \bff{0}{-\kappa_+}{0}{\mt}\Big].
\label{sol3}
\eqa
The explicit expressions shown in \eqns{sol1}{sol3} are solutions of 
\eqn{wiMM}, therefore proving the validity of the U(1) massive-massive
Ward identity. The proof is straightforward, although the FORM program
that derives WI = 0 generates of the order of $100,000$ in some intermediate 
step.

\subsection{Fixed-width scheme}

In the previous section we have shown that the massive-massive Ward identity
is satisfied in the Fermion-Loop scheme when vertices are added. In many
numerical evaluations, where one would like to save as much as possible of CPU 
time, the implementation of the Fermion-Loop is very time-consuming. 
Therefore, one would like to have a gauge-preserving scheme which is as
simple as possible from the point of view of building an event generator.
For conserved currents this scheme is given by the use of a fixed width for
vector bosons in all channel, not only annihilation but also scattering.
It is a nonsense, from the point of view of a full-fledged field-theoretical
formulation, but it has been shown~\cite{kn:bhf2} that the numerical 
differences, with respect to the Fermion-Loop scheme, are tiny at LEP~2 
energies and also for small, but not vanishing, electron's scattering angle.

We have taken the complete, massive-massive, Ward identity and have neglected
all vertex corrections. In this case one can show that the Ward identity is
satisfied with the following choice:
\bqa
P^{-1}_s &=& - x_2ys + M^2 - i\,M\Gamma_{\ssW},  \nl
P^{-1}_t &=& - Yys + M^2 - i\,M\Gamma_{\ssW},  \nl
M^2(p^2) &=& M^2, \quad {\mbox for} \quad p^2 = - s' \quad \mbox{and} \quad
p^2 = - \kappa_+,
\eqa
where $\Gamma_{\ssW}$ is the $\wb$, on-shell, total width.
Therefore, we have a formulation of the fixed-width scheme that works also for
non-zero external fermion masses.

\section{Approximations.}

A full implementation of the Fermion-Loop scheme can be achieved without
having to rely on any sort of approximation. Of course, one has to compute
all form-factors occurring in the vertices and this is, notoriously,
a lengthy procedure.

One of the beautiful aspects of computing the Ward identity is that, in the
final answer for vertices, all Gram determinants disappear and a full
scalarization of the answer has been successfully achieved. This is not
the case when we need the vertices for the amplitude, since all powers
of Gram determinants, up to the third one, will appear.

For this reason we start investigating into the numerical impact of mass
corrections. Having proved the relevant Ward identity for the fully
massive case, we have been able to show that masses do not spoil any 
cancellation mechanism. In other words, the $1+5+4$ diagrams in the
$t$-channel preserve gauge invariance in the Fermion-Loop scheme with all
masses kept explicitly. Therefore, inside this set of diagrams we are allowed
to investigate the numerical relevance of masses and to neglect some,
if convenient.

The dominant effect is due to the collinear region where the mass in the
electron line acts as a regulator,
\bq
\frac{1}{4}\,L_{\mu\nu}W^{\mu\nu} = 
W_1\,\Big[ 2\,\frac{\mes}{\lpar Xys\rpar^2} - \frac{1}{Xys}\Big] +
2\,W_2\,\frac{y-1}{Xy^3s} + \ord{X}.
\eq
It is clear that terms proportional to $\me$ can be safely neglected inside
$W_{1,2}$ and, therefore, we can take the incoming positron as massless
without effecting the numerical precision of the result. The quark masses
are only relevant, from a numerical point of view, for the multi-peripheral
diagrams. In the amplitude we must keep all terms proportional to
\bq
{1\over {\lpar Q_- - k\rpar^2 + \mus}}\, \qquad
{\mus\over {\Big[\lpar Q_- - k\rpar^2 + \mus\big]^2}},
\eq
and similar ones for the down quark. Once we neglect $\me$, apart from the
photon flux-function, the only terms proportional to the quark masses
arise from the internal propagators of the $u,d$-quarks and not from
couplings. Propagators in the multi-peripheral diagrams are, therefore,
the only place where we would like to keep quark masses. But one
can easily see that, in the complete Ward identity, all terms proportional
to the quark propagators drop out, therefore, we can write the Ward identity
for $\me = \muq = \md = 0$. In this case we derive
\bq
\Gamma^P_l(\mu)\,\Gamma^P_q(\mu)\,\Big[ x_2\,\frac{\Pi_s}{16\,\pi^2} -
Y\,\frac{\Pi_t}{16\,\pi^2}\Big],
\eq
from transitions, and
\bq
- \Gamma^P_l(\mu)\,\Gamma^P_q(\mu)\,{{W^{\ssW\ssW}_0}\over {16\,\pi^2}},
\eq
from vertices, that indeed add up to zero.

To summarize, the Fermion-Loop scheme with all external fermion masses set to 
zero, but the electron mass in the photon flux-function and the quark masses in
the quark propagators of the multi-peripheral diagrams, preserves gauge 
invariance.

From a numerical point of view, this is all we need in evaluating the 
single-$\wb$ process. Using the Fermion-Loop, as a gauge preserving scheme, 
has, moreover, the bonus of automatically setting the correct scale for all
running coupling constants.

\section{Cancellation of the ultraviolet divergences.}

When we assume that the $\wb$-currents are treated as conserved ones, then
ten operators are enough to describe the $\ph\wb\wb$ vertex, even for
proving the U(1) Ward identity. We will use
\bqa
W^1_{\mu\alpha\beta} &=& 4\,{\cal N}\,\Big[
\drii{\alpha}{\beta}\,Q_{+\mu}+\drii{\mu}{\beta}\,Q_{-\alpha} -
\drii{\mu}{\alpha}\,Q_{\beta}\Big],  \nl
W^{2,3}_{\mu\alpha\beta} &=& 2\,{\cal N}\,\Big[
\drii{\mu}{\beta}\,Q_{-\alpha} \mp \drii{\mu}{\alpha}\,Q_{\beta}\Big],  \nl
W^4_{\mu\alpha\beta} &=& 2/,{\cal N}^3\,Q_{+\mu}\,Q_{-\alpha}\,Q_{\beta}, \nl
W^5_{\mu\alpha\beta} &=& {\cal N}\,\Big[ \varepsilon(Q_+,\mu,\alpha,\beta)-
\varepsilon(Q_-,\mu,\alpha,\beta)\Big],  \nl
W^6_{\mu\alpha\beta} &=& -{\cal N}\,\Big[ \varepsilon(Q_+,\mu,\alpha,\beta)+
\varepsilon(Q_-,\mu,\alpha,\beta)\Big],  \nl
W^7_{\mu\alpha\beta} &=& -{\cal N}^3\,\varepsilon(Q_+,Q-,\alpha,\beta)\,
Q_{\mu},  \nl
W^8_{\mu\alpha\beta} &=& 2\,{\cal N}^3\,Q_{-\mu}\,Q_{-\alpha}\,Q_{\beta},  \nl
W^9_{\mu\alpha\beta} &=& 2\,{\cal N}\,\drii{\alpha}{\beta}\,Q_{-\mu},  \nl
W^{10}_{\mu\alpha\beta} &=& {\cal N}^3\,\varepsilon(Q_+,Q_-,\alpha,\beta)\,
Q_{+\mu}.
\eqa
We will write the vertex as
\bq
W^{\ph\ssW\ssW}_{\mu\alpha\beta} = \frac{g^3\stw}{16\,\pi^2}\,
(Xys)^{1/2}\,\asums{i=1,10}\,I^{\ph\ssW\ssW}_i\,W^i_{\mu\alpha\beta}.
\eq
For one generation, $l,\nu_l,u,d$, of massless fermions we obtain
\bqa
   I^1 &=&
      x_2\,y\,s \, \lpar  -C_{32}-C_{33}+2\,C_{34} \rpar
\nl
{}&+&
      X\,y\,s \, \lpar  C_{12}+C_{22}-C_{33}+C_{34} \rpar
      +Y\,y\,s \, \lpar  C_{21}+C_{22}-2\,C_{23}+C_{31}\right.
\nl
{}&-&\left.
     2\,C_{33}+C_{34} \rpar
     - \frac{2}{3}+2\,C_{24}-2\,C_{35}+2\,C_{36},
\nl\nl
   I^2 &=&
      x_2\,y\,s \, \lpar  -2\,C_{22}+2\,C_{23}+3\,C_{32}+3\,C_{33}-6\,C_{34} \rpar
\nl
{}&+&
      X\,y\,s \, \lpar  -C_{12}-3\,C_{22}+2\,C_{23}+3\,C_{33}-3\,C_{34} \rpar
\nl
{}&+&
      Y\,y\,s \, \lpar  -2\,C_{11}+2\,C_{12}-5\,C_{21}-3\,C_{22}+8\,C_{23} \rpar
\nl
{}&-&
      3\,C_{31}+6\,C_{33}-3\,C_{34} 
      +2\,C_{24}+6\,C_{35}-6\,C_{36},
\nl\nl
   I^3 &=&
      x_2\,y\,s \, \lpar  -2\,C_{22}+2\,C_{23}-C_{32}+C_{33} \rpar
\nl
{}&+&
      X\,y\,s \, \lpar  C_{12}+C_{22}+2\,C_{23}+C_{33}+C_{34} \rpar
      +Y\,y\,s \, \lpar  -C_{21}+C_{22}\right.
\nl{}&-&\left.
         C_{31}+C_{34} \rpar + 2\,C_{24}+2\,C_{35}+2\,C_{36},
\nl\nl
   I^4 &=&
      8\,X\,y\,s \, \lpar  -\,C_{22}+\,C_{23}+\,C_{33}-\,C_{34} \rpar,
\nl\nl
   I^{5,6,7} &=& 0, 
\nl\nl
   I^8 &=&
      X\,y\,s \, \lpar  -4\,C_{12}-8\,C_{22}-4\,C_{23}-8\,C_{34} \rpar,
\nl\nl
   I^9 &=&
      x_2\,y\,s \, \lpar  2\,C_{12}+2\,C_{23}-2\,C_{32}+2\,C_{34} \rpar
\nl
{}&+&
      X\,y\,s \, \lpar  2\,C_{12}+2\,C_{22}+2\,C_{23}+2\,C_{34} \rpar
\nl
{}&+&
      Y\,y\,s \, \lpar  -2\,C_{11}-2\,C_{21}+2\,C_{22}-2\,C_{23}-2\,C_{33}
      + 2\,C_{34} \rpar
\nl
{}&-&
      \frac{2}{3}+4\,C_{24}+4\,C_{36}
\nl\nl
   I^{10} &=& 0,
\eqa
where all $C_{ij}$-functions are computed with zero internal masses.
If we compute the ultraviolet divergent part of the vertex, for one massless
generation, then the following results will emerge:
\bq
W^{\ph\ssW\ssW}_{\mu\alpha\beta}\mid_{_{\rm UV}} = \frac{g^3\stw}{16\,\pi^2}\,
(Xys)^{1/2}\,\frac{2}{3}\,\Ddr\,\Big[W^1_{\mu\alpha\beta} +
W^9_{\mu\alpha\beta}\Big],
\eq
but, for conserved currents, i.e. $Q_{+\beta} = 0, Q_{+\alpha} = - Q_{-\alpha}$,
we get
\bq
W^1_{\mu\alpha\beta} + W^9_{\mu\alpha\beta} = 2\,{\cal N}\,
V^0_{\mu\alpha\beta}.
\eq
Therefore, the ultraviolet divergent part of the vertex is proportional to
the lowest order. This has an important consequence, in view of the fact
that inside \eqn{g2fact} there is a remaining factor $1/g^2$.
The combination of bare quantities and of one-loop corrections,
\bq
\frac{1}{g^2} + \frac{1}{8\,\pi^2}\,\lpar I_1 + I_9\rpar\mid_{_{\rm UV}},
\label{overren}
\eq
appearing in the total amplitude, is ultraviolet finite.

For the $t-b$ doublet result we need two different configurations of
internal masses, $(\mt,0,\mt)$ and $(0,\mt,0)$. With
\bq
\tcft{ij} = C_{ij}\lpar 0,\mt,0\rpar\,\qquad
\tcftt{ij} = C_{ij}\lpar \mt,0,\mt\rpar,
\eq
and
\bq
\tcfp{ij} = \tcftt{ij} + \frac{1}{2}\,\tcft{ij}, \qquad
\tcfm{ij} = \tcftt{ij} - \frac{1}{2}\,\tcft{ij},
\eq
the resulting expressions for the coefficients are as follows:
\bqa
   I_1 &=&
      x_2 y s   \,\Big[ -\frac{1}{2} \tcfp{32}-\frac{1}{2} \tcfp{33}+
      \tcfp{34} \Big]
\nl{}&+&
      \frac{1}{2} X y s   \,\Big[ \tcfp{12}+\tcfp{22}-\tcfp{33}+\tcfp{34} \Big]
\nl{}&+&
      Y y s   \,\Big[ \frac{1}{2} \tcfp{21}+\frac{1}{2} \tcfp{22}-\tcfp{23}+
      \frac{1}{2} \tcfp{31}-\tcfp{33}+\frac{1}{2} \tcfp{34} \Big]
\nl{}&+&
      \frac{1}{2} \mts   \,\Big[ \tcftt{11}-\tcftt{12} \Big]
      +\frac{1}{2} \mts \tcftt{0} 
\nl{}&-&
      \frac{1}{2}+\tcfp{24}-\tcfp{35}+\tcfp{36},
\nl\nl
   I_2 &=&
      x_2 y s   \,\Big[ -\tcfp{22}+\tcfp{23}+\frac{3}{2} \tcfp{32}+\frac{3}{2} 
      \tcfp{33}-3 \tcfp{34} \Big]
\nl{}&+&
      X y s   \,\Big[ -\frac{1}{2} \tcfp{12}-\frac{3}{2} \tcfp{22}+\tcfp{23}
         +\frac{3}{2} \tcfp{33}-\frac{3}{2} \tcfp{34} \Big]
\nl{}&+&
      Y y s   \,\Big[ -\tcfp{11}+\tcfp{12}-\frac{5}{2} \tcfp{21}-\frac{3}{2} 
       \tcfp{22}
\nl{}&+&
     4 \tcfp{23} -\frac{3}{2} \tcfp{31}+3 \tcfp{33}-\frac{3}{2} \tcfp{34} \Big]
\nl{}&-&
      \frac{1}{2} \mts   \,\Big[  \tcftt{11}-\tcftt{12} \Big]
       -\frac{1}{2} \mts \tcftt{0} 
      +\tcfp{24}+3 \tcfp{35}-3 \tcfp{36}
\nl\nl
   I_3 &=&
      x_2 y s   \,\Big[ -\tcfp{22}+\tcfp{23}-\frac{1}{2} \tcfp{32}+
      \frac{1}{2} \tcfp{33} \Big]
\nl{}&+&
      X y s   \,\Big[ \frac{1}{2} \tcfp{12}+\frac{1}{2} \tcfp{22}+
       \tcfp{23}+\frac{1}{2} \tcfp{33}+\frac{1}{2} \tcfp{34} \Big]
\nl{}&+&
      \frac{1}{2} Y y s   \,\Big[ -\tcfp{21}+\tcfp{22}-\tcfp{31}+\tcfp{34} \Big]
      +\frac{1}{2} \mts   \,\Big[ \tcftt{11}+\tcftt{12} \Big]
\nl{}&+&
      \frac{1}{2} \mts \tcftt{0} 
      +\tcfp{24}+\tcfp{35}+\tcfp{36}
\nl\nl
   I_4 &=&
      4 X y s   \,\Big[ -\tcfp{22}+\tcfp{23}+\tcfp{33}-\tcfp{34} \Big]
\nl\nl
   I_5 &=&
      x_2 y s   \,\Big[ -\tcfm{32}+\tcfm{33} \Big]
      +X y s   \,\Big[ \tcfm{12}+\tcfm{22}+2 \tcfm{23}+\tcfm{33}+\tcfm{34} \Big]
\nl{}&+&
      Y y s   \,\Big[ -2 \tcfm{11}+2 \tcfm{12}-3 \tcfm{21}+\tcfm{22}+2 \tcfm{23} 
      -\tcfm{31}+\tcfm{34} \Big]
\nl{}&+&
      \mts   \,\Big[ \tcftt{11}+\tcftt{12} \Big]
      +\mts \tcftt{0} 
      +6 \tcfm{24}+6 \tcfm{35}+6 \tcfm{36}
\nl\nl
   I_6 &=&
      x_2 y s   \,\Big[ -\tcfm{32}-\tcfm{33}+2 \tcfm{34} \Big]
\nl{}&+&
      X y s   \,\Big[ -\tcfm{12}+\tcfm{22}-2 \tcfm{23}-\tcfm{33}+\tcfm{34} \Big]
\nl{}&+&
      Y y s   \,\Big[ \tcfm{21}+\tcfm{22}-2 \tcfm{23}+\tcfm{31}-2 \tcfm{33}+
      \tcfm{34} \Big]
\nl{}&-&
      \mts   \,\Big[  \tcftt{11}-\tcftt{12} \Big]
      -\mts \tcftt{0} 
\nl{}&-&
      \frac{1}{3}-2 \tcfm{24}-6 \tcfm{35}+6 \tcfm{36}
\nl\nl
   I_7 &=&
      4 X y s   \,\Big[ \tcfm{12}+\tcfm{23} \Big]
\nl\nl
   I_8 &=&
      2 X y s   \,\Big[ -\tcfp{12}-2 \tcfp{22}-\tcfp{23}-2 \tcfp{34} \Big]
\nl\nl
   I_9 &=&
      x_2 y s   \,\Big[ \tcfp{12}+\tcfp{23}-\tcfp{32}+\tcfp{34} \Big]
\nl{}&+&
      X y s   \,\Big[ \tcfp{12}+\tcfp{22}+\tcfp{23}+\tcfp{34} \Big]
\nl{}&+&
      Y y s   \,\Big[ -\tcfp{11}-\tcfp{21}+\tcfp{22}-\tcfp{23}-\tcfp{33}+
         \tcfp{34} \Big]
\nl{}&-&
       \mts   \tcftt{12} 
      -\frac{1}{2}+2 \tcfp{24}+2 \tcfp{36}
\nl\nl
   I_{10} &=&
      4 X y s   \,\Big[ \tcfm{12}+\tcfm{23} \Big]
\label{tbcoeff}
\eqa
When computing the $t-b$ contribution to the vertex we find that the 
ultraviolet divergence is $\mt$-independent. One should remember that the
whole calculation is organized in the following way: if we denote the vertex 
schematically by $\Vvert$, then
\bq
\Vvert = \asums{\ff} V_{\ff}({\hbox{massless}}) + 
\Bigl[\Vvert_{\ft}({\hbox{massive}})-\Vvert_{\ft}({\hbox{massless}})\Bigr].
\label{hyperc}
\eq
The sum in \eqn{hyperc} combines with the $1/g^2$-term to cancel the
ultraviolet divergent term while the rest is a subtracted term that is
ultraviolet finite. 
When we consider the massive-massive case, there are $24$ form-factors.
Computing the divergent part we find
\bqa
W^{\ph\ssW\ssW}_{\mu\alpha\beta}\mid_{_{\rm UV}} &=& \frac{g^3\stw}{16\,\pi^2}\,
(Xys)^{1/2}\,\frac{1}{3}\,\Ddr\,\Big[W^1_{\mu\alpha\beta} -
2\,W^3_{\mu\alpha\beta}  \nl
{}&+& 2\,W^5_{\mu\alpha\beta} -
4\,W^6_{\mu\alpha\beta} - 2\,W^7_{\mu\alpha\beta} -
2\,W^8_{\mu\alpha\beta}\Big],
\eqa
where the operators are given in \eqn{operators}. However, we easily derive the
following relation
\bqa
4\,{\cal N}\,V^0_{\mu\alpha\beta} &=&
W^1_{\mu\alpha\beta} - 2\,W^3_{\mu\alpha\beta} 
+ 2\,W^5_{\mu\alpha\beta}  \nl
{}&-& 4\,W^6_{\mu\alpha\beta} - 2\,W^7_{\mu\alpha\beta} - 
2\,W^8_{\mu\alpha\beta},
\eqa
showing, again, factorization of the divergence into the lowest order.
For a massive internal world there are also $\ph\wb\phi$ and $\ph\phi\phi$
vertices. One finds again factorization of ultraviolet divergences, for instance
\bqa
W^{\ph\ssW\phi}_{\mu\beta} &=& i\,g^3\stw\,
{\cal W}^{\ph\ssW\phi}_{\mu\beta},  \nl
W^{\ph\ssW\phi}_{\mu\beta}\mid_{_{\rm UV}} &=& \frac{i\,g^3\stw}{16\,\pi^2}\,
\frac{\mts}{M}\,\frac{3}{2}\,\Ddr\,\drii{\mu\beta}.
\eqa
The corresponding lowest order vertex gives $i\,g\stw M\drii{\mu}{\beta}$. The
two combine into the following expression:
\bq
C_{\mu\beta} = \frac{M^2}{g^2}\,\drii{\mu}{\beta} + 
M\,{\cal W}^{\ph\ssW\phi}_{\mu\beta},
\eq
and 
\bqa
C_{\mu\beta}\mid_{_{\rm UV}} &=& \sW\,\Big[
\frac{1}{G_{\ssW}} + \frac{1}{16\,\pi^2}\,\frac{f_{\ssW}(\sW)}{\sW}
\Big]_{_{\rm UV}} + \frac{1}{16\,\pi^2}\,\mts\,\frac{3}{2}\,\Ddr  \nl
{}&=& \frac{1}{16\,\pi^2}\,\Big[ f_{\ssW}(\sW)_{_{\rm UV}} + 
\frac{3}{2}\,\mts\,\Ddr\Big] = 0.
\eqa
Similar results hold for the other combinations, $\ph\phi\wb$ and
$\ph\phi\phi$.

These observations, completing the argument shown in \eqn{overren}, prove that 
the amplitude is ultraviolet finite in the Fermion-Loop scheme.

As we have shown, the $\varepsilon$-terms do not cancel in \eqn{tbcoeff}.
They only do for a complete massless generation, $l\,\nu_l\,u\,d$.
Indeed, it is very easy to show that the $\varepsilon$-terms are proportional 
to the hyper-charge so that in the total they cancel, but they cancel only if
we add the full content of the three fermionic generations while keeping all 
the fermions (including the top quark) at zero mass. With a massive top quark
they do not cancel and give an $\mt$-dependent contribution. 

\section{Re-summations in the neutral sector.}

Dyson re-summation of transitions in the neutral sector are also available and
we will proceed to their construction.
First, we define indices $i,j,\dots =1,2$ with $1 \equiv \zb$ and
$2 \equiv \ph$. The transitions are given by
\bq
S = \frac{g^2}{16\,\pi^2}\,\Sigma, \qquad \Sigma^{ij}_{\mu\nu} =
\Sigma^{ij}_0\,\drii{\mu}{\nu} + \Sigma^{ij}_1\,p_{\mu}p_{\nu}.
\eq
The re-summed propagators are obtained in the usual way, namely
\bqa
\bDelta^{ij}_{\mu\nu} &=& \Delta^{ij}\,\drii{\mu\nu} + 
\Delta^{il}\,S^{lk}_{\mu\nu}\,\Delta^{kj} + \dots 
\nl
{}&=& \Delta^{il}\,\drii{\mu}{\alpha}\,X^{lj}_{\alpha\nu}, \quad
X = \lpar 1 - S\,\Delta\rpar^{-1}.
\eqa
It is convenient to write
\bq
\lpar 1 - S\,\Delta\rpar_{\mu\nu} = K\,\drii{\mu}{\nu} + H\,p_{\mu}p_{\nu},
\eq
with $K$ and $H$ matrices defined by
\bqa
K =
\lpar
\ba{lr}
1-S^{\ssZ\ssZ}_0\Delta_{\ssZ} & -S^{\ssZ\ph}_0\Delta_{\ph} \\
-S^{\ph\ssZ}_0\Delta_{\ssZ}  & 1-S^{\ph\ph}_0\Delta_{\ph}
\ea
\rpar
\qquad
H = -
\lpar
\ba{lr}
S^{\ssZ\ssZ}_1\Delta_{\ssZ} & S^{\ssZ\ph}_1\Delta_{\ph} \\
S^{\ph\ssZ}_1\Delta_{\ssZ} & S^{\ph\ph}_1\Delta_{\ph}
\ea
\rpar
\eqa
Let $X = I\,\drii{\mu}{\nu} + J\,p_{\mu}p_{\nu}$, then one can easily find
a solution of the following form:
\bqa
I &=& K^{-1}, \qquad J = - \lpar K + p^2\,H\rpar^{-1}\,H\,K^{-1}, \nl
X_{\mu\nu} &=& \Big[ \drii{\mu}{\nu} - \lpar K+p^2\,H\rpar^{-1}\,H\,
p_{\mu}p_{\nu}\Big]\,K^{-1}.
\eqa
The re-summed propagators are
\bq
\bDelta^{ij}_{\mu\nu} = \Delta^i\,X^{ij}_{\mu\nu},
\eq
where we have found that
\bq
X_{\mu\nu} = K^{-1}\,P_{\mu\nu} + T^{-1}\,L_{\mu\nu}.
\eq
The Lorentz-tensors appearing in the previous equation are defined by
\bq
P_{\mu\nu} = \drii{\mu}{\nu} - {{p_{\mu}p_{\nu}}\over {p^2}}, \qquad
L_{\mu\nu} =  {{p_{\mu}p_{\nu}}\over {p^2}},
\eq
the remaining matrices being expressed as
\bqa
K^{ij} &=& \delta^{ij} - S^{il}_0\,\Delta_{lj}, \quad H^{ij} = - S^{il}_1\,
\Delta_{lj},  \nl
T^{ij} &=& K^{ij} + p^2\,H^{ij} = \delta^{ij} - t^{il}\,\Delta_{lj}.
\eqa
These results can be simplified, as we have done in the charged sector, by
using Ward identities for two-point functions, i.e.
\bqa
i\,p^{\mu}\,S^{\ph\ph}_{\mu\nu} &=& i\,t^{\ph\ph}\,p_{\nu} = 0,  \nl
i\,p^{\mu}\,S^{\ph\ssZ}_{\mu\nu} &=& i\,t^{\ph\ssZ}\,p_{\nu} = 0,  \nl
i\,p^{\mu}\,S^{\ssZ\ph}_{\mu\nu} &=& i\,t^{\ssZ\ph}\,p_{\nu} = 0,  
\eqa
giving $t^{\ph\ph} = t^{\ph\ssZ} = t^{\ssZ\ph} = 0$. Therefore the matrix 
$T^{-1}$ is diagonal, with entries
\bq
T^{-1}_{\ssZ\ssZ} = {1\over {1-t^{\ssZ\ssZ}\,\Delta_{\ssZ}}}, \qquad
T^{-1}_{\ph\ph} = 1.
\eq
The re-summed photon propagator becomes
\bq
\bDelta^{\ph\ph}_{\mu\nu} = \Big[
p^2 - S^{\ph\ph}_0 - {{\lpar S^{\ph\ssZ}_0\rpar^2}\over 
{p^2+\bzms-S^{\ssZ\ssZ}_0}}\Big]\,P_{\mu\nu} + \frac{1}{p^2}\,L_{\mu\nu}.
\label{ggresum}
\eq
The re-summed photon propagator, in the 't Hooft-Feynman gauge, satisfies
the same relation as the bare propagator,
\bq
p^{\mu}\,\Delta^{\ph\ph}_{\mu\nu} = p^{\mu}\,\bDelta^{\ph\ph}_{\mu\nu} =
\frac{p_{\nu}}{p^2}.
\eq
The $\zb-\ph$ re-summed transition is
\bq
\bDelta^{\ph\ssZ}_{\mu\nu} = {{S^{\ph\ssZ}_0}\over 
{\lpar p^2+\bzms-S^{\ssZ\ssZ}_0\rpar,\lpar p^2-S^{\ph\ph}_0\rpar - \lpar 
S^{\ph\ssZ}_0\rpar^2}}\,P_{\mu\nu}.
\label{gzresum}
\eq
It is convenient to re-express the results of \eqns{ggresum}{gzresum} in
terms of running couplings. We obtain
\bqa
\bDelta^{\ph\ph}_{\mu\nu} &=& \Big\{ - \frac{e^2(p^2)}{e^2\,p^2} +
\frac{\stws}{\ctws}\,\Big[ 1 - \frac{s^2(p^2)}{\stws}\Big]^2\,G_{\ssZ}(p^2)
\Big\}\,P_{\mu\nu} + \frac{L_{\mu\nu}}{p^2},  \nl
\bDelta^{\ph\ssZ}_{\mu\nu} &=& \frac{\stw}{\ctw}\,\Big[ 1 -
\frac{s^2(p^2)}{\stws}\Big]\,G_{\ssZ}(p^2)\,P_{\mu\nu}.
\eqa
The propagator-function $G_{\ssZ}$ is given in \eqn{defG}. For fermionic loops 
there is no transition $\ph-\hkn$, while
\bq
S^{\mu}_{\phi\ssZ} = \frac{g^2}{16\,\pi^2\ctws}\,\Sigma_{\phi\ssZ}\,ip^{\mu},
\quad \Sigma_{\phi\ssZ} = - \frac{1}{2}\,\frac{\mts}{M}\,\bff{0}{p^2}{\mt}{\mt}.
\eq
In our conventions, we denote with a subscript $\phi$ the $\hkn$-field.
Therefore, for non-zero $\mt$, we have to complete the diagonalization.
Let us introduce indices $a,b,\dots = 1,\dots,5$ with
\bq
S_{55} = S_{\phi\phi}, \quad S_{\mu 5} = S^{\mu}_{\ssZ\phi}, \quad
S_{5\nu} = S^{\nu}_{\phi\ssZ}.
\eq
One should remember that the transitions $S_{\vb\vb}$ have already been
re-summed in $\bDelta^{\ssZ\ssZ}_{\mu\nu}$ etc, therefore the additional
re-summation gives
\bqa
\tDelta_{ab} &=& \Delta_{ab} + \Delta_{ac}\,S_{cd}\,\Delta_{db} + \dots =
\Delta_{ac}\,X_{cb},  \nl
X_{ab} &=& \lpar 1 - S\,\Delta\rpar^{-1}_{ab}.
\eqa
In this equation one should use
\bq
\Delta_{\mu\nu} \equiv \bDelta^{\ssZ\ssZ}_{\mu\nu}, \qquad 
\Delta_{\phi\phi} \equiv \Delta_{\phi}.
\eq
The matrix to be inverted is
\bqa
\lpar
\ba{lr}
\drii{\mu}{\nu} & +i\,S_{\ssZ\phi}\,\Delta_{\phi}\,p^{\mu}  \\
-i\,S_{\ssZ\phi}\,p^{\mu}\,\bDelta^{\ssZ\ssZ}_{\mu\nu} & 1 - S_{\phi\phi}\,
\Delta_{\phi}
\ea
\rpar
\eqa
however, it is easily found that the following property holds:
\bq
p^{\mu}\,\bDelta^{\ssZ\ssZ}_{\mu\nu} = \Delta_{\ssZ}\,T^{-1}_{\ssZ\ssZ}\,p^{\nu}
= {1\over {p^2+\bzms - t^{\ssZ\ssZ}}}\,p^{\nu}.
\eq
There are additional Ward identities in the neutral sector, corresponding to
two-point functions, that give
\bqa
{}&{}& i\,p^{\mu}\,S^{\mu\nu}_{\ssZ\ssZ} + \bzm\,S_{\phi\ssZ}\,i\,p^{\nu} = 
\lpar t^{\ssZ\ssZ} + \bzm\,S_{\phi\ssZ}\rpar\,i\,p^{\nu} = 0,
\eqa
\bqa
{}&{}& i\,p_{\mu}\,S^{\mu\nu}_{\ssZ\ssZ}\,\lpar -i\,p_{\nu}\rpar +
i\,p_{\mu}\,\bzm\lpar - S_{\phi\ssZ}\rpar\,i\,p^{\mu} 
\nl
{}&-& i\,p_{\mu}\,\bzm\lpar  S_{\phi\ssZ}\rpar\,i\,p^{\mu} +
\bzms\,S_{\phi\phi}
\nl
{}&=& p^2\,t^{\ssZ\ssZ} +2\,p^2\,\bzm\,S_{\phi\ssZ} + \bzms\,S_{\phi\phi} = 0.
\eqa
The solution to these conditions is
\bq
S_{\phi\ssZ} = - \frac{\bzm}{p^2}\,S_{\phi\phi}, \qquad
t^{\ssZ\ssZ} = \frac{\bzms}{p^2}\,S_{\phi\phi},
\eq
allowing to express everything in terms of the $\hkn-\hkn$ transition.
In this way our matrix becomes
\bqa
\lpar
\ba{lr}
\drii{\mu}{\nu} & i\,b\,p_{\mu}  \\
-i\,b'\,p_{\nu}  & c
\ea
\rpar
\eqa
with entries given by
\bqa
b &=& - \frac{\bzm}{p^2}\,S_{\ssZ\phi}\,\Delta_{\ssZ}, \qquad 
b' = - \frac{\bzm}{p^2}\,{{S_{\ssZ\phi}}\over
{p^2+\bzms - t^{\ssZ\ssZ}}},  \nl
c &=& 1 - S_{\phi\phi}\,\Delta_{\ssZ},
\eqa
where we have used the fact that in the 't Hooft-Feynman gauge $\Delta_{\ssZ} = 
\Delta_{\phi} = 1/(p^2+\bzms)$.
After inversion of the matrix we obtain
\bqa
{1\over {c-bb'p^2}}\,
\lpar
\ba{lr}
\lpar c-bb'p^2\rpar\,\drii{\mu}{\nu} + bb'\,p_{\mu}p_{\nu} & -i\,b\,p_{\mu}  \\
+i\,b'\,p_{\nu}  & 1
\ea
\rpar
\eqa
This result allows us to write down the remaining re-summed propagators. They 
are:
\bqa
\tDelta^{\ssZ\ssZ}_{\mu\nu} &=& \Big[
p^2 + \bzms - S^{\ssZ\ssZ}_0 - {{\lpar S^{\ssZ\ph}_0\rpar^2}\over
  {p^2 - S^{\ph\ph}_0}}\Big]^{-1}\,P_{\mu\nu}
\nl
{}&+& {{p^2\lpar p^2 + \bzms - S_{\phi\phi}\rpar}\over
{\lpar p^2 + \bzms\rpar^2\,\lpar p^2 - S_{\phi\phi}\rpar
}}\,L_{\mu\nu},
\nl\nl
\tDelta_{\phi\phi} &=& {
{p^2\,\lpar p^2+\bzms\rpar - \bzms\,S_{\phi\phi}}\over
{\lpar p^2+\bzms\rpar^2\,\lpar p^2 - S_{\phi\phi}\rpar}}
\nl\nl
\tDelta^{\mu}_{\ssZ\phi} &=& - \tDelta^{\mu}_{\phi\ssZ} =
\,\,i\,p^{\mu}\,\bzm\,
{{S_{\phi\phi}}\over {\lpar p^2+\bzms\rpar^2\,\lpar p^2 - 
S_{\phi\phi}\rpar}}.
\eqa
\subsection{The running $\zb$ mass.}

With the results for transitions in the neutral sector, we can compute the 
amplitude for $e^+(p_+)\,e^-(p_-) \to f(q_-)\,\barf(q_+)$. There are seven
diagrams contributing, when we do not neglect the fermion masses. Accordingly,
the amplitude is
\bqa
M &=& g^2 Q_f\stws\,\gapu{\mu} \otimes \gapu{\mu}\,\bDelta^{\ph\ph}_{\mu\nu}
\nl
{}&+&
\frac{g^2\stw}{2\,\ctw}\,\gapu{\mu} \otimes \gapu{\nu}\,\lpar \tcif\,\gdp -
2\,Q_f\,\stws\rpar\,\bDelta^{\ph\ssZ}_{\mu\nu}
\nl
{}&-&
\frac{g^2 Q_f\,\stw}{2\,\ctw}\,\gapu{\mu}\,\lpar - \frac{1}{2}\,\gdp +
2\,\stws\rpar \otimes \gapu{\nu}\,\bDelta^{\ssZ\ph}_{\mu\nu}
\nl
{}&-&
\frac{g^2}{4\,\ctws}\,\gapu{\mu}\,\lpar - \frac{1}{2}\,\gdp + 2\,\stws\rpar
\otimes \gapu{\nu}\,\lpar \tcif\,\gdp - 2\,Q_f\,\stws\rpar\,
\tDelta^{\ssZ\ssZ}_{\mu\nu}
\nl
{}&-&
\frac{g^2}{2\,\ctw}\,\tcif\,\frac{\mf}{M}\,\gapu{\mu}\,\lpar - 
\frac{1}{2}\,\gdp + 2\,\stws\rpar \otimes \gfd\,\bDelta^{\ssZ\phi}_{\mu}
\nl
{}&+&
\frac{g^2}{2\,\ctw}\,\frac{\me}{2\,M}\,\gfd \otimes
\gapu{\mu}\,\lpar \tcif\,\gdp - 2\,Q_f\,\stws\rpar\,\tDelta^{\phi\ssZ}_{\mu}
\nl
{}&+&
\frac{g^2}{2}\,\tcif\,\frac{\me\mf}{M^2}\,\gfd \otimes \gfd\,\tDelta^{\phi\phi},
\eqa
where $\tcif$ is the third component of isospin.
We extract from the first four terms the part of the vector-vector propagators
proportional to $\drii{\mu}{\nu}$ and obtain the familiar $\ph \oplus \zb$
exchanges with running couplings. This part of the amplitude will be denoted 
by $M_{\delta\delta}$ and we write
\bq
M = M_{\delta\delta} + M_{\rm extra}.
\eq
From the use of the Dirac equation we find
\bqa
\sla{p} &\otimes& \sla{p} \,=\, \sla{p} \,\otimes\, \sla{p}\,\gdp \,=\, 
\sla{p}\,\gdp \,\otimes\, \sla{p} \,=\, 0,  \nl
\sla{p}\,\gdp &\otimes& \sla{p}\,\gdp \,=\, 4\,\me\mf\,\gfd \,\otimes\, \gfd,
\nl
\sla{p}\,\gdp &\otimes& \gfd \,=\, -2\,i\,\me\,\gfd \,\otimes\, \gfd, \qquad
\gfd \,\otimes\, \sla{p}\,\gdp \,=\, 2\,i\,\mf\,\gfd \,\otimes\, \gfd.
\eqa
Therefore, collecting the various terms, we obtain the following expression
\bqa
M_{\rm extra} &=& g^2\,\me\mf\,\tcif\,\gfd \otimes \gfd\,\Big\{
\frac{1}{2\,\ctws}\,\Big[ -\,G_{\ssZ} + 
{{p^2\lpar p^2+\bzms-S_{\phi\phi}\rpar}\over 
{\lpar p^2+\bzms\rpar^2\,\lpar p^2-S_{\phi\phi}\rpar}}\Big]\,\frac{1}{p^2}
\nl
{}&+& \frac{1}{\ctw}\,\frac{\bzm}{M}\,{{S_{\phi\phi}}\over
{\lpar p^2+\bzms\rpar^2\,\lpar p^2-S_{\phi\phi}\rpar}} +
\frac{1}{2\,M^2}\,{{p^2\,\lpar p^2+\bzms-\frac{\bzms}{p^2}\,S_{\phi\phi}\rpar}
\over {\lpar p^2+\bzms\rpar^2\,\lpar p^2-S_{\phi\phi}\rpar}}\Big\}.
\nl
\eqa
\begin{guess}
This result can be simplified considerably if we introduce a running 
$\zb$-mass. We define
\end{guess}
\bq
M_{\rm extra} = \frac{g^2}{2\,\ctws}\,\me\mf\,\tcif\,\gfd \otimes \gfd\,
{{G_{\ssZ}}\over {\bzms(p^2)}},
\eq
so that the entire amplitude can be written as the Born-like sum of $\ph$
and $\zb$ exchange, but with $g^2 \to g^2(p^2), \stws \to s^2(p^2)$ etc
and with an {\em effective} $\zb$ propagator given by
\bq
\Delta^{\zb\zb,\rm eff}_{\mu\nu} = G_{\ssZ}(p^2)\,\Big[
\drii{\mu}{\nu} + {{p_{\mu}p_{\nu}}\over {\bzms(p^2)}}\Big].
\eq
For a massless internal world we obtain a rather simple result:
\bqa
{1\over {\bzms(p^2)}} &=& \frac{1}{p^2}\,\Big[ {1\over {\bzms\,G_{\ssZ}(p^2)}} -
1\Big]  \nl
{}&=& \frac{g^2(\sZ)}{g^2(p^2)}\,\frac{c^2(p^2)}{c^2(\sZ)}\,\frac{1}{\sZ}.
\eqa
which satisfies the relation $\bzms(\sZ) = \sZ$. 

As expected the $\zb$ and $\wb$ running masses satisfy the same relation as 
the bare masses,
\bq
{1\over {\bzms(p^2)}} = {{c^2(p^2)}\over {M^2(p^2)}},
\eq
while complex poles satisfy
\bq
\sZ = \frac{g^2(\sZ)}{g^2(\sW)}\,\frac{\sW}{c^2(\sZ)}.
\eq
The extension to non-zero $\mt$ is straightforward.

\section{Projection into the Imaginary Fermion-Loop.}

All relevant Ward identities are linear in the vertex functions and the
inverse propagators. As a consequence the real and imaginary parts
fulfill the Ward identities separately.
Therefore, a simplified minimal approach to incorporate the finite width
while ensuring both U(1) and  SU(2) gauge invariance consists in only taking
into account the imaginary parts of fermionic corrections.

Having the full Fermion-Loop at our disposal we can project the result into
this minimal version. The relevant formulas are listed below. First, the
two-point functions. With $\sman=-\pmoms$ we write:
\bqa
\Imb\bff{0}{\pmoms}{0}{0} &=& \pi\thf{\sman}, 
\quad
\Imb\bff{0}{\pmoms}{0}{\mlone} = 
\pi\lpar 1- \frac{\mlones}{\sman}\rpar\,\thf{\sman-\mlones}, 
\nl
\Imb\bff{0}{\pmoms}{\mlone}{\mlone} &=& 
\pi\lpar 1- 4\frac{\mlones}{\sman}\rpar^{1/2}\,\thf{\sman-4\mlones}, 
\eqa
Furthermore, when $\pmoms = - s_p$ and $s_p$ is a complex pole, the relevant 
functions become as follows:
\bqa
\bff{0}{-s_p}{0}{0} &=& \Ddr + 2 -
\Big[ \ln\lpar\frac{-s_p}{\tHss}\rpar -2\,i\,\pi\Big], \nl
\bff{0}{-s_p}{0}{m} &=& \Ddr + 2 - \ln\frac{m^2}{\tHss}
\lpar 1 - \frac{m^2}{s_p}\rpar\,\Big[ \ln\lpar 1 - \frac{s_p}{m^2}\rpar  \nl
{}&-& 2\,i\,\pi\,\thf{M^2-m^2}\Big],  \nl
\bff{0}{-s_p}{m}{m} &=& \Ddr + 2 - \ln\frac{m^2}{\tHss} - 
\beta\,\Big[ \ln\frac{\beta+1}{\beta-1} - 2\,i\,\pi\,
\thf{M^2-4\,m^2}\Big],
\nl
\eqa
where $M^2 = \Reb\lpar s_p\rpar$ and $\beta^2 = 1 - 4\,m^2/s_p$.
After scalarization the vertices are combinations of $\sbff{0}$-functions and 
of $\scff{0}$-functions. The imaginary part of a $\scff{0}$-function, in the 
general case, is not particularly simple but it is obtained from its
complete expression~\cite{kn:book}. An alternative way of deriving it is
to remember that the imaginary part follows from the sum over all cuts
of the scalar three-point function. In the generalization of the Fermion-Loop
and in proving the Ward identity one should remember to sum over the three 
cuts, i.e., the imaginary part of the triangle $\vb\wb\wb$ ($\vb = \zb,\ph$) 
is the sum of all cuts over intermediate states. 
Here we give one explicit example where we consider the function
$\cff{0}{\pones}{\ptwos}{P^2}{\mone}{\mtwo}{\mtre}$.
The cuts give
\bqa
C_{13} &=& \intmomi{4}{q}\,\thf{-q_0}\,\thf{q_0+P_0}\,
{{\delta\lpar q^2+m^2_1\rpar\,\delta\lpar\lpar
q+P\rpar^2+m^2_3\rpar}\over {\lpar q+p_1\rpar^2+m^2_2}},  \nl
C_{23} &=& \intmomi{4}{q}\,\thf{q_0+p_{10}}\,\thf{-q_0-P_0}\,
{{\delta\lpar\lpar q+p_1\rpar^2+m^2_2\rpar\,
\delta\lpar\lpar q+P\rpar^2+m^2_3\rpar}
\over {q^2+m^2_1}},  \nl
C_{12} &=& \intmomi{4}{q}\,\thf{q_0}\,\thf{-q_0-p_{10}}\,
{{\delta\lpar q^2+m^2_1\rpar\,\delta\lpar\lpar q+p_1\rpar^2+m^2_2\rpar}
\over {\lpar q+P\rpar^2+m^2_3}},
\eqa
where the subscript $ij$ specify the cut: $C_{ij}$ is the diagram where 
the cut propagators correspond to internal masses $m_i, m_j$.
If we introduce the following quantities:
\bqa
a^2 &=& \frac{1}{4\,s}\,\lambda\lpar -s,p^2_1,p^2_2\rpar,  \nl
b &=& - \frac{1}{2}{{-s+p^2_1-p^2_2}\over {\sqrt{s}}},  \nl
c &=& \frac{1}{2\,\sqrt{s}}\,\lpar m^2_3-m^2_1-s\rpar,  \nl
d^2 &=& \frac{1}{4\,s}\,\lambda\lpar s,m^2_1,m^2_3\rpar,
\eqa
with $P^2 = -s$ and $\lambda$ the K\'allen $\lambda$-function, then we 
obtain
\bq
C_{13} = \frac{\pi}{4\,a\sqrt{s}}\,\thf{s-\lpar m_1+m_3\rpar^2}\,
\ln{{m^2_2-m^2_1 + p^2_1 -2\,cb + 2\,ad}\over {m^2_2-m^2_1+p^2_1-2\,cb - 
2\,ad}},
\eq
where the $\theta$-function reflects the positivity of 
$\lambda(s,m^2_1,m^2_3)$ and the constraint $s \,\ge\, \mid m^2_1-m^2_3\mid$.
Similar results can be derived for the other two cuts.
Note, however, that once the massive top contributions are included this 
scheme is in fact not much easier than the complete Fermion-Loop scheme.

\section{Conclusions.}

The Fermion-Loop scheme developed in~\cite{kn:bhf1} and refined in
\cite{kn:bhf2} makes the approximation of neglecting all masses for
the incoming and outgoing fermions in the processes $e^+e^- \to n\,$fermions.

In this paper we have given the construction of a complete, massive-massive,
Fermion-Loop scheme, i.e., a scheme for incorporating the finite-width effects 
in the theoretical predictions for tree-level, LEP~2 and beyond, processes.
Therefore, we have generalized the scheme to incorporate external non-conserved
currents. 

We work in the 't Hooft-Feynman gauge and create all relevant
building blocks, namely the vector-vector, vector-scalar and scalar-scalar
transitions of the theory, all of them one-loop re-summed. The loops,
entering the scheme, contains fermions and, as done before in~\cite{kn:bhf2},
we allow for a non-zero top quark mass inside loops. There is a very simple
relation between re-summed transitions and running parameters, since
Dyson re-summation is most easily expressed in terms of running coupling
constants and running sinuses.

In our generalization, we have found particularly convenient to introduce
additional running quantities. They are the running masses of the vector
bosons, $\bzms(p^2) = M^2(p^2)/c^2(p^2)$, formally connected to the location
of the $\wb$ and $\zb$ complex poles.

After introducing these running masses, it is relatively simple to prove that
all $\Smat$-matrix elements of the theory assume a very simple structure.
Coupling constants, sinuses and masses are promoted to running quantities
and the $\Smat$-matrix elements retain their Born-like structure, with
running parameters instead of bare ones, and vector-scalar or scalar-scalar
transitions disappear if we employ unitary-gauge--like vector boson propagators
where the masses appearing in the denominator of propagators are the running 
ones.

Renormalization of ultraviolet divergences has been easily extended to the 
massive-massive case by showing that all ultraviolet divergent parts of
the one-loop vertices, $\ph\wb\wb, \ph\wb\phi, \ph\phi\wb$ and
$\ph\phi\phi$ for instance, are proportional to the lowest order part.
Therefore, the only combinations that appear are of the form $1/g^2 + 
\vb\vb\vb\,$vertex or $M^2/g^2 + \vb\vb\phi\,$ vertex etc.
All of them are, by construction, ultraviolet finite.

Equipped with our generalization of the Fermion-Loop scheme, we have been
able to prove the fully massive U(1) Ward identity which guarantees that
our treatment of the single-$\wb$ processes is the correct one. As a 
by-product of the method, the cross-section for single-$\wb$ production
automatically evaluates all channels at the right scale, without having
to use ad hoc re-scalings and avoiding the approximation of a unique scale
for all terms contributing to the cross-section. 

The generalization of the Fermion-Loop scheme goes beyond its, most
obvious, application to single-$\wb$ processes and allows for a gauge
invariant treatment of all $e^+e^- \to n\,$fermion processes with a
correct evaluation of the relevant scales. Therefore, our scheme can
be applied to several other processes like $e^+e^- \to \zb\ph^*$ and,
in general to all $e^+e^- \to 6\,$fermion processes, with the inclusion
of top quarks.

\end{document}